\documentclass{aa}

\usepackage{lscape}
\usepackage{graphicx}
\usepackage{rotating}
\usepackage{array}
\usepackage{amsfonts}
\usepackage{aalongtable}

\renewcommand{\d}{\, {\rm d}}

\newcommand{\astroph}[1]{\textbf{[#1]}}

\usepackage[sort]{natbib}
\usepackage{txfonts}

\begin{document}
\title{MAMBO Mapping of Spitzer c2d Small Clouds and
  Cores\thanks{Based on observations with the IRAM 30m-telescope.}}

  \author{
    Jens Kauffmann\inst{1,2,3}
    \and
    Frank Bertoldi\inst{4}
    \and
    Tyler L.\ Bourke\inst{3}
    \and
    Neal J.\ Evans II\inst{5}
    \and
    Chang Won Lee\inst{3,6}
  }
  
  \offprints{J.\ Kauffmann, jkauffmann@cfa.harvard.edu}
  
  \institute{
    Max-Planck-Institut f\"ur Radioastronomie, Auf dem H\"ugel 69,
    D-53121 Bonn, Germany
    \and
    Initiative in Innovative Computing at Harvard
    University, 60 Oxford Street, 02138 Cambridge, MA, USA
    \and
    Harvard-Smithsonian Center for Astrophysics, 60 Garden Street,
    02138 Cambridge, MA, USA
    \and
    Argelander Institut f\"ur Astronomie, Universit\"at Bonn,
    Auf dem H\"ugel 71, D-53121 Bonn, Germany
    \and
    Department of Astronomy, University of Texas at Austin,
    1 University Station C1400, 78712 Austin, TX, USA
    \and
    Korea Astronomy and Space Science Institute, 61-1 Hwaam-dong,
    Yusung-gu,
    Daejeon 305-348, Korea
  }
  
  \date{Received xxx; accepted xxx}

\abstract{}{
  To study the structure of nearby ($< 500 ~ \rm pc$)
  dense starless and star-forming cores with the particular goal to
  identify and understand evolutionary trends in core
  properties, and to explore the nature of Very Low
  Luminosity Objects ($\le 0.1 \, L_{\odot}$; VeLLOs).
}{
  Using the MAMBO bolometer array, we create maps
  unusually sensitive to faint (few mJy per beam) extended
  ($\approx 5 \arcmin$) thermal dust continuum emission at
  $1.2 ~ \rm mm$ wavelength. Complementary information on embedded
  stars is obtained from Spitzer, IRAS, and 2MASS. \astroph{The
    journal article will come with FITS files of the maps.}
}{
  Our maps are very rich in structure, and we characterize
  extended emission features (``subcores'') and compact intensity
  peaks in our data separately to pay attention to this complexity. We
  derive, e.g., sizes, masses, and aspect ratios for the subcores, as
  well as column densities and related properties for the
  peaks. Combination with archival infrared data then enables the
  derivation of bolometric luminosities and temperatures, as well as
  envelope masses, for the young embedded stars.
}{
  Starless and star-forming cores occupy the same parameter space in
  many core properties; a picture of dense core
  evolution in which any dense core begins to actively form stars once
  it exceeds some fixed limit in, e.g., mass, density, or both, is
  inconsistent with our data. A concept of
  \emph{necessary conditions} for star formation appears to provide a
  better description: dense cores fulfilling certain conditions
  \emph{can} form stars, but they do not need to, respectively have
  not done so yet. Comparison of
  various evolutionary indicators for young stellar objects in
  our sample (e.g., bolometric temperatures) reveals inconsistencies
  between some of them, possibly
  suggesting a revision of some of these indicators. Finally, we
  challenge the notion that VeLLOs form in cores not expected to
  actively form stars, and we present a first systematic study
  revealing evidence for structural differences between starless and
  candidate VeLLO cores.
}


\maketitle
%

\section{Introduction\label{sec-survey:introduction}}
Stars form from dense gas ($\gg 10^5 ~ \rm cm^{-3}$). Such gas is,
e.g., found in discrete
nearby ($\lesssim 500 ~ \rm pc$) cold ($\approx 10 ~ \rm K$)
small-scale ($\lesssim 0.1 ~ \rm pc$) condensations referred
to as dense cores \citep{myers1983:dense_cores}. These are thought to
be the sites where low-mass stars form \citep{myers1983:nh3}. Their
properties provide the initial conditions for star formation. It is
thus necessary to understand the physical state of dense cores in
order to be able to understand the star formation process in detail.

Some of the many physical parameters of dense cores are their sizes,
masses, and densities. Given that thermal dust continuum emission at
$\approx 1 ~ \rm mm$ wavelength is a better mass tracer than molecular
emission lines (e.g., \citealt{tafalla2002:depletion}), and that
modern bolometer cameras enable complete maps of the dust
emission from dense cores to be made within a short time, dust
emission maps are a
prime tool to study the above core parameters. Dust emission surveys thus
enable a systematic overview of dense core properties.

In recent years several surveys studied dense cores in large (up to
several $10 ~ \rm pc$) and massive (up to several $10^3 \, M_{\odot}$)
complexes of molecular clouds (\citealt{motte1998:ophiuchus};
\citealt{johnstone2000:rho_ophiuchi, johnstone2001:orion_b,
  johnstone2004:extinction_treshold_ophiuchus, johnstone2006:orion_b,
  johnstone2006:orion_a}; \citealt{hatchell2005:perseus};
\citealt{enoch2006:perseus,enoch2007:cloud_comparison};
\citealt{young2006:ophiuchus}; \citealt{stanke2006:rho_ophiuchi}).
Most cores in these clouds are in groups and clusters (e.g.,
\citealt{enoch2007:cloud_comparison}). Although it is believed that
most stars in the galaxy form in such environments
\citep{magnani1995:high_latitude_t-tauri_stars}, two issues make it
difficult to study the fundamental physics of star formation in such
environments. First, cores can be confused in crowded regions forming
clusters; it becomes difficult to identify and study individual cores
(see, e.g., \citealt{motte1998:ophiuchus}; see also
\citealt{ward-thompson2006:cores_ppv} for illustrative examples).
Second, cores might interact in crowded regions; more parameters than
relevant in isolated cores influence the evolution of dense cores.
These problems can be avoided by studying dense cores in relative
isolation (\citealt{clemens1988:small_clouds};
\citealt{bourke1995:bhr}) instead of cores hosted in larger (up to
$\gtrsim 100 ~ \rm pc$) complexes of giant molecular clouds
\citep{blitz1993:molecular_clouds}.

The present study therefore focuses on isolated cores. We have
observed both starless cores and cores
with embedded stars, covering a large range in size, mass, central
density and (supposedly) evolutionary status. This creates a unique
database for detailed studies of dependencies between various dense
core parameters and their relation to dense core evolution.

\subsection{The Need for a new Dust Emission Survey}
Several surveys of isolated cores with and without active star
formation have been conducted in the past $\approx 15$ years (e.g.,
\citealt{motte2001:protostars}, \citealt{visser2002:lynds_clouds},
\citealt{kirk2005:scuba_survey},
\citealt{young2006:c2d_scuba}, \citealt{wu2007:c2d_sharc-ii}). They
have covered a large number of dense cores in a broad variety of
physical states.  While these surveys in principle should yield a
comprehensive database on the properties of dense cores, this is in
fact true only to a limited extent. Many existing studies were
relatively insensitive to faint emission (because of noise) and
structure on scales beyond $1 \arcmin ~ {\rm to} ~ 2 \arcmin$ (due to
artifacts, e.g., due to `jiggle maps').

State-of-the-art bolometer cameras nowadays allow for the reliable
imaging of faint large-scale ($\gtrsim 5 \arcmin$) extensions of
sources initially --- if at all --- only detected in their brightest
intensity peaks. This urged us to conduct a new comprehensive dust
emission survey towards isolated dense cores at $\approx 1 ~ \rm mm$
wavelength using up-to-date instrumentation. For this we use the
Max-Planck Millimetre Bolometer (MAMBO) arrays
\citep{kreysa1999:mpifr_bolometers} at the IRAM 30m-telescope. This
combination of telescope and receiver was probably the most sensitive
facility for mapping of extended dust emission available before the
recent emergence of LABOCA (when considering the levels of extended
artifacts and statistical noise). The research presented here is the
first census with MAMBO that covers $\ge 10$ starless dense cores.

A further motivation of our survey is the demand for complementary
data on dense cores for the recent Spitzer Space Telescope imaging
surveys of dense cores. Such data is needed to, e.g., put the
properties of the stellar content probed by Spitzer in context with
dense core properties like density and mass; the latter can usually
not be derived from the Spitzer data. In particular the Spitzer Legacy
Program ``From Molecular Cores to Planet Forming Disks'' (AKA.\
``Cores to Disks'' or ``c2d''; \citealt{evans2003:c2d}), which imaged
82 nearby ($\le 500 ~ \rm pc$) isolated dense cores as part of
its agenda, stimulated a number of coordinated surveys of the c2d
cores covering many spectral windows. Our survey --- the c2d
MAMBO survey --- is one of four new dust continuum emission surveys of
isolated cores carried out in the c2d framework (with
\citealt{young2006:c2d_scuba}, \citealt{wu2007:c2d_sharc-ii}, and
Brede et al., in prep., being the other three).

\subsection{Specific Aim of the c2d MAMBO Survey\label{sec-survey:aim}}
Dust emission data can be used to address a number of open issues in
star formation research. In general, stars are thought to form
from very dense molecular cores ($\rm H_2$ densities
$\gtrsim 10^5 ~ \rm cm^{-3}$). In particular, the spatial mass
distribution is thought to govern the stability of dense cores against
gravitational collapse (e.g., \citealt{mckee1999:mpp}). One might thus
hope to derive criteria to be fulfilled to allow star formation in
dense cores by comparing the mass distributions of starless cores with
those of cores actively forming stars. This could lead to a better
understanding of the underlying star formation physics. The c2d MAMBO
survey is ideally suited for such comparative studies since it probes
the mass distribution in cores with and without ongoing star formation
at high sensitivity (see Sec.\ \ref{sec-survey:star-form_ability} for
details).

Also, the evolution of young stellar objects (or ``protostars'';
hereafter YSOs), i.e., stars that are surrounded by significant
amounts of the matter from which they did or do form, can be studied
with dust continuum emission data. These allow the investigation of the
relation between the broadband spectral properties, the structure of
the circumstellar envelope, and present models of evolutionary schemes
(see Sec.\ \ref{sec-survey:evol_protostars} for details).

Our sample includes four candidate Very Low Luminosity
Objects (VeLLOs; see \citealt{kauffmann2005:vellos} and
\citealt{difrancesco2006:ppv}), enigmatic objects
of apparent protostellar nature but unusually low luminosity that were
recently discovered by c2d (\citealt{young2004:l1014},
\citealt{bourke2005:l1014-outflow}, \citealt{dunham2006:iram_04191}; cf.\
\citealt{andre1999:iram04191}). They are defined as objects with
internal luminosities (i.e., not including the dense core luminosity
from interstellar heating due to, e.g., the interstellar radiation
field) $\le 0.1 \, L_{\odot}$ that are embedded in dense cores. Some
of these objects have been interpreted as very young ($\lesssim 10^4$
years; \citealt{andre1999:iram04191}) YSOs of stellar final mass
(i.e., $\ge 0.08 \, M_{\odot}$).  Based on their low accretion rates,
others are thought to be young objects in the process of growing to
substellar final mass (\citealt{young2004:l1014},
\citealt{bourke2005:l1014-outflow}; see
\citealt{dunham2006:iram_04191} for a summarizing discussion). The c2d
MAMBO survey is the first study with a sensitive and homogeneous data
set on several starless, YSO, and VeLLO natal cores. This allows
comparative studies between these core types to better understand how
VeLLO natal cores differ from others (see Sec.\
\ref{sec-survey:form_and_evol_vellos} for details).

\subsection{Structure of the present Study}
In Sec.\ \ref{sec-survey:sampel_and_obs} we begin with an introduction
to our sample, provide a summary of the observations, and detail our
data reduction. This includes a description of a new iterative data
reduction scheme designed to mitigate problems affecting the
reconstruction of maps of weak extended emission. Section
\ref{sec-survey:results} provides a first discussion of the source
properties and an overview of the source identification and
quantification strategies. YSO properties from supplemental data are
derived in Sec.\ \ref{sec-survey:protostellar_data}. The main
discussion of the data follows in Sec.\ \ref{sec-survey:analysis},
where we in particular study the structure and evolution of dense
cores in general (Sec.\ \ref{sec-survey:core_evolution}), YSO cores
(Sec.\ \ref{sec-survey:evol_protostars}), and VeLLOs (Sec.\
\ref{sec-survey:form_and_evol_vellos}). We summarise our study in
Sec.\ \ref{sec-survey:summary}. Appendix \ref{appendix:dust_em_prop}
discusses the standard dust emission properties adopted by the c2d
collaboration to derive the mass distribution from observed dust
emission maps.

\section{Sample, Observations, and Data Reduction\label{sec-survey:sampel_and_obs}}
\subsection{Sample Selection}
Our sample of 38 cores (Table \ref{tab-survey:c2d-MAMBO_sampel})
is drawn from the initial target list of the c2d
Spitzer survey of isolated dense cores \citep{evans2003:c2d}. This
list contains dense cores within about $400 ~ \rm pc$ from the sun
that are smaller than $\approx 5 \arcmin$ and were mapped in dense gas
tracers before 
the start of the c2d surveys. These cores are nearby enough to
allow for the detection of intrinsically faint embedded sources. Their
small apparent size allows them to be mapped in a reasonable
time. This sample was drawn from the compilations and surveys of
\citet[ for $\rm NH_3$ maps]{jijina1999:database},
\citet[ for CS and $\rm N_2H^+$ maps]{lee2001:contr_survey}, and
\citet[ for $\rm N_2H^+$ maps]{caselli2002:n2h+},
and completed by adding individual cores. Due to time
constraints not all sources in our sample were finally
observed by Spitzer as part of c2d. Some further cores have or will be
covered as part of the Spitzer Gould Belt
survey\footnote{http://www.cfa.harvard.edu/gouldbelt/index.html}, or
have been covered in other programs (see Table
\ref{tab-survey:c2d-MAMBO_sampel} for details).

The initial c2d target list contains about 150 cores. At the start of
the c2d survey, MAMBO was the
most efficient bolometer camera available for sensitive mapping of
large fields. Therefore, those cores in this list that were expected
or known to have faint or extended dust continuum emission, or both,
and are visible from Spain,
were mapped by MAMBO. The other cores were observed in other
surveys with SCUBA on the JCMT \citep{young2006:c2d_scuba}, 
SHARC~\textsc{ii} on the CSO \citep{wu2007:c2d_sharc-ii}, and with SIMBA on
the SEST (Brede et al., in prep.). We obtained usable MAMBO
data for 38 cores, which are listed in Table 
\ref{tab-survey:c2d-MAMBO_sampel}. Of these cores 21 are starless
to our present knowledge, 4 contain VeLLO candidates (out of which 2,
those in L1014 and L1521F, can be considered proven VeLLOs), and 13 contain
YSOs (see Sec.\ \ref{sec-survey:ass_protostars} for a discussion of the
association between YSOs and dense cores). Spitzer data is
available for 32 cores (i.e., 3/4) of the MAMBO sample (see Table
\ref{tab-survey:c2d-MAMBO_sampel} for Spitzer observation references).
The MAMBO sample covers 25 out of the 82 cores observed by c2d, and
thus roughly 1/3 of the total c2d sample. In total 13 out of the 17
YSO and VeLLO candidate cores observed by us using MAMBO were targeted
by Spitzer. However, only for 8 of these (less than half) there is c2d
data and therefore homogeneous Spitzer photometry.

\begin{table*}
\caption{Dense cores covered by the c2d MAMBO survey. The table lists,
  from left to right, the core name adopted within the c2d
  collaboration, names for the cores adopted by other authors, the
  core distance, the type of the core (i.e., starless, YSO,
  or VeLLO), the approximate galactic environment of the dense
  core, and whether Spitzer data is available for the core. The last
  column gives a
  rough description of the morphology of the subcores (i.e.,
  extended substructures; see Sec.\ \ref{sec-survey:id_and_quant})
  within each core   using the c2d standard
  morphology keywords (\citealt{enoch2006:perseus}; see Sec.\
  \ref{sec-survey:prop_overview} for a
  description).\label{tab-survey:c2d-MAMBO_sampel}}
\begin{footnotesize}
\begin{center}
\begin{tabular}{llllllllll}
\hline \hline
\rule{0ex}{3Ex}c2d Name & Other Names & Distance$^a$ & Type$^b$ & 
  Region$^c$ & Spitzer Data$^d$ & Morphology Keywords$^e$\\
  & & pc\vspace{1ex}\\ \hline
L1355      & & $200 \pm 50^1$ & s.l.\ & Cepheus Flare & c2d &
  elongated, extended\\
L1521B-2   & B217B& $140 \pm 10^2$ & s.l.\ & Taurus & c2d &
  multiple, elongated, extended\\
L1521F     & & $140 \pm 10^2$ & Ve.\ & Taurus & c2d &
  elongated, extended\\
L1521-2    & & $140 \pm 10^2$ & s.l.\ & Taurus & c2d &
  round, extended\\
L1524-4    & & $140 \pm 10^2$ & s.l.\ & Taurus & c2d &
  multiple, round to elongated, extended\\
B18-1      & TMC-2A & $140 \pm 10^2$ & s.l.\ & Taurus & c2d &
  multiple, round to elongated, extended\\
TMC-2      & L1529 & $140 \pm 10^2$ & s.l.\ & Taurus & c2d &
  multiple, elongated, extended\\
B18-2      & & $140 \pm 10^2$ & s.l.\ & Taurus & c2d &
  multiple, elongated, extended\\
B18-4  & TMC-3 & $140 \pm 10^2$ & YSO & Taurus & IRAC, MIPS &
  multiple, elongated, extended\\
B18-5  & & $140 \pm 10^2$ & s.l.\ & Taurus & c2d &
  multiple, elongated, extended\\
TMC-1C     & B220 & $140 \pm 10^2$ & s.l.\ & Taurus & c2d &
  multiple, elongated, extended\\
TMC-1      & L1534 & $140 \pm 10^2$ & s.l.\ & Taurus & c2d &
  multiple, elongated, extended\\
L1507A     & & $140 \pm 10^2$ & s.l.\ & Taurus & c2d &
  multiple, round to elongated, extended\\
L1582A     & B32 & $ 400 \pm 40^3$ & YSO & $\lambda$ Orionis & IRAC, MIPS &
  multiple, elongated, extended\\
IRAS05413  & HH212-mm & $450 \pm 50^4$ & YSO & Orion B & IRAC, MIPS &
  multiple, round to elongated, extended\\
L1622A     & & $450 \pm 50^4$ & s.l.\ & Orion East & IRAC, MIPS &
  multiple, elongated, extended\\
L183       & L134N & $110 \pm 10^5$ & s.l.\ & north of Ophiuchus &
IRAC, MIPS & multiple, round to elongated, extended\\
L438       & CB119 & $270 \pm 50^6$ & s.l.\ & Aquila Rift & c2d &
  round, extended\\
L492       & CB128 & $270 \pm 50^6$ & s.l.\ & Aquila Rift & c2d &
  elongated, extended\\
CB188      & L673-1 & $300 \pm 100^7$ & YSO & Aquila Rift, Cloud B & c2d &
  elongated, extended\\
L673-7     & & $300 \pm 100^7$ & Ve.\ & Aquila Rift, Cloud B & c2d &
  multiple, round to elongated, extended\\
L675       & CB193 & $300 \pm 100^7$ & s.l.\ & Aquila Rift, Cloud B & c2d &
  elongated, weak\\
L1100      & CB224 & $400 \pm 50^8$ & YSO & northern Cyg OB7 & &
  elongated, extended\\
L1041-2    & & $400 \pm 50^8$ & YSO & northern Cyg OB7 & &
  multiple, round to elongated, extended\\
L1148      & L1147 & $325 \pm 25^9$ & Ve.\ & Cepheus Flare & c2d &
  multiple, elongated, extended\\
L1155E     & L1158 & $325 \pm 25^9$ & s.l.\ & Cepheus Flare & c2d &
  multiple, elongated, extended\\
L1082C-2   & & $400 \pm 50^{10}$ & s.l.\ & Cyg OB7 & &
  elongated, extended\\
L1082C     & & $400 \pm 50^{10}$ & YSO & Cyg OB7 & &
  elongated, extended\\
L1082A     & B150 & $400 \pm 50^{10}$ & YSO & Cyg OB7 & &
  multiple, round to elongated, extended\\
L1228      & & $200 \pm 50^{11}$ & YSO & Cepheus Flare & c2d &
  round, extended\\
Bern48     & RNO129 & $200 \pm 50^{11}$ & YSO & Cepheus Flare & c2d &
  round, extended\\
L1172A     & & $288 \pm 25^9$ & YSO & Cepheus Flare & IRAC, MIPS &
  multiple, round, extended\\
L1177      & CB230 & $288 \pm 25^{9}$ & YSO & Cepheus Flare & IRAC &
  round, extended\\
L1021      & & $250 \pm 50^{12}$ & s.l.\ & Cyg OB7 & c2d &
  elongated, extended\\
L1014      & & $250 \pm 50^{12}$ & Ve.\ & Cyg OB7 & c2d &
  elongated, extended\\
L1103-2    & L1106 & $250 \pm 50^{12}$ & s.l.\ & Cyg OB7 & &
  elongated, extended\\
L1251A     & & $300 \pm 50^{13}$ & YSO & Cepheus Flare & c2d &
  multiple, round to elongated, extended\\
L1197      & & $300 \pm 50^{14}$ & s.l.\ & Cepheus Flare & c2d &
  elongated, extended\\
\hline
\end{tabular}
\end{center}
\rule{0ex}{3Ex}Notes:\\
a) Standard c2d distances (Bourke, priv.\ comm.), derived by reddening
methods or association (in position-velocity space) to objects of
known distance. The individual references are: [1]
\citet{obayashi1998:l1333}, via association; [2]
\citet{kenyon1994:tau_auriga}; [3] \citet{murdin1977:l_orionis}; [4]
\citet{genzel1989:orion}, via association to L1630; [5]
\citet{franco1989:high-latitude_distances}; [6]
\citet{straizys2003:distances}; [7]
\citet{herbig1983:hh_proper-motions}, via association to L673; [8]
\citet{dobashi1994:cygnus}, and association to L1041; [9]
\citet{straizys1992:distances}, and association with L1172; [10] via
reddening, following \citet{bourke1995:bhr}; [11]
\citet{kun1998:cep_flare}; [12] \citet{pagani1996:b164}, via
association to B164; [13] \citet{kun1993:l1251}; and [14]
\citet{yonekura1997:cepheus_cassiopeia}, via association to L1192/L1200.\\
b) The abbreviations mark starless cores (`s.l.'; these have no known
star within the core boundaries defined in Sec.\
\ref{sec-survey:id_and_quant}), YSO cores (`YSO'; YSOs within the core
boundaries), and VeLLO cores (`Ve.'; VeLLOs within the core
boundaries). The cores L1521-2, B18-1, TMC-1, and L1622A are excluded
from the YSO category since the YSOs are off the main core body. See
Sec.\ \ref{sec-survey:ass_protostars} for a discussion of the
association between YSOs and dense cores.\\
c) The designation of regions roughly follows
\citet{dame1987:co_survey}, \citet{maddalena1986:ori_mon},
\citet{yonekura1997:cepheus_cassiopeia}, and \citet{dobashi1994:cygnus}.\\
d) Cores with data from the c2d project \citep{evans2003:c2d} are
labeled `c2d'. For other cores existing Spitzer imaging data is
indicated in the table, and the Spitzer program numbers and PIs are:
GO 3584 (Padgett) and GO 20302 (Andre) for B18-4; GO 20339 (Stauffer)
for L1582A; GO 3315 (Noriega-Crespo) and GTO 47 (Fazio) for IRAS05413;
GTO 47 (Fazio) for L1622A; GTO 94 (Lawrence) and GTO 53 (Rieke) for
L183; L 30574 (Allen)
and IOC 717 (Rieke) for L1172A; and GTO 124 (Gehrz) for L1177.\\
e) The c2d standard keywords are adopted from the c2d Bolocam survey
by \citet{enoch2006:perseus}. See Sec.\ \ref{sec-survey:prop_overview}
for a description.
\end{footnotesize}
\end{table*}

The observed cores reside in very different environments, ranging
from regions of isolated low-mass star formation, like Taurus,
over more turbulent regions, like the Cyg OB7 molecular cloud complex
and the Cepheus Flare, to the high-mass star-forming site of Orion.
In this respect the c2d MAMBO sample does not constitute a homogeneous
sample, but a cross section of the different star-forming clouds in
the solar neighborhood. Correspondingly our sample includes
cores from near ($\approx 100 ~ \rm pc$) to intermediate
($\approx 400 ~ \rm pc$) distances.

\subsection{Observations}
Continuum observations of the 1.2~mm thermal dust emission were done
with the IRAM 30m-telescope on Pico Veleta (Spain) using the
37-receiver MAMBO-1 (projected array diameter of $120 \arcsec$)
and 117-receiver MAMBO-2 bolometer ($240 \arcsec$ diameter) cameras of
the Max-Planck-Institut f\"ur Radioastronomie
\citep{kreysa1999:mpifr_bolometers}. A total of 122 usable maps was
taken between the summer of 2002 and the winter of 2003/2004 in the
framework of a flexible observing pool (Table
\ref{tab-survey:observations}). Only 6 maps were acquired using the
MAMBO-1 array. The weather conditions were good, with zenith optical
depths between 0.1 and 0.3 for most of the time, and above this for
$\approx 20\%$ of the time. Some of the maps were affected by clouds.
All but 13 maps (i.e., 11\%) were taken at an elevation above
$40 \degr$. This yields typical line-of-sight optical depths much
below 0.5. Two maps are affected by strong anomalous refraction
(\citealt{altenhoff1987:anom_refrac}; one map on the southern part of
L1041-2, one map on L1148).

\begin{table}
\caption{MAMBO map details. For each core the products `number of
  maps' $\times$ `array type' list how often a core was observed
  with which array (`37' for MAMBO-1 and `117' for MAMBO-2). Several
  products are given if a dense core was observed with
  both arrays. The last column lists the minimum noise level in a map
  after smoothing to $20 \arcsec$ resolution.\label{tab-survey:observations}}
\begin{center}
\begin{tabular}{llllllllll}
\hline \hline
\rule{0ex}{3Ex}c2d Name & MAMBO Maps & Minimum Noise Level\\
  & & mJy per $11 \arcsec$ beam\vspace{1ex}\\ \hline
L1355      &   2 $\times$ 117 & 1.6\\
L1521B-2   &   5 $\times$ 117 & 1.1\\
L1521F     &   2 $\times$ 117 & 2.0\\
L1521-2    &   4 $\times$ 117 & 1.3\\
L1524-4    &   2 $\times$ 117 & 1.4\\
B18-1      &   4 $\times$ 117 & 1.1\\
TMC-2      &   4 $\times$ 117 & 0.9\\
B18-2      &   2 $\times$ 117 & 1.3\\
B18-4      &   3 $\times$ 117$^a$ & 1.1\\
B18-5      &   3 $\times$ 117$^a$ & 1.1\\
TMC-1C     &   6 $\times$ 117 & 1.2\\
TMC-1      &   6 $\times$ 117 & 1.7\\
L1507A     &   5 $\times$ 117 & 1.6\\
L1582A     &   3 $\times$ 117 & 1.3\\
IRAS05413  &   3 $\times$ 117 & 1.2\\
L1622A     &   7 $\times$ 117 & 1.1\\
L183       &   4 $\times$ 117 & 1.4\\
L438       &   2 $\times$ 117 & 1.2\\
L492       &   3 $\times$ 117 & 0.8\\
CB188      &   1 $\times$ 37, 2 $\times$ 117 & 0.8\\
L673-7     &   2 $\times$ 117 & 1.3\\
L675       &   1 $\times$ 37, 2 $\times$ 117 & 0.9\\
L1100      &   1 $\times$ 37, 2 $\times$ 117 & 1.3\\
L1041-2    &   4 $\times$ 117 & 1.4\\
L1148      &   5 $\times$ 117 & 1.1\\
L1155E     &   2 $\times$ 117 & 1.5\\
L1082C-2   &   2 $\times$ 117 & 1.8\\
L1082C     &   2 $\times$ 117 & 1.5\\
L1082A     &   4 $\times$ 117 & 1.1\\
L1228      &   3 $\times$ 117 & 1.3\\
Bern48     &   1 $\times$ 37, 1 $\times$ 117 & 1.0\\
L1172A     &   2 $\times$ 117 & 1.5\\
L1177      &   1 $\times$ 117 & 1.3\\
L1021      &   3 $\times$ 117 & 1.0\\
L1014      &   2 $\times$ 37 & 1.3\\
L1103-2    &   3 $\times$ 117 & 1.4\\
L1251A     &   3 $\times$ 117 & 1.2\\
L1197      &   6 $\times$ 117 & 0.9\\
\hline
\end{tabular}
\end{center}
\rule{0ex}{3Ex}Notes: a) B18-4 and B18-5 were observed together
in one field.
\end{table}

The beam size on the sky was $11 \arcsec$, and the effective frequency
$250 ~ \rm GHz$ with half sensitivity limits at 210 and
$290 ~ \rm  GHz$. Pointing and focus position were usually checked before 
and after each map. The pointing and focus corrections were usually
below $3 \arcsec$ and $0.3 ~ \rm mm$, respectively. The zenith optical
depth was typically measured with a skydip at least once within an
hour from start or end of a map. The sources were observed with the
standard on-the-fly technique, where the telescope secondary was
chopping by $\approx 40 \arcsec$ to $\approx 70 \arcsec$ at a rate of
2~Hz parallel to the scanning direction of the telescope. The
telescope was constantly scanning in azimuth for up to $90 ~ \rm s$ at
a speed of $6 ~ {\rm to} ~ 8 \arcsec ~ \rm s^{-1}$ before turning
around, except for one map of
L1521B-2 for which the telescope moved in azimuth and elevation
(then using the ``rotated'' chopping secondary). Where
possible, each source was mapped
with varying scanning directions (in equatorial coordinates) and chop
throws. Several times per
week Mars and Uranus were observed for absolute flux
calibration. ``Secondary'' calibrators of constant but not
a priori known flux density were observed every few hours. The scatter
in their retrieved flux densities suggests a relative map-to-map
calibration uncertainty $< 15\%$, suggesting an absolute
uncertainty of order 20\%.

Part of the data were already presented by
\citet[][ for L1521F]{crapsi2004:l1521f},
\citet[][ for TMC-2 and L492]{crapsi2005:survey},
\citet[][ for L1014]{young2004:l1014}, and
\citet[][ for TMC-1C]{schnee2007:tmc-1_temperature}.
The raw data for these sources are included in this work and are
partially complemented by additional maps\footnote{Some of these data
  was kindly provided by A.\ Crapsi.}.

\subsection{Data Reduction\label{sec-survey:data_reduction}}
\subsubsection{Classical Limitations \& A new
  Strategy\label{sec-survey:new_strategy}}
The reduction of bolometer data for extended sources is still an
evolving art, in particular when dealing with faint emission. The
basic problem is that the brightness of the sky usually exceeds the
intensity of the astronomical sources by 2 to 4 orders of magnitude
(e.g., at $1.2 ~ \rm mm$ wavelength the intensity from a
$300 ~ \rm K$ atmosphere of optical depth $\ge 0.1$ exceeds the
intensity from interstellar dust at $10 ~ \rm K$ with an $\rm H_2$
column density $\le 10^{23} ~ \rm cm^{-2}$ by a factor $\ge 640$). The
sky brightness (and if possible also its rapid fluctuations known as
``skynoise'', that contribute to the noise) thus has to be subtracted
during the data reduction. This subtraction unfortunately can
introduce artifacts to the derived maps.

Specifically, classical data reduction algorithms filter a significant
fraction of the source flux when sources have a size comparable to the
size of the bolometer array (relevant when subtracting skynoise), a
size similar to an individual bolometer map (i.e., before mosaicing;
relevant when subtracting the mean sky brightness as, e.g, the minimum
intensity in the map), or both. (Note that these fundamental problems
also affect maps from modern total-power bolometers like
LABOCA\footnote{http://www.apex-telescope.org/bolometer/laboca/},
SCUBA-2\footnote{http://www.jach.hawaii.edu/JCMT/continuum/scuba2.html},
and
Bolocam\footnote{http://casa.colorado.edu/~aikinr/research/bolocam.html}.
A detailed analysis of this problem for Bolocam has been
given by \citet{enoch2006:perseus}.) This is the case for a large
fraction of the sources studied in our sample. We therefore implement
an innovative data reduction scheme that partially mitigates classical
reconstruction problems.

The central element of our data reduction algorithm is the subtraction
of the expected signal due to the astronomical source (based on a
source model) from the raw data before determining and then
subtracting from it the mean sky brightness and skynoise. This
processed raw data is then turned into a map. The sum of this
``residual'' map (since source emission is subtracted) and the source
model yields an improved source model that can be used in further
iterations. This approach has the advantage that the resulting source
model suffers from lesser artifacts due to skynoise and sky
subtraction than if no existing knowledge of the source structure was
used during the subtraction. The initial source model for the
iterative scheme comes from a map created using classical (i.e.,
biased) algorithms. It can be shown analytically that, in the ideal
case without noise, sources much larger than in the classical approach
can be recovered. In our iterative scheme, the residual
intensities decrease with a factor $2^{-n}$, where $n$ is the number
of iterations. To be specific, full recovery of the source structure
requires the source diameter, $d_{\rm source}$, to be smaller than the
effective map size,
\begin{equation}
d_{\rm source} <
\ell_{\rm scan} + \ell_{\rm chop} \, ,
\label{eq:reconstr-1}
\end{equation}
where $\ell_{\rm scan}$ and $\ell_{\rm chop}$ are (following IRAM
terminology) the scan length and the chop throw (of the secondary
mirror), respectively. Slightly larger sources can still be recovered,
but convergence is slower than $2^{-n}$. This size can be increased by
mosaicing. Classical methods, on the contrary, require for full
recovery (with $d_{\rm array}$ being the array diameter)
\begin{equation}
d_{\rm source} <
\ell_{\rm scan} + \ell_{\rm chop} - d_{\rm array}
\label{eq:reconstr-2}
\end{equation}
when not subtracting skynoise, and have the additional constraint
\begin{equation}
d_{\rm source} < d_{\rm array}
\label{eq:reconstr-3}
\end{equation}
if doing so. Mosaicing will furthermore bias intensities. Experiments
with simulated raw data including noise show that the simulated maps
are consistent with the above ideal analysis.

In a formal sense, the improvements due to our approach thus mean that
observations are only limited by Eq.\ (\ref{eq:reconstr-1}), while
conventional approaches are limited by Eqs.\ (\ref{eq:reconstr-2}) and
(\ref{eq:reconstr-3}). The gain in mapping parameter space is
illustrated in Fig.\ \ref{fig-survey:bolo_summary} in Appendix
\ref{appendix:data_reduction}. There, we also present a direct
comparison between maps derived using conventional methods to those
from our approach (Fig.\ \ref{fig-survey:bolo_demo}).

For typical MAMBO map sizes
$\ell_{\rm scan} = 360 \arcsec ~ {\rm to} ~ 540 \arcsec$ and chop throws
$\ell_{\rm chop} = 60 \arcsec$ sources with diameters
$d_{\rm source} < 6 \arcmin ~ {\rm to} ~ 9 \arcmin$ can thus be
recovered without significant bias when using our new approach. For
comparison, when using the larger MAMBO-2 array, classical algorithms
would only recover sources with a diameter
$d_{\rm source} < 3 \arcmin ~ {\rm to} ~ 6 \arcmin$, where skynoise
filtering sets the additional constraint that
$d_{\rm source} < 4 \arcmin$ even for much longer scan lengths.

\subsubsection{Implementation}
Our iterative data reduction scheme is implemented using the MOPSI
software package, which is developed and maintained by R.\ Zylka (IRAM
Grenoble); for L1521B-2 we used MOPSIC, an upgrade of MOPSI also
able to handle data obtained using the ``rotated'' wobbler.
These packages provide subroutines to estimate and remove the
correlated skynoise, to reconstruct maps from chopped bolometer data
using the EKH algorithm \citep{emerson1979:ekh}, to simulate chopped
intensity difference maps for given source models, and many other
tools. These routines are combined to carry out our iterative data
reduction scheme.

During the first iteration no skynoise is removed. The chopped
intensity difference maps derived for each epoch are co-added and then
restored using EKH reconstruction. In the resulting map the area with
emission is then 
marked by hand. The intensity for the area outside the area with
emission is set to zero, as is done for all pixels with negative
intensity inside the area with emission. The modified intensity map is
then smoothed to an effective resolution of $20 \arcsec$, except for
visually selected areas with strong emission, where the data is
smoothed to $14 \arcsec$ resolution. This smoothing increases the
signal-to-noise ratio for emission more extended than the beam. This
map is then taken as the source model for the
second iteration.

In the further iterations the simulated chopped intensity difference
map for the source model is first subtracted from the raw data for
every single observation. Then the correlated skynoise is estimated
and subtracted in these residual maps.  We use a correlation
exclusion radius of $50 \arcsec$ for all sources with data from the 37
receiver MAMBO-1 camera (i.e., for each array receiver a neighborhood
of $50 \arcsec$ radius is excluded when estimating skynoise, since
this reduces intensity biases), and
$100 \arcsec$ otherwise. To remove offsets due to sky intensity and
instabilities, a constant intensity (i.e., a baseline of order 0) is
then subtracted from each receiver signal for every scan lag. A
co-added intensity map including all observations is then derived
using EKH restoration. This map is then added to the source model, and
a source model for the next iteration is derived as done above. In
total 10 iterations are performed. Note that the smoothing of the
source model does not reduce the resolution in the reconstructed maps:
the source map derived at the end of every iteration is corrected for
differences between the actual and model source structure due to
smoothing since these show up in the residual maps finally added
to the source model.

The final maps presented here are usually smoothed to an effective
resolution of $20 \arcsec$, though they are available for resolutions
down to $11 \arcsec$. This smoothing improves the signal-to-noise
level for emission more extended than the beam. In order to facilitate
comparison with maps of different resolutions the data are calibrated
in mJy per $11 \arcsec$ beam. Further weight maps give the variation
of the effective integration time per pixel across the map. They are
calculated during the data reduction and used to calculate noise maps.
The noise increases with increasing distance from the map center due
to a decrease in exposure time towards the map boundaries.

\section{Results\label{sec-survey:results}}
\subsection{Maps}
Figure \ref{fig-survey:maps} presents the intensity maps of the
dense cores covered
by our MAMBO observations. The minimum noise level in the maps ranges
from $0.8 ~ {\rm to} ~ 2.0 ~ \rm mJy$ per $11 \arcsec$ beam and is
listed for each
source in Table \ref{tab-survey:observations}. To facilitate
comparison, all maps are drawn with the same contouring levels and to
the same angular scale.


In most sources the maps reveal extended emission on scales of several
arc-minutes, and in many fields the presence of compact emission with
a full width at half maximum of only a few beams. Compared with many
other bolometer surveys that often detect only the most prominent
intensity peaks, our MAMBO maps are very rich in structure.

In spite of our careful data reduction, some maps are still affected
by artifacts such as stripes in the scanning direction and extended
areas of negative intensities. Sources significantly affected by this
are given lesser weight in our analysis. These are in particular
objects with obvious scanning stripes in their boundaries (e.g., L183
C4; see below for object names), i.e., series of intensity peaks in
the scanning direction separated by the chop throw, those located at
map boundaries (e.g., L183 C6), or with a low intensity contrast to
the surrounding at the object boundary (e.g., L1082C-2 C1; often due
to artifacts).

\astroph{The journal article will be accompanied by the maps in FITS
  format.}

\subsection{Source Identification and
  Quantification\label{sec-survey:id_and_quant}} The MAMBO maps reveal
extended emission and compact emission peaks with no direct
correspondence between these. We therefore separately identify and
quantify extended features, which we term `subcores', emission peaks
in these subcores, and additional `significant peaks' that fulfill some
significance criterion and are not necessarily located within the
subcore boundaries.

For some properties it is not possible to derive their uncertainties
by Gaussian error propagation. In these cases we run Monte-Carlo
simulations with artificial noise with properties identical to the
actual one to estimate the uncertainties. Uncertainties are then
derived as the standard deviation of the derived values. To reduce
the noise level for the extended emission, all quantities derived for
the cores and peaks are derived from maps smoothed to $20\arcsec$
resolution unless noted otherwise.

\subsubsection{Extended Structures}
\paragraph{Identification}
Our source identification method aims at a formal definition of what a
human investigator would intuitively identify as a source. Our
criteria are therefore not mathematically exact, but adopted from a
series of trial-and-error experiments.

To identify sources we generate signal-to-noise ratio (SNR) maps for a
beamsize of $20 \arcsec$, which are then smoothed to a spatial
resolution of $60\arcsec$. Regions above an SNR of 2 in these maps are
taken as source candidates. Obvious artifacts at the rim of the maps
are excluded. The remaining area above an SNR of 2 is then
hand-divided into `subcores'. For this we search the $60 \arcsec$ SNR
maps for saddle points and draw core boundaries by following the
gradient. This scheme is conceptually similar to, but not identical
with, the ``CLUMPFIND'' algorithm by \citet{williams1994:clumpfind}.
Our scheme thus assigns all emission above an SNR of 2 to some
`subcore'.

\paragraph{Quantification}
Table \ref{tab-survey:cores-general} lists general information on the
identified subcores. For each subcore we list the centroid position,
defined as the mean position of the pixels in each subcore, where the
pixels are not weighted by their intensity.  The quoted uncertainties
reflect the influence of statistical noise, while systematic
contributions are not folded in. For each field mapped the subcores
are labeled by a `C', followed by a number in order of increasing
right ascension. In the tables a star marks those cores associated
with a YSO (see below). Cores with uncertain properties due to an
unreliable reconstruction are marked by square brackets.

We then list the area assigned to the subcore, $A$, and the
corresponding effective radius,
\begin{equation}
r_{\rm eff} = (A / \pi)^{1/2} \, .
\end{equation}
Furthermore, we derive the flux densities, $F$, and the corresponding
masses, $M$, for the total subcore's area (subscripts `tot') and for
the area above 50\% of the peak intensity for the respective subcore
(subscripts '50\%'). The masses are derived from the flux densities as
described in Sec.\ \ref{sec-survey:mass_estimates}. Finally, Table 
\ref{tab-survey:cores-general} lists the maximum SNR for each subcore,
$S^{\rm beam}_{\rm max}$. The uncertainties of $F_{\rm tot}$ and
$M_{\rm tot}$ are derived by Gaussian error propagation of the
measured noise level. For $F_{50\%}$ and $M_{50\%}$ the uncertainties
are estimated by the Monte-Carlo experiments described above.

Table \ref{tab-survey:cores-geometry} lists geometrical information on
the identified cores. This is done by fitting ellipsoids
to the area above 50\% of the peak intensity for the respective core
(no weighting by intensity is applied)\footnote{This is implemented by
  using the IDL routine FitEllipse by David Fanning.}. We give the
major and minor axis, $a$ and $b$, their ratio, and the position angle
(east of north). The filling factor, $f$, gives the fraction of the
fitted ellipse filled with emission above 50\% peak intensity. Low
filling factors indicate cases in which ellipses
are a bad fit to the observed intensity distributions. We also
list the effective diameter for the area above 70\% of the peak
intensity, $r_{70\%}$ \citep{crapsi2005:survey}. No attempt has been
made to separately handle
subcores in which the emission from two or more peaks contributes to
$r_{70\%}$. In these cases $r_{70\%}$ is overestimated by an uncertain
amount.


\subsubsection{Peaks}
\paragraph{Identification}
In order to quantify all intensity peaks in our maps we study the peak
position of every identified subcore. In addition we search each map
manually for `significant peaks'. These have an SNR of at least 4 and
can be spatially separated from already identified peaks. To identify
them we search the map for local intensity peaks with an SNR $\ge 4$
and draw the surrounding contour at the peak intensity minus three
times the local noise level. Such a peak is taken to be significant if
no brighter peak is situated in this contour. In the area outside
identified cores we also require that there is also no fainter local
peak within the contour. This additional restriction excludes most
artifacts.

\paragraph{Quantification}
Peak properties are listed in Table \ref{tab-survey:peaks}. The quoted
uncertainties reflect the influence of statistical noise, while
systematic contributions are not folded in. For each mapped field
peaks are labeled by a `P', followed by a number in order of
increasing right ascension. Uncertain peaks, which are not well
reconstructed, are marked by square brackets. These were identified by
searching the maps for peaks close to map boundaries, and those
affected by scanning stripes.

Besides the peak
position we give the subcore to which the peak belongs, if applicable.
The peak intensity, $F^{\rm beam}_{\rm max}$, is listed besides the
corresponding peak $\rm H_2$ column density and visual extinction,
$N({\rm H_2})$ and $A_V$. These are derived using the c2d standard
assumptions on dust properties, as outlined in Appendix 
\ref{appendix:dust_em_prop}, but they are only listed for starless
cores for which
there is no internal heating. In YSO cores the dust
temperature exceeds $10 ~ \rm K$ and needs to be calculated separately
for each source (Sec.\ \ref{sec-survey:mass_estimates}). Also given are the
flux density and the corresponding mass within an aperture of
$4 \, 200 ~ \rm AU$ radius, $F_{4200 \rm AU}$ and $M_{4200 \rm AU}$,
as derived in Sec.\ \ref{sec-survey:mass_estimates}. This aperture of
constant physical diameter is a better measure of the mass
distribution than the central column density since, for close enough
sources, beam smearing does not play a role. The specific radius of
$4 \, 200 ~ \rm AU$ is somewhat arbitrary;
\citet{motte2001:protostars} used it before, given that it matches
$30 \arcsec$ at the distance of Taurus ($\approx 140 ~ \rm pc$) and is
about the radius of an envelope's volume available for YSO formation
(Sec.\ \ref{sec-survey:sf-efficiency}). We finally list the
flux density for the c2d standard apertures of $20\arcsec$,
$40\arcsec$, $80\arcsec$, $120\arcsec$ diameter (e.g.,
\citealt{young2006:c2d_scuba} and \citealt{wu2007:c2d_sharc-ii}). In
deviation to
most other parts of this work the flux densities for the
apertures are derived from unsmoothed
maps since the smoothed maps sometimes have a resolution comparable to
the aperture size.


\subsubsection{Mass and Column Density
  Estimates\label{sec-survey:mass_estimates}}
Mass and column density estimates for starless cores are derived from
the observed flux densities assuming the c2d standard dust emission
properties, i.e., a dust temperature of $10 ~ \rm K$ and an opacity of
$0.0102 ~ \rm cm^{2} ~ g^{-1}$ (per gram of ISM) at MAMBO's effective
observing wavelength. Appendix \ref{appendix:dust_em_prop} presents a
detailed discussion of the conversion between mass and dust emission.
The related discussion of uncertainties in temperature and dust
opacities suggest a systematic uncertainty for mass estimates of an
order of a factor two. Dust near YSOs is, however, heated by the YSO
radiation, and for later YSO stages likely to be more coagulated than
interstellar dust. Therefore, we adopt higher dust temperatures and
opacities when deriving masses for YSO cores. The temperature
gradients also prevent us from deriving meaningful estimates of the
peak column density of YSO cores; therefore we do not quote
values.

The dust temperature near YSOs can be estimated from analytic
models of temperature profiles for YSO envelopes cooling via
dust emission. For opacities with power-law dependence on
wavelength, $\kappa \propto \lambda^{-\beta}$, where $\beta$ is the
emissivity spectral index, the dust temperature profile in a
YSO envelope only heated radiatively by the star is
\begin{equation}
T_{\rm d}^{\rm rad}(r) = 38.4 ~ {\rm K}
\left( \frac{r}{100 ~ \rm AU} \right)^{-q}
\left( \frac{L}{L_{\odot}} \right)^{q/2} \, ,
\label{eq-survey:quant-1}
\end{equation}
where $q = 2 / (4 + \beta)$, $L$ is the luminosity, and $r$ is the
distance from the star \citep{terebey1993:disks_and_envelopes}. The
numerical values hold for $\beta = 1$, which is consistent with
observations (e.g., \citealt{belloche2006:ngc1333-iras4a}) and assumed
in the following. Rearrangement allows for an estimate of the radius
at a particular dust temperature,
\begin{equation}
r^{\rm rad}(T_{\rm d}) = 2890 ~ {\rm AU}
\left( \frac{T_{\rm d}}{10 ~ \rm K} \right)^{-1/q}
\left( \frac{L}{L_{\odot}} \right)^{1/2} \, .
\end{equation}
However, in absence of nearby stars, the interstellar dust is heated
by the interstellar radiation field and in collisions with gas (which
is heated by cosmic rays and the UV-part of the interstellar radiation
field), which produces a typical equilibrium temperature of order
$10 ~ \rm K$ (e.g., \citealt{goldsmith2001:temperatures}; see
\citealt{evans2001:em_models}, \citealt{galli2002:struc_and_stab}, and
\citealt{young2004:formaldehyde} for actual core temperature
models). The dust temperature cannot drop below this value, so that
\begin{equation}
T_{\rm d}(r) =
\max(10 ~ {\rm K}, T_{\rm d}^{\rm rad}[r]) \, .
\end{equation}
For this temperature profile \citet{belloche2006:ngc1333-iras4a}
derive the mass-weighted dust temperature. This assumes optically thin
dust emission, which applies to the majority of the mass. Only the
densest and hottest parts of an envelope might be optically thick,
which leads to an insignificant bias towards slightly overestimated average
dust temperatures. For this calculation we assume the
density profile in a subcore to be roughly described by a
power-law, i.e.\ $\varrho \propto r^{-p}$, where $\varrho$ is the
density and $r$ is the distance from peak center. Then the
mass-weighted dust temperature within an aperture of radius $R$ is
\begin{equation}
\left< T_{\rm d} \right > =
\left\{
\begin{array}{ll}
  \frac{3 - p}{3 - q - p} T_{\rm d}^{\rm rad}(R) , &
  \hspace{-0.5cm}
  {\rm if} ~ T_{\rm d}^{\rm rad}(R) \ge 10 ~ {\rm K}\, ,\\
  \left(
  \frac{q}{3 - p - q}
  \left[ \frac{r^{\rm rad}(10 ~ {\rm K})}{R} \right]^{3 - p}
  + 1 \right) \cdot 10 ~ {\rm K} , & {\rm otherwise} \, .
\end{array}
\right.
\label{eq-survey:quant-2}
\end{equation}
For YSO peaks and subcores we use this average temperature to
derive masses from the flux densities of the whole subcore, of the
area above 50\% peak intensity, and of an aperture of
$4 \, 200 ~ \rm AU$ radius. In these cases the aperture radius $R$ is
set to be the effective core radius of the whole core, or the
geometric mean of the major and minor axis at 50\% peak intensity, or
$4 \, 200 ~ \rm AU$, respectively. We assume a power-law exponent of
$p = 2$ for the density profile (i.e., profile for a singular
isothermal sphere; \citealt{shu1977:self-sim_collapse}). The mean dust
temperatures are thus overestimated for density profiles shallower
than $\varrho \propto r^{-2}$. The YSO luminosities
are taken to be identical to the bolometric luminosities derived in
Sec.\ \ref{sec-survey:protostar_properties}. For the VeLLOs
associated with L1521F P1, L1148 P1 (a VeLLO candidate), and L1014 P1,
however, we use the
better constrained ``internal luminosities'' derived by
\citet{bourke2006:l1521f}, Kauffmann et al.\ (in prep.) and
\citet{young2004:l1014}. Given the observed luminosities, the
temperatures assumed for class I sources are $(15 \pm 3) ~ \rm K$,
while for class 0 sources they reach from 10 to $24 ~ \rm K$ and
$21 ~ \rm K$ is derived for the only class II source (in L1021 P1;
see below for YSOs and their classes).

The uncertain density profile leads to an uncertain mass estimate. We
consider a range in the exponent $p$ from 2 to $3/2$, as expected for
YSO envelopes \citep{shu1977:self-sim_collapse}, in order to
gauge the uncertainties in the mass estimate. This shows that the
aperture averaged temperatures in power-law envelopes, and
correspondingly the derived masses, are uncertain by 30\% and less.

For the dust near class II YSOs we adopt an
opacity of $0.02 ~ \rm cm^2 ~ g^{-1}$ because of an expected enhanced
coagulation
(\citealt{motte2001:protostars},
\citealt{ossenkopf1994:opacities}). For the other YSO and
starless cores we adopt the standard c2d opacity, which is
$0.0102 ~ \rm cm^2 ~ g^{-1}$ at MAMBO's observing wavelength.

Circumstellar disks might also contribute to the millimetre continuum
emission. Their contribution to apertures with radii
$\ge 4 \, 200 ~ \rm AU$ is estimated to be $\lesssim 10\%$
for YSOs of the classes 0 and I \citep{motte2001:protostars}. The
emission from the disks does therefore not significantly bias our
envelope mass estimates for the youngest YSOs. Class II YSOs, in
contrast, are not expected to have envelopes (e.g.,
\citealt{andre2000:pp_iv} and references therein); their emission is
dominated by compact ($\lesssim 1 \, 000 ~ \rm AU$) disks with a
temperature above the one estimated from Eq.\
(\ref{eq-survey:quant-2}). Our procedure will thus overestimate the
mass for YSOs with significant disk emission.

\subsection{Overview of Source Properties\label{sec-survey:prop_overview}}
We use the morphology keywords adopted by \citet{enoch2006:perseus} to
describe our maps. These are summarised in Table
\ref{tab-survey:c2d-MAMBO_sampel} and can be employed to compare the
dense core morphologies revealed by our maps with those found in other
c2d bolometer surveys.

All dense cores that were observed are detected in our maps,
i.e., the maximum SNR in the map is at least 4. For one source
only `weak' emission is detected, meaning a peak SNR equal to or below 5;
all other sources are brighter. In total 21 cores (55\% of all cores)
have at least two peaks separated by less than $3 \arcmin$, and we
consider them to be `multiple'. All cores are `extended', as they
contain at least one subcore with an equivalent radius exceeding $30
\arcsec$. We take all subcores with a major-to-minor axis ratio
exceeding 1.2 to be `elongated', and the others to be `round'. Then 6
cores (16\%) only contain round subcores, 21 cores (55\%) only contain
elongated subcores, and 9 cores (24\%) contain both round and
elongated subcores.

Figure \ref{fig-survey:histograms} gives an overview over the sizes, masses,
column densities, and ellipticities of well reconstructed
subcores. These have typical values in the range
$(1 ~ {\rm to} ~ 6) \cdot 10^4 ~ {\rm AU} =
0.05 ~ {\rm to} ~ 0.30 ~ {\rm pc}$ for the effective radius,
$0.5 ~ {\rm to} ~ 20 \, M_{\odot}$ for the subcore masses,
$0.1 ~ {\rm to} ~ 1.0 \, M_{\odot}$ for the mass within
$4 \, 200 ~ \rm AU$ from the peak, and major-to-minor axes ratios at
50\% peak intensity ranging from 1 to about 4. Implications from these
properties for the physical state and evolution of dense cores and
YSOs are discussed in Sec.\ \ref{sec-survey:analysis}.

\section{Supplemental YSO Data\label{sec-survey:protostellar_data}}
For the c2d MAMBO survey to be fully exploited, it needs to be
complemented with information on the YSOs covered by the maps. Most of
our analysis below is based on IRAS data since c2d finally observed
only half of the YSO (\& VeLLO) cores in our sample (i.e, 8 out of 17)
and a homogeneous characterization is not possible based on c2d Spitzer
data only. Also, we only use c2d photometry for Spitzer source
characterization to guarantee homogeneity, implying that some images in
the Spitzer archive cannot be used for the present study. All YSOs
visible in Spitzer images are, however, either characterized by IRAS
or Spitzer, with the singular exception of a Spitzer source in L1582A
not covered by c2d data and not detected by IRAS.

\subsection{Associated IRAS, 2MASS, and Spitzer
  Sources\label{sec-survey:ass_protostars}}
Table \ref{tab-survey:IRAS_not-detected} lists sources from the IRAS
Point Source Catalogue (PSC; \citealt{beichman1988:iras_supplement})
and IRAS Faint Source Catalogue (FSC;
\citealt{moshir1990:iras_faint_source_cat}) not associated with
intensity peaks in the MAMBO maps, and Table
\ref{tab-survey:IRAS_detected} lists those that are associated. The
sources are considered to be associated with subcores and peaks
detected by MAMBO if the separation between an IRAS source and a MAMBO
subcore or peak is less than the uncertainty of the separation at the
$2 \sigma$ level (i.e., the IRAS uncertainty ellipse plus the MAMBO
pointing error). These sources are most likely physically associated
with the dense cores and are assumed to be young stars.

\setcounter{table}{6}
\begin{table}
\caption{IRAS sources in the MAMBO maps that are not detected as MAMBO
  emission
  peaks. For every core in the c2d MAMBO survey the table lists
  sources from the IRAS Point Source Catalogue and the IRAS Faint
  Source Catalogue (preceded by an `F'), and the subcores in which the
  IRAS sources are located, if applicable. Objects from the IRAS Point
  Source and Faint Source catalogues present in both catalogues are
  connected. The last column gives the
  spectral index between 12 and $25 ~ \rm \mu m$
  wavelength.\label{tab-survey:IRAS_not-detected}}
\begin{center}
\begin{tabular}{llllllllll}
\hline \hline
\rule{0ex}{3Ex}Core & IRAS Source & Asso.\ &
$\alpha^{25 ~ \rm \mu m}_{12 ~ \rm \mu m}$\vspace{1.7ex}\\\hline
L1521-2
 &   F04262+2654  & C1 & $ 0.27 \pm 0.43 $\\
L1524-4
 &  04274+2420  & & $ -0.85 \pm 0.27 $\\
 &  $\lfloor$ F04274+2420  & & $ -0.83 \pm 0.19 $\\
B18-1
 &  04292+2427  & & $ < -1.5 $\\
 &  $\lfloor$ F04291+2427  & & $ < -1.9 $\\
TMC-2
 &  04294+2413  & & $ -0.9 \pm 0.15 $\\
 &  $\lfloor$ F04294+2413  & & $ -0.89 \pm 0.11 $\\
B18-4
 &  04326+2405  & C1 & $ < -1.68 $\\
 &  $\lfloor$ F04326+2405  & & $ < -0.75 $\\
TMC-1C
 &  04380+2553  & & $ -0.51 \pm 0.15 $\\
 &  $\lfloor$ F04380+2553  & & $ -0.76 \pm 0.12 $\\
TMC-1
 &   F04383+2549  & & $ < -1.12 $\\
 &  04392+2529  & & $  $\\
L1582A
 &   F05290+1229  & & $ -0.51 \pm 0.17 $\\
L1622A
 &  05517+0151  & & $ -2.14 \pm 0.16 $\\
 &  $\lfloor$ F05517+0151  & & $ -2.25 \pm 0.15 $\\
 &  05519+0148  & [C2] & $  $\\
 &  05519+0157  & & $  $\\
L438
 &  18116$-$0707  & & $ < -1.33 $\\
L492
 &  18130$-$0341  & & $  $\\
 &  18132$-$0350  & & $ -1.82 \pm 0.15 $\\
CB188
 &  19180+1127  & & $ < -1.31 $\\
L675
 &  19217+1103  & & $ -1.81 \pm 0.25 $\\
L1148
 &  20395+6714  & & $  $\\
 &  20410+6710$^a$  & C2 & $  $\\
L1228
 &   F20598+7728  & & $ -0.26 \pm 0.34 $\\
L1103-2
 &  21399+5632  & & $ > -1.47 $\\
L1251A
 &   F22282+7454  & & $ < -2.54 $\\
\hline
\end{tabular}
\end{center}
\rule{0ex}{3Ex}Notes: a) Probably an artifact since no corresponding
Spitzer source exists.
\end{table}

\begin{sidewaystable*}
\begin{footnotesize}
\caption{Properties of MAMBO-detected IRAS sources. For every core in
  the c2d MAMBO survey the table lists dust emission peaks and, if
  applicable, the related subcore associated with an IRAS source
  (Sec.\ \ref{sec-survey:ass_protostars} discusses the identification of
  associated sources; FSC sources are preceded by an `F').
  Associated 2MASS sources are
  listed too. The table further lists the spectral index between 12
  and $25 ~ \rm \mu m$ wavelength, the bolometric temperature and
  luminosity, the submillimetre-to-bolometric luminosity ratio, the
  mass within a peak-centered aperture of $4 \, 200 ~ \rm AU$ radius,
  and the infrared SED class.\label{tab-survey:IRAS_detected}}
\begin{center}
\begin{tabular}{llllllllllllllllllll}
\hline \hline
\rule{0ex}{4Ex}Core & Asso.\ & IRAS Source & 2MASS Source &
 $\alpha^{25 ~ \rm \mu m}_{12 ~ \rm \mu m}$ &
 $T_{\rm bol}^{\rm IRAS}$ &  $L_{\rm bol}^{\rm IRAS}$ &
 $L_{\rm submm}^{\rm IRAS} / L_{\rm bol}^{\rm IRAS}$ &
 $M^{\rm YSO}_{\rm 4200 \rm AU}$ & Class\\
 & & & & & K & $L_{\odot}$ & & $M_{\odot}$\vspace{1ex}\\\hline
B18-1
 & [C2], [P5] &  04292+2422  &  04321540+2428597  & $ 0.84 \pm 0.13 $
 & $< 602$ & $> 1.09$ & $> 0.014$ & $> 0.179$ &  I \\
 & & \&  F04292+2422  \\
B18-4
 & C1, P1 &  04325+2402  &  04353539+2408194  & $ > 3.76 $ & $< 83$ & $
0.42 ~ {\rm to} ~ 0.79  $ & $ 0.053 ~ {\rm to} ~ 0.099 $ & $
0.522 ~ {\rm to} ~ 0.589  $ &  I \\
 & & \&  F04325+2402  \\
TMC-1C
 & P3 &  04385+2550  &  04413882+2556267  & $ 0.41 \pm 0.13 $ & $> 596$ & $
0.23 ~ {\rm to} ~ 0.47  $ & $ 0.007 ~ {\rm to} ~ 0.015 $ & $
0.042 ~ {\rm to} ~ 0.047  $ &  I \\
 & & \&  F04385+2550  \\
TMC-1
 & C2, P2 &  04381+2540  &  04411267+2546354  & $ 1.52 \pm 0.14 $ & $ 131 $ & $
0.67 \pm 0.02  $ & $ 0.043 \pm 0.017 $ & $ 0.379 \pm 0.010 $ &  I \\
 & & \&  F04381+2540  \\
IRAS05413
 & C2, P2 &  05412-0105  &  05434630-0104439  & $ > 0.25 $ & $> 67$ & $
0.52 ~ {\rm to} ~ 7.26  $ & $ 0.008 ~ {\rm to} ~ 0.115 $ & $
0.308 ~ {\rm to} ~ 0.597  $ &  0 \\
 & & \&  F05411-0106  \\
 & C3, P3 &  05413-0104  &  none  & $ > -0.70 $ & $< 55$ & $
\approx 10.5  $ & $\approx 0.017$ & $\approx 0.95$ &  0 \\
CB188
 & C1, P1 &  19179+1129  &  19201494+1135400  & $ > 0.88 $ & $< 248$ & $
1.74 ~ {\rm to} ~ 2.14  $ & $ 0.015 ~ {\rm to} ~ 0.018 $ & $
0.272 ~ {\rm to} ~ 0.288  $ &  I \\
L1100
 & C1, P1 &  20355+6343  &  none  & $ 0.70 \pm 0.18 $ & $< 103$ & $
1.22 ~ {\rm to} ~ 2.13  $ & $ 0.023 ~ {\rm to} ~ 0.041 $ & $
0.445 ~ {\rm to} ~ 0.515  $ &  I \\
 & & \&  F20355+6343  \\
L1041-2
 & C4, P4 &  20361+5733  &  none  & $ > 3.12 $ & $< 60$ & 
$ 3.91 ~ {\rm to} ~ 5.07 $ & $  0.025 ~ {\rm to} ~ 0.033 $ &
$ 0.900 ~ {\rm to} ~ 0.966 $ &  0 \\
 & & \&  F20361+5733  \\
L1148
 & C1, P1 &  F20404+6712  &  20405664+6723047  & $ > -0.07 $ &
$ > 110 $ & $
0.06 ~ {\rm to} ~ 0.40  $ & $ 0.023 ~ {\rm to} ~ 0.165 $ & $
0.115 ~ {\rm to} ~ 0.146  $ &  I \\
L1082C
 & C1, P1 &  20503+6006  &  none  & $ > 1.05 $ & $ > 62 $ & $ 0.32 ~ {\rm to} ~ 1.73
$ & $  0.026 ~ {\rm to} ~ 0.141 $ & $ 0.402 ~ {\rm to} ~ 0.576 $ &  0\\
 & & \&  F20503+6007  \\
L1082A
 & C1, P1 &  20520+6003  &  20531346+6014425  & $ 0.39 \pm 0.20 $ & $ <198 $ & $
1.00 ~ {\rm to} ~ 2.49  $ & $ 0.018 ~ {\rm to} ~ 0.044 $ & $
0.388 ~ {\rm to} ~ 0.491  $ &  I \\
 & C3, P5 &  20526+5958  &  none  & $  $ & $ < 119 $ &
$ 0.84 ~ {\rm to} ~ 1.36 $ & $
0.038 ~ {\rm to} ~ 0.061  $ & $ 0.538 ~ {\rm to} ~ 0.602 $ &  I\\
L1228
 & C1, P1 &  20582+7724  &  20571294+7735437  & $ 0.35 \pm 0.11 $ & $ 307 $ & $
2.27 \pm 0.04  $ & $ 0.035 \pm 0.008 $ & $ 0.782 \pm 0.007 $ &  I \\
 & & \&  F20582+7724  \\
Bern48
 & C1, P1 &  21004+7811  &  20591408+7823040  & $ -0.17 \pm 0.10 $ & $ 628 $ & $
11.05 \pm 0.20  $ & $ 0.008 \pm 0.001 $ & $ 0.483 \pm 0.003 $ &  I \\
 & & \&  F21004+7811  \\
L1172A
 & C2, P2 &  21017+6742  &  21022122+6754202  & $ > -0.13 $ & $ < 100 $ & $
0.37 ~ {\rm to} ~ 0.59  $ & $ 0.049 ~ {\rm to} ~ 0.077 $ & $
0.370 ~ {\rm to} ~ 0.403  $ &  I\\
 & & \&  F21017+6742  \\
L1177
 & C1, P2 &  21169+6804  &  21173862+6817340  & $ 1.26 \pm 0.27 $ &
$ < 127 $ & $ 2.50 ~ {\rm to} ~ 3.32  $ & $ 0.038 ~ {\rm to} ~ 0.051 $ & $
1.030 ~ {\rm to} ~ 1.114  $ &  I \\
 & & \&  F21168+6804  \\
L1021
 & P1 &  21197+5046  &  21212751+5059475  & $ -0.82 \pm 0.13 $ & $ > 990 $ & $
4.09 ~ {\rm to} ~ 6.91  $ & $ 0.001 ~ {\rm to} ~ 0.002 $ & $
0.021 ~ {\rm to} ~ 0.024  $ &  II \\
L1251A
 & C2, P3 &  22290+7458  &  22300004+7513578  & $ > 0.68 $ & $ > 88 $ & $
0.12 ~ {\rm to} ~ 0.75  $ & $ 0.033 ~ {\rm to} ~ 0.216 $ & $
0.284 ~ {\rm to} ~ 0.380  $ &  I \\
 & & \&  F22290+7458  \\
\hline
\end{tabular}
\end{center}
\end{footnotesize}
\end{sidewaystable*}

We include IRAS sources of any quality. This in principle could lead
to considerable bias of our sample, since also extragalactic sources
and late type stars would be picked up by these criteria. However,
only a single IRAS source projected onto well detected emission is not
detected as a MAMBO intensity peak (i.e., in L1521-2 C1; see Table
\ref{tab-survey:IRAS_not-detected}); all other IRAS sources that lie
within subcore boundaries but are not detected as MAMBO peaks are
either detected as extensions in the MAMBO intensity contours (in
B18-4 C1), projected onto noisy parts of a MAMBO map (in L1622A C2),
or of dubious quality and discarded anyway (in L1148 C2).

The criterion of association is relaxed for IRAS05413 P2, L1100 P1,
and L1082C P1. For these sources inconsistencies between the IRAS PSF
and FSC indicate positional errors of order $1 \arcmin$, and we
associate them with neighboring MAMBO peaks since these would be
unusually bright and compact for starless peaks. For IRAS 20410+6710,
which is projected onto L1148 C2, our more sensitive Spitzer maps show
no corresponding source in the MIPS bands. We therefore consider this
IRAS source to be an artifact.

For those IRAS sources associated with dust emission peaks
it is possible to significantly improve the accuracy of their position
by adopting the position of the MAMBO peak. This then allows for a
search of 2MASS counterparts of these sources. We do so by searching for
2MASS sources less than $10 \arcsec$ away from the MAMBO peak. If
several 2MASS sources are found we assume the one closest to the MAMBO
peak to be the counterpart. The identified counterparts are listed in
Table \ref{tab-survey:IRAS_detected}. If no counterpart is found, 2MASS
upper limits are taken to be similar to those of nearby 2MASS
sources with upper limits.

This combined data yields a spectral coverage with data
near $1 ~ \rm \mu m$ wavelength (from 2MASS), a well sampled range
from $12 ~ {\rm to} ~ 100 ~ \rm \mu m$ (from IRAS), and information at
$1 \, 200 ~ \rm \mu m$ (from MAMBO). For consistency with
previous work (e.g., \citealt{andre1999:iram04191} and
\citealt{motte2001:protostars}), and to avoid
problems of distance bias, we use dust emission flux densities for the
$4 \, 200 ~ \rm AU$ aperture to study the spectral energy
distributions.\medskip

\noindent In Table \ref{tab-survey:SST_protostars} we also present a
list of Spitzer point sources from the 3rd c2d data delivery\footnote{See
  http://ssc.spitzer.caltech.edu/legacy/c2dhistory.html for a detailed
documentation.} associated with MAMBO dust emission peaks.
To identify associated objects we search
our maps for Spitzer sources
that are detected as point sources at $24 ~ \rm \mu m$ wavelength,
have a $24 ~ \rm \mu m$ flux $\ge 2 ~ \rm mJy$, and are offset by less
than $20 \arcsec$ from a dust emission peak. We use the $24 ~ \rm \mu m$
data for our search since in this band Spitzer reliably picks up
sources with infrared emission apparently in excess of the
photospheric emission.

The choice of the flux and offset cuts is
guided by previous knowledge about sources in our sample. The
$24 ~ \rm \mu m$ flux cut for our search must be $\le 2 ~ \rm mJy$,
since manual inspection of our maps reveals YSOs of such flux in
our sample (e.g., SSTc2d J223105.6+751337 in L1251A). The offset cut
must be $\ge 10 \arcsec$, larger than offsets recently observed for
YSOs in our sample (L1014-IRS; see
\citealt{huard2006:l1014}). Using the c2d processed source catalogues
(documented as part of c2d's 4th data delivery)
for the area surveyed by the Spitzer Wide-Area Infrared Extragalactic
survey (SWIRE; \citealt{lonsdale2003:swire}), with our criteria we expect to
find 1.07 chance alignments between extragalactic background objects
and one of the about 110 dust emission peaks in our MAMBO maps.

\begin{sidewaystable*}
\begin{footnotesize}
\caption{Properties of MAMBO detected c2d Spitzer sources. Like Table
  \ref{tab-survey:IRAS_detected}, but now listing sources from Spitzer point
  source catalogues and giving the spectral index between $2.2$ and
  $24 ~ \rm \mu m$ wavelength. VeLLO candidates are marked in the
  class designation.
\label{tab-survey:SST_protostars}}
\begin{center}
\begin{tabular}{llllllllllllllllllll}
\hline \hline
\rule{0ex}{3Ex}Field & Asso.\ & c2d Spitzer Source & 2MASS Source &
 $\alpha_{2.2 ~ \rm \mu m}^{24 ~ \rm \mu m}$ &
 $T_{\rm bol}^{\rm SST}$ & $L_{\rm bol}^{\rm SST}$ &
 $L_{\rm submm}^{\rm SST} / L_{\rm bol}^{\rm SST}$ &
 $M^{\rm YSO}_{\rm 4200 \rm AU}$ & Class\\
 & & & & & K & $L_{\odot}$ & & $M_{\odot}$\vspace{1ex}\\\hline
L1521F & C1, P1 & SSTc2d J042839.0+265135  &  none  & $ 1.56 \pm 0.08
$ & $ < 30 $ & $
0.06 ~ {\rm to} ~ 0.08  $ & $ 0.400 ~ {\rm to} ~ 0.545 $ & $
0.729 ~ {\rm to} ~ 0.749  $ &  0, VeLLO \\
B18-1 & [C2], [P5] & SSTc2d J043215.4+242859 & 04321540+2428597 & $-0.25 \pm 0.06$ &
  \multicolumn{3}{l}{no IRAC data available}\\
TMC1-1C & P3 & SSTc2d J044138.8+255627  &  04413882+2556267  & $ -0.04 \pm 0.05 $ & $ 681
$ & $  0.37 \pm 0.03 $ & $ 0.008 \pm 0.023 $ & $ 0.022 \pm 0.005 $ &
II \\
TMC-1 & C2, P2 & SSTc2d J044112.7+254635 & 04411267+2546354 & $0.96 \pm 0.06$ &
  \multicolumn{3}{l}{no IRAC data available}\\
CB188 & C1, P1 & SSTc2d J192014.9+113540  &  19201494+1135400  & $ 0.25 \pm 0.05 $ & $ 433
$ & $  1.12 \pm 0.09 $ & $ 0.024 \pm 0.022 $ & $ 0.322 \pm 0.007 $ &  I \\
L673-7 & C1, P1 & SSTc2d J192134.8+112123  &  none  & $ 0.97 \pm 0.08 $ & $ < 55 $ & $
0.03 ~ {\rm to} ~ 0.06  $ & $ 0.284 ~ {\rm to} ~ 0.573 $ & $
0.375 ~ {\rm to} ~ 0.394  $ &  0, VeLLO \\
L1148 & C1, P1 & SSTc2d J204056.7+672305  &  20405664+6723047  & $ 0.73 \pm 0.05 $ &
$ < 140 $ & $  0.05 ~ {\rm to} ~ 0.10 $ & $ 0.074 ~ {\rm to} ~ 0.162 $ & $
0.139 ~ {\rm to} ~ 0.148  $ &  I, VeLLO\\
L1228 & C1, P1 & SSTc2d J205712.9+773544  &  20571294+7735437  & $ 0.14 \pm 0.05 $ & $ 388
$ & $  1.82 \pm 0.13 $ & $ 0.042 \pm 0.027 $ & $ 0.832 \pm 0.007 $ &  I \\
Bern48 & C1, P1 & SSTc2d J205914.0+782304  &  20591408+7823040  & $ -0.42 \pm 0.10 $ & $ 742
$ & $  8.35 \pm 0.51 $ & $ 0.009 \pm 0.013 $ & $ 0.265 \pm 0.002 $ &
II \\
L1014 & C1, P1 & SSTc2d J212407.5+495909  &  none  & $ 0.33 \pm 0.07 $ & $ < 163 $ & $
0.04 ~ {\rm to} ~ 0.15  $ & $ 0.118 ~ {\rm to} ~ 0.411 $ & $
0.318 ~ {\rm to} ~ 0.357  $ &  I, VeLLO \\
L1251A & C2, P3 & SSTc2d J222959.5+751404  &  22300004+7513578  & $ 0.51 \pm
0.05 $ & $ < 197 $ & $  0.11 ~ {\rm to} ~ 0.35 $ & $ 0.060 ~ {\rm to} ~ 0.187 $ & $
0.327 ~ {\rm to} ~ 0.382  $ &  I \\
 & C3, P4 & SSTc2d J223031.8+751409  &  none  & $ 0.57 \pm 0.08 $ & $ < 44 $ & $
0.17 ~ {\rm to} ~ 0.30  $ & $ 0.170 ~ {\rm to} ~ 0.301 $ & $
0.831 ~ {\rm to} ~ 0.900  $ &  0 \\
 & C4, P5 & SSTc2d J223105.6+751337  &  none  & $ -0.23 \pm 0.07 $ & $ < 63 $ & $
0.08 ~ {\rm to} ~ 0.17  $ & $ 0.201 ~ {\rm to} ~ 0.421 $ & $
0.630 ~ {\rm to} ~ 0.683  $ &  0 \\
\hline
\end{tabular}
\end{center}
\end{footnotesize}
\end{sidewaystable*}

Table \ref{tab-survey:spitzer_non-c2d} furthermore gives a list of
prominent YSOs detected in Spitzer data not taken by c2d, defined as
previously known IRAS YSOs. This excludes sources in L1622A, since
these are not well covered by our MAMBO maps. One source in L1582A
that has apparent outflow nebulosity and was not detected by IRAS is
also included. We cannot, however, associate a MAMBO peak to this
star, and we therefore have to exclude it from the further
analysis. Manual inspection of the images did not reveal further
probable YSO candidates. No c2d photometry exists for these images,
and the Spitzer source properties are not further discussed in the
following, except when refining YSO positions (Sec.\
\ref{sec-survey:offsets}). Including the object in L1582A, the
combined Spitzer data thus shows 7 YSOs in cores previously believed
to be starless (Table \ref{tab-survey:spitzer_new}). Four out of these
(including the VeLLO candidate in L673-7) are reported here for the
first time.

\begin{table}
\caption{Prominent YSOs detected in Spitzer
  data not taken by c2d, and related IRAS sources, if
  existing.\label{tab-survey:spitzer_non-c2d}}
\begin{center}
\begin{tabular}{llllllllll}
\hline \hline
\rule{0ex}{3Ex}Core & IRAS Counterpart\vspace{1ex}\\ \hline
B18-4 & 04325+2402\\
L1582A$^a$ & none\\
IRAS05413 & 05412-0105\\
 & 05413-0104\\
L1172A & 21017+6742\\
L1177 & 21169+6804\\
\hline
\end{tabular}
\end{center}
\rule{0ex}{3Ex}Notes: a) MIPS source at 05~32~02.9, +12~31~05 (J2000.0)
\end{table}

\begin{table}
  \caption{Cores in the c2d MAMBO survey with YSOs uncovered by c2d
    that were believed to be starless at the beginning of the c2d
    survey. Comments on outflows refer to jet-like extended
    ($\gg 1\arcmin$) features in IRAC images.
\label{tab-survey:spitzer_new}}
\begin{center}
\begin{tabular}{llllllllll}
\hline \hline
\multicolumn{2}{l}{\rule{0ex}{3Ex}MAMBO Core} & Comments\vspace{1ex}\\ \hline
L1521F & C1 & VeLLO, \citet{bourke2006:l1521f}\\
L1582A & C2$^a$ & no c2d Spitzer data, possible outflow\\
L673-7 & C1$^a$ & VeLLO, Dunham et al., in prep.\\
L1148 & C1 & VeLLO, Kauffmann et al.\ \citeyear{kauffmann2005:vellos}
  and in prep.\\
L1014 & C1 & VeLLO, \citet{young2004:l1014}\\
L1251A & C3$^a$ & new class 0, prominent outflow\\
 & C4$^a$ & new class 0, possible outflow\\
\hline
\end{tabular}
\end{center}
\rule{0ex}{3Ex}Notes: a) reported here for the first time
\end{table}

The data on the MAMBO-associated Spitzer sources is complemented by
2MASS data where possible. Given the nominal positional uncertainties,
2MASS sources within $2 \arcsec$ from Spitzer sources are assumed to
be counterparts of these. For Spitzer sources without 2MASS
counterparts upper limits to their flux densities in the 2MASS bands
are derived from upper limits for nearby 2MASS sources not detected in
all filters. The source emission is thus probed in the
$1 ~ {\rm to} ~ 70 ~ \rm \mu m$ wavelength range (by 2MASS and Spitzer;
no c2d Spitzer
data is available at $160 ~ \rm \mu m$) and at $1 \, 200 ~ \rm \mu m$
(from MAMBO for the $4 \, 200 ~ \rm AU$ aperture).\medskip

\noindent Note that probably many more Spitzer sources are associated
with the MAMBO cores observed in our study. However, here we are only
interested in sources directly associated with features (i.e., peaks)
seen in our MAMBO maps. The discussion of the other sources --- which
are expected to belong to the YSO classes II and III, since they lack
significant millimetric dust emission --- is deferred to a later
paper. Also note that the emission of some YSOs might be spatially
confused with dense core emission. For example, the PSC source
04326+2405 in B18-4 (marked by the only circle) might manifest as an
extension in the MAMBO dust emission contours, which however cannot be
uniquely separated from the dense core emission.

\subsection{YSO Properties\label{sec-survey:protostar_properties}}
Below we characterise the sources associated with dust emission peaks.
All our target cores were covered by the extensive but insensitive
IRAS and 2MASS surveys. These data thus allows for a homogeneous
quantification of all sources in our survey that are brighter than a
few times $0.1 \, L_{\odot}$. Fainter sources could only be detected
by Spitzer, from which data exists only for a fraction of our dense
cores.

Given the differences in the spectral bands probed by these
instruments, it is difficult to compare a source only detected by
2MASS and IRAS to one only detected by 2MASS and Spitzer. In order to
explore and suppress related biases in estimates of source properties,
we therefore analyse the Spitzer and IRAS data separately. To better
distinguish results from the different approaches, in the following
properties are labeled by the data source used to derive them
(superscripts `IRAS' and `SST', the latter for the Spitzer Space
Telescope).

\subsubsection{Estimates from IRAS and 2MASS Data}
Following \citet{lada1987:ass_to_protostars}, the spectral properties
of young stars can be characterised by the spectral index between
$12$ and $25 ~ \rm \mu m$ wavelength,
\begin{equation}
\alpha^{25 ~ \rm \mu m}_{12 ~ \rm \mu m} =
\frac{
  \log (
  {12 ~ \rm \mu m} \cdot F_{25 ~ \rm \mu m}
  /
  [{25 ~ \rm \mu m} \cdot F_{12 ~ \rm \mu m}]
  )
}{
  \log ({25 ~ \rm \mu m} / {12 ~ \rm \mu m})
} \, .
\end{equation}
Spectral indices are listed in Tables
\ref{tab-survey:IRAS_not-detected} and \ref{tab-survey:IRAS_detected}
for all IRAS sources in the MAMBO maps, if detected in these bands.
They roughly probe whether the spectral energy distribution (SED) is
dominated by photospheric or envelope emission and are sometimes used
to classify observed YSO SEDs within evolutionary schemes
\citep{lada1987:ass_to_protostars}.

For the IRAS sources associated with MAMBO peaks for which 2MASS
data is available, we calculate the bolometric temperature
defined by \citet{myers1993:bol_temperatures},
\begin{equation}
T_{\rm bol} = \frac{\zeta(4)}{4\zeta(5)}
\frac{h \langle \nu \rangle}{k_{\rm B}} \, ,
\label{eq-survey:def-t_bol}
\end{equation}
where
$\langle \nu \rangle =
\int_0^{\infty} \nu F_{\nu} \d \nu \left/
\int_0^{\infty} F_{\nu} \d \nu \right.$
is the flux-weighted mean frequency, $\zeta$ is the Riemann zeta
function, and $h$ and $k_{\rm B}$ are Planck's and Boltzmann's
constant. We integrate across the SED by interpolation between
observed 2MASS and IRAS bands. For this we use piecewise power laws
matching the flux densities, as indicated in Fig.\
\ref{fig-survey:example_sed}. At $500 ~ \rm \mu m$ wavelength these
connect to a power-law representing a modified blackbody of opacity
$\propto \lambda^{-2}$ and $15 ~ \rm K$ temperature that is tuned to
fit the MAMBO data; piecewise powerlaws are adopted to hold between
$\lambda = 500 ~ \rm \mu m$ and $1200 ~ \rm \mu m$ (which is MAMBO's
central wavelength), and $\lambda = 1200 ~ \rm \mu m$ and $\infty$
(adopting a flux density distribution $\propto \lambda^{-4}$). For
opacity-modified blackbodies of $10 ~ {\rm to} ~ 20 ~ \rm K$
temperature this approximates the actual flux density at
$\lambda = 500 ~ \rm \mu m$ within a factor of 1.5. The
related uncertainty on integrated
properties is much lower though, since the submillimetre part of the
SED usually only contributes a few percent to the total luminosity
(see, e.g., Table \ref{tab-survey:IRAS_detected}). The
integration extends from the 2MASS bands at $\approx 1 ~ \rm \mu m$ to
infinity. Unlike \citeauthor{myers1993:bol_temperatures} we treat all
flux density upper limits like actual detections. Therefore, the
derived bolometric temperatures must be interpreted with some caution,
if a source is not detected in some of the bands. In these cases lower
and upper limits are assigned to the bolometric temperature, depending
on whether adopting flux densities below the upper limits would
increase or decrease the calculated mean frequency,
respectively.\medskip

\setcounter{figure}{1}
\begin{figure}
\includegraphics[height=\linewidth,bb=148 55 468 608,clip,angle=-90]{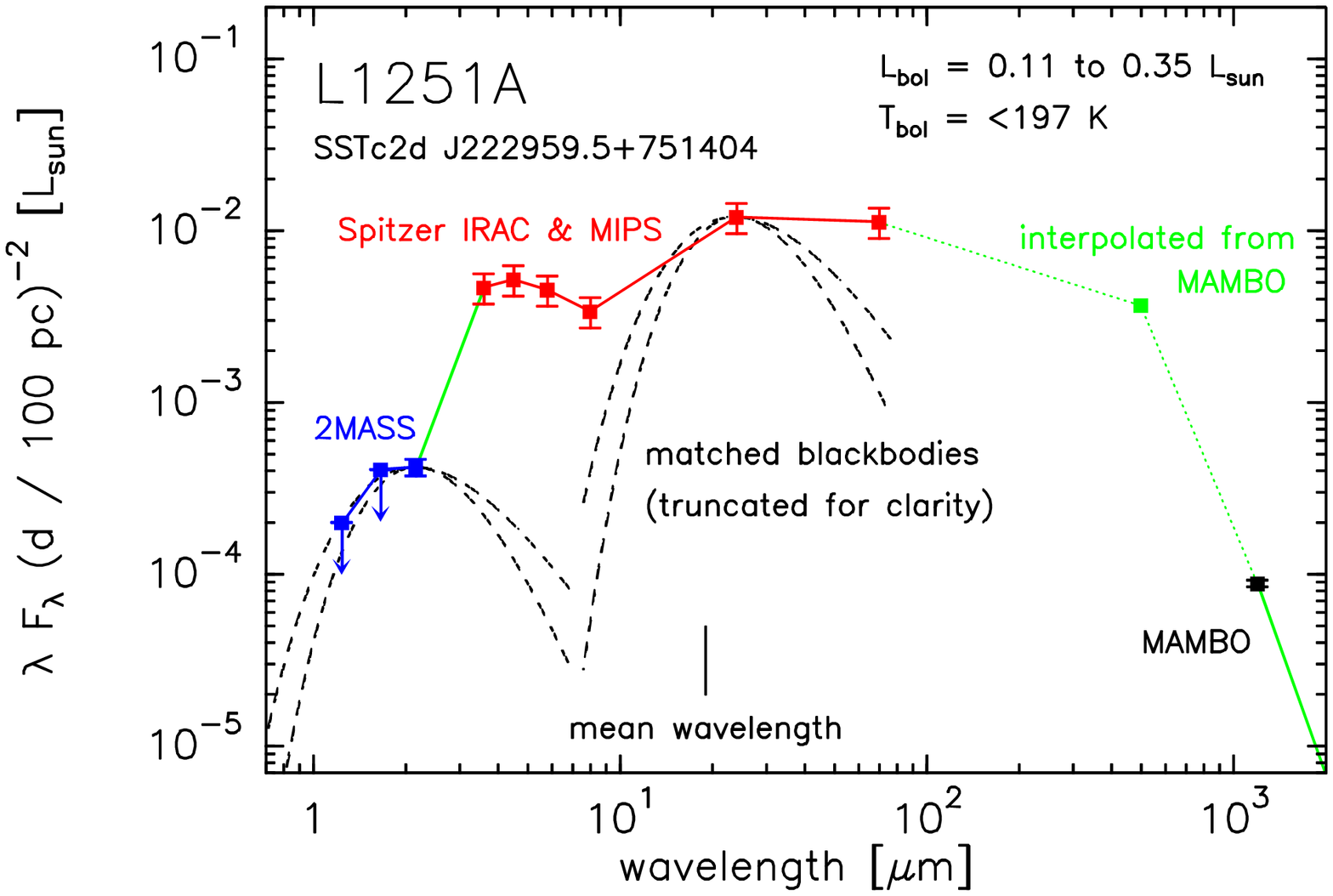}\\
\includegraphics[height=\linewidth,angle=-90]{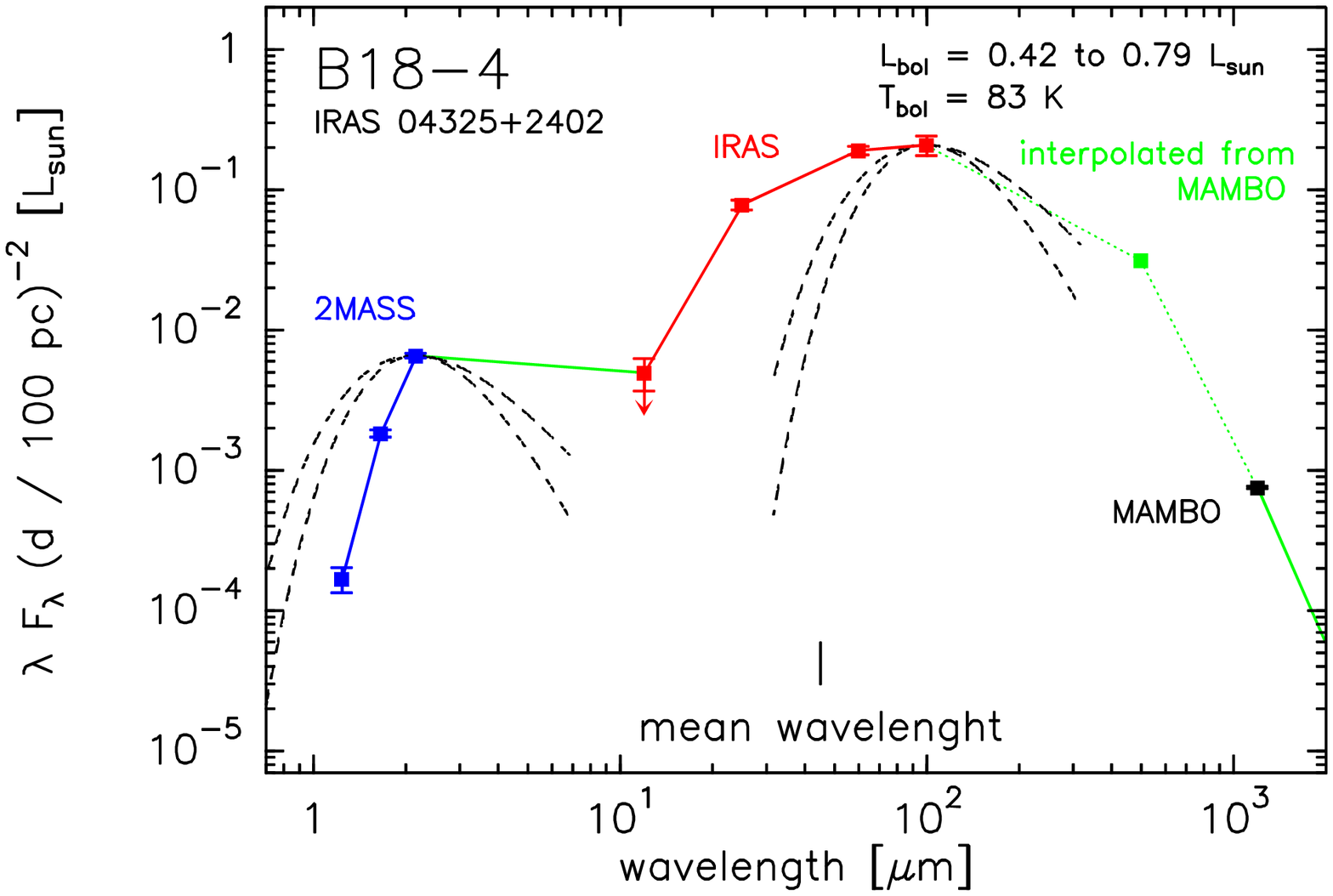}
\caption{Example SEDs for YSOs in the c2d
  MAMBO sample. The \emph{top panel} shows the infrared star SSTc2d
  J222959.5+751404 associated with the P3 peak in the C2 subcore of
  L1251A, while the \emph{bottom panel} shows the data for IRAS
  04325+2402 associated with the P1 peak in B18-4. The SEDs are
  composed from 2MASS, Spitzer or IRAS, and MAMBO data, as indicated by
  \emph{labels}. Also indicated are the interpolated datapoints at
  $500 ~ \rm \mu m$ that is calculated from the MAMBO flux density
  assuming a dust opacity $\propto \lambda^{-2}$ and a dust
  temperature of $15 ~ \rm K$. \emph{Straight lines} indicate the
  flux density interpolation used to calculate the bolometric
  luminosity and temperature indicated in the \emph{upper right
    corners} (as well as the \emph{mean wavelengths} calculated as
  $c / \langle \nu \rangle$ from Eq.\ [\ref{eq-survey:def-t_bol}] and
  indicated at the frame bottom). Only
  \emph{limits} can be derived on $T_{\rm bol}$ and $L_{\rm bol}$
  because of some upper limits in the 2MASS data; a conservative lower
  limit to the luminosity is set by the sum of opacity-modified
  blackbodies matched to the SED at short and long wavelength
  (\emph{dashed lines}, drawn for opacities scaling
  $\propto \nu^{0 ~ {\rm to} ~ 2}$; see text for
  details). The ``sign'' of the bolometric temperature limits is set
  by the mean
  wavelength lying at longer wavelength than all photometric
  bands with upper limits.\label{fig-survey:example_sed}}
\end{figure}

\noindent Based on the derived bolometric temperature the YSOs are
separated into four infrared classes. Following \citet{chen1995:tbol,
  chen1997:five_clouds} all stars with $T_{\rm bol} < 70 ~ \rm K$
are considered to be class 0 sources. Those with
$70 \le T_{\rm bol} / {\rm K} < 650$ belong to class I,
for $650 \le T_{\rm bol} / {\rm K} < 2880$ they belong to class II, and
class III sources have $T_{\rm bol} \ge 2880 ~ \rm K$. In case of
upper or lower limits to the bolometric temperature, the class
corresponding to the derived limit is given. This treatment of
uncertainties in class assignments is somewhat
unsatisfying, however it does not influence our later analysis that is
based on the numerical value of $T_{\rm bol}$.

Sources in a given class (i.e., with similar bolometric temperature)
are believed to be in a similar evolutionary stage
(\citealt{chen1995:tbol} and \citealt{chen1997:five_clouds}; see
\citealt{young2005:yso_models} for illustrative examples). In their
evolution from a deeply embedded object to a star surrounded by a
remnant dust disk, YSOs are though to pass through phases 0 to III in
increasing order. Thus class 0 objects would represent the youngest
YSOs and class III objects would be in a stage just prior to the main
sequence. Almost all of the MAMBO-detected stars covered by the
present survey are in classes 0 and I; only one class II and no class
III object are among them.\medskip

\noindent From the interpolated SED we derive an estimate for the
bolometric luminosity,
\begin{equation}
L_{\rm bol} = 4 \pi d^2 \int_0^{\infty} F_{\nu} \d \nu \, ,
\end{equation}
for IRAS sources associated with MAMBO dust emission peaks. If the
source is detected in a few bands only, an upper limit to the
bolometric luminosity is calculated by integrating across the flux
density upper limits, while a lower limit is given by the sum of the
maximum 2MASS ``monochromatic luminosity'', $\nu F_{\nu}$, plus the
maximum detected $\nu F_{\nu}$ in the bands of longer wavelength (see
Fig.\ \ref{fig-survey:example_sed} for an illustration). The
latter builds on the fact that for
blackbodies modified by an opacity scaling with frequency as
$\nu^{0 ~ {\rm to} ~ 2}$, the bolometric luminosity is
the peak value of $\nu F_{\nu}$ times a factor 1.05
to 1.36. In a similar fashion we derive the luminosity in the
submillimetre wavelength range,
\begin{equation}
L_{\rm submm} = 4 \pi d^2 \int_0^{c / 350 ~ \rm \mu m}
F_{\nu} \d \nu \, ,
\end{equation}
where $c$ is the speed of light. For both luminosities we also quote
uncertainties due to statistical noise, while we do not consider
systematic contributions. The aforementioned uncertainties in the
submillimetre SED near $500 ~ \rm \mu m$ wavelength do, however,
render the submillimetre luminosity uncertain by a factor of order 2;
no precision measurements of it from our data is possible.

\subsubsection{Estimates from Spitzer and 2MASS
  Data\label{sec-survey:estimates_spitzer}}
The YSO properties derived from Spitzer and 2MASS data (Table
\ref{tab-survey:SST_protostars}) are calculated using
methods similar to those
adopted for the combined IRAS and 2MASS data. The spectral sampling at
wavelengths $\gtrsim 25 ~ \rm \mu m$
is, however, much worse than for the combined 2MASS and IRAS dataset,
resulting in more uncertain estimates for cold sources
(with SED peaks at wavelengths $\gtrsim 25 ~ \rm \mu m$, i.e.\ with
$T_{\rm bol} \lesssim 160 ~ \rm K$).
The spectral index for the 2MASS and Spitzer bands,
$\alpha_{2.2 ~ \rm \mu m}^{24 ~ \rm \mu m}$, is derived by fitting the
monochromatic luminosities in the $2.2 ~ {\rm to} ~ 24 ~ \rm \mu m$
wavelength range by a power law. It is, thus, not directly comparable to
the spectral index derived from IRAS data.

Four stars in our sample might qualify as VeLLOs, given that they
appear to be embedded in dense cores and have internal luminosities
likely below $0.1 \, L_{\odot}$. This group includes the sources
associated with L1521F P1, L1148 P1, and L1014 P1, which have been
subject to detailed studies (\citealt{bourke2006:l1521f};
\citealt{kauffmann2005:vellos}, and in prep.;
\citealt{young2004:l1014}). These studies confirm the spectral
properties found here. In particular, based on detailed radiative
transfer SED modeling, they support the prevalence of internal
luminosities $< 0.1 \, L_{\odot}$. This is an important factor, since
the SED analysis presented here has bad spectral sampling at long
wavelength, leading to uncertain properties for cool sources. It is
also not suited to single out the internal luminosity due to an
embedded VeLLO. Note that SED modeling of the source in L673-7 has
not been presented so far, leading to a more uncertain status of this
source.

Given the evidence for YSO outflows from scattered light nebulosity
and outflows seen in CO, the stars in L1521F P1, L1014 P1, L1148 P1,
and L673-7 P1 are proven to be associated with the respective cores
and are not unrelated background objects (see the above references and
additionally Dunham et al., in prep., for L673-7, and
\citealt{bourke2005:l1014-outflow} for L1014). A recent search for
YSOs with internal luminosities $< 0.1 \, L_{\odot}$ by Dunham et al.\
(in prep.) lists three of these sources in their ``group 1'' of
confirmed low-luminosity YSOs; L1448-IRS is listed in their ``group
3'' of good, but unconfirmed, candidates for low-luminosity YSOs. The
star in L1251A P5 however, for which our above analysis suggests a low
luminosity, is not likely to be a VeLLO since its luminosity limits
appear to be too high.

\subsubsection{Properties chosen for the Analysis}
The above analysis yields two different sets of properties, one based
on Spitzer data, one based on IRAS. For the bright sources of a few
$0.1 \, L_{\odot}$, the bolometric temperatures and luminosities estimated from
IRAS data are used for the further YSO analysis. This is the
preferred option because IRAS data is available for all our cores, and
because Spitzer fails to detect a significant fraction of the
YSO emission at long wavelengths (see Sec.\
\ref{sec-survey:temp_and_lumi}, leading to biased estimates. We
abstain from constructing SEDs combining flux densities from IRAS and
Spitzer in order to better control mission-related systematic trends
in the data. The IRAS data might, on the other hand,
suffer from contamination of unresolved nearby sources. Such
contamination is not apparent at a significant level in any of the
cases where we have data from both IRAS and Spitzer though.

Fainter sources are usually only detected by Spitzer. Thus, Spitzer
data is used for the study of the sources in L673-7 P1, and L1251A P4
and P5. For L1521F P1, L1148 P1, and L1014 P1 results from the more
involved Spitzer data analysis by \citet{bourke2006:l1521f}, Kauffmann
et al.\ (in prep.), and \citet{young2004:l1014} are used.

\section{Analysis\label{sec-survey:analysis}}
In this section we exploit the survey data in order to better
understand the state and evolution of starless and YSO dense cores. We
discuss the general core properties before we turn to the discussion
of specific issues. Throughout the following discussion we exclude
starless peaks and subcores with uncertain properties, unless noted
otherwise.

\subsection{General Core Properties\label{sec-survey:morphology}}
\begin{figure*}
  \begin{tabular}{cc}
    \includegraphics[width=0.47\linewidth,bb=35 12 380 262,clip]{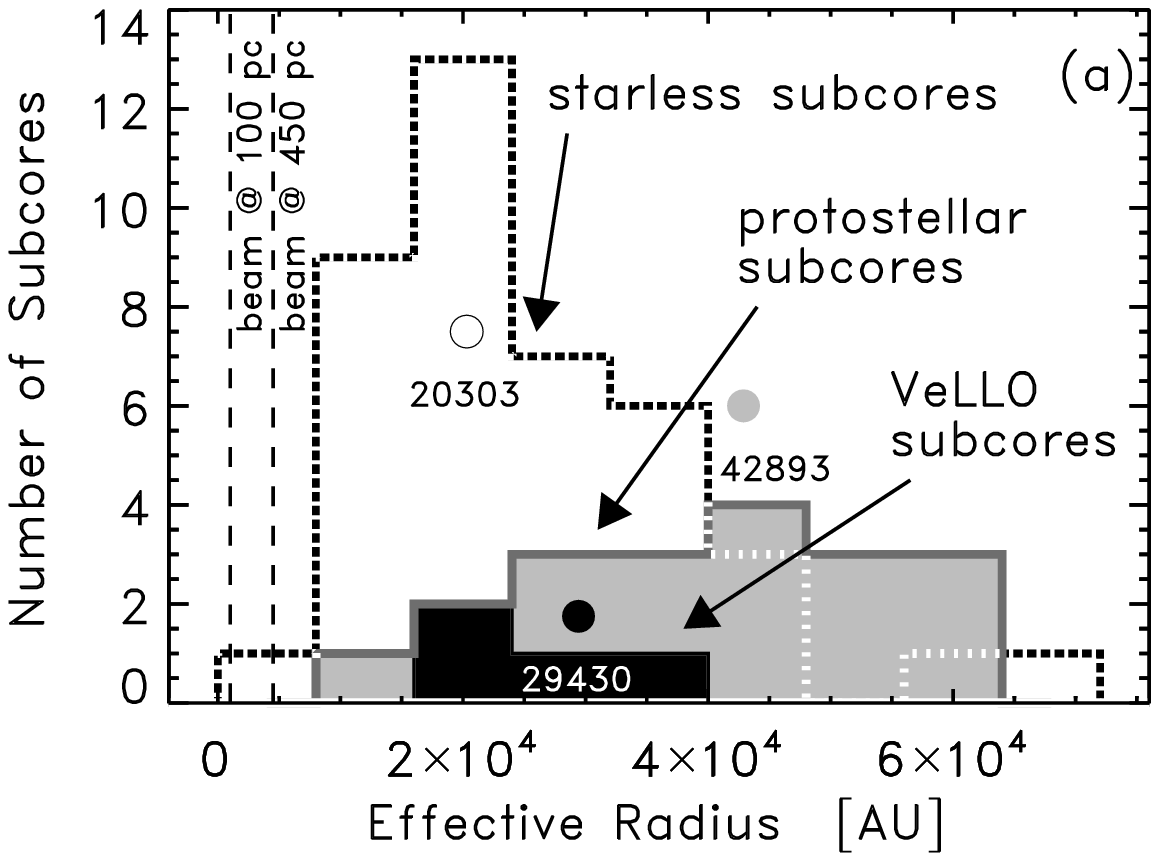} &
    \includegraphics[width=0.47\linewidth,bb=35 12 380 262,clip]{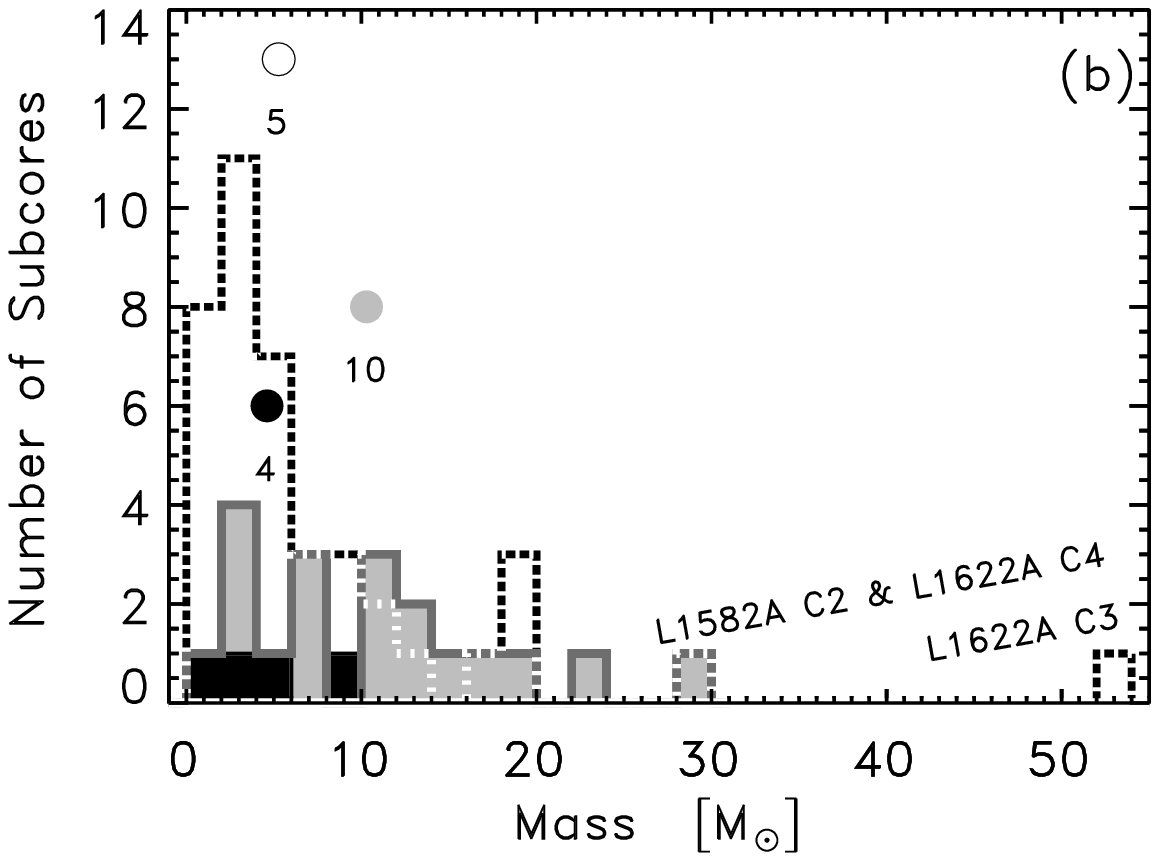}\\
    \includegraphics[width=0.47\linewidth,bb=35 12 380 262,clip]{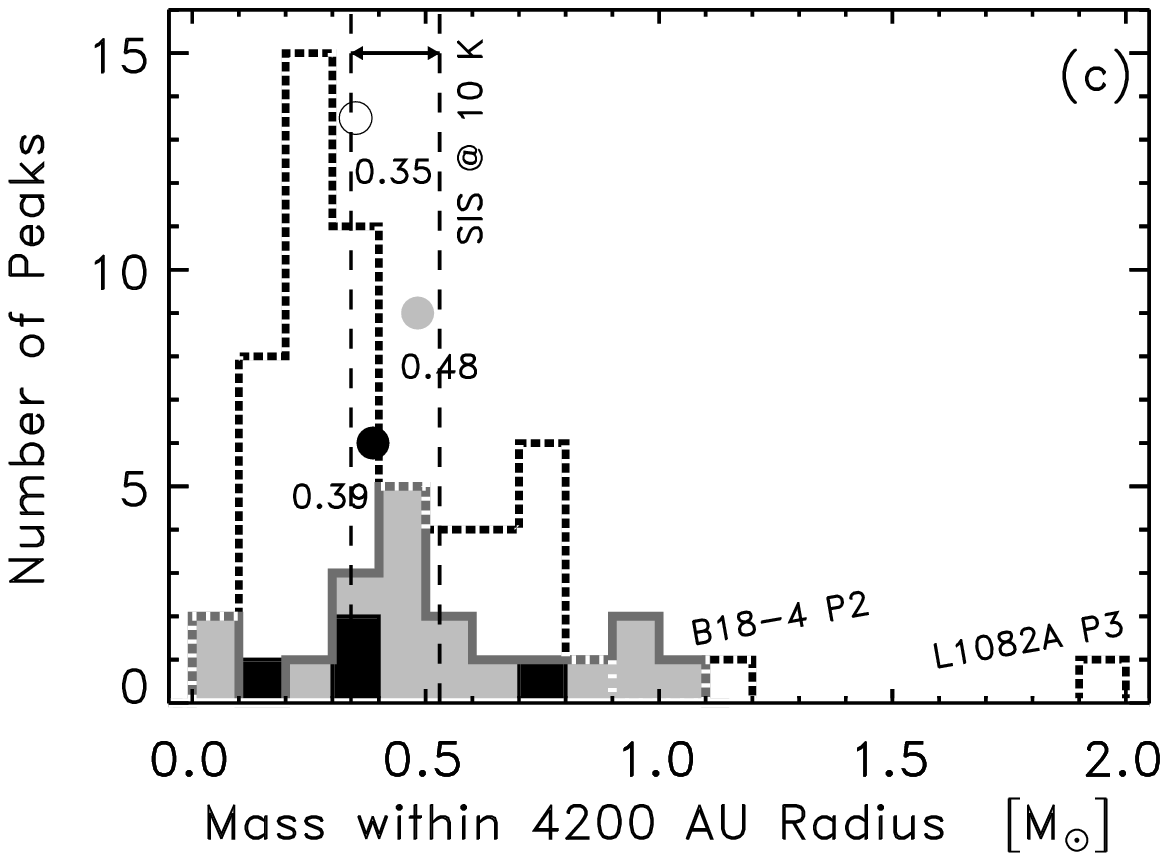} &
    \includegraphics[width=0.47\linewidth,bb=35 12 380 262,clip]{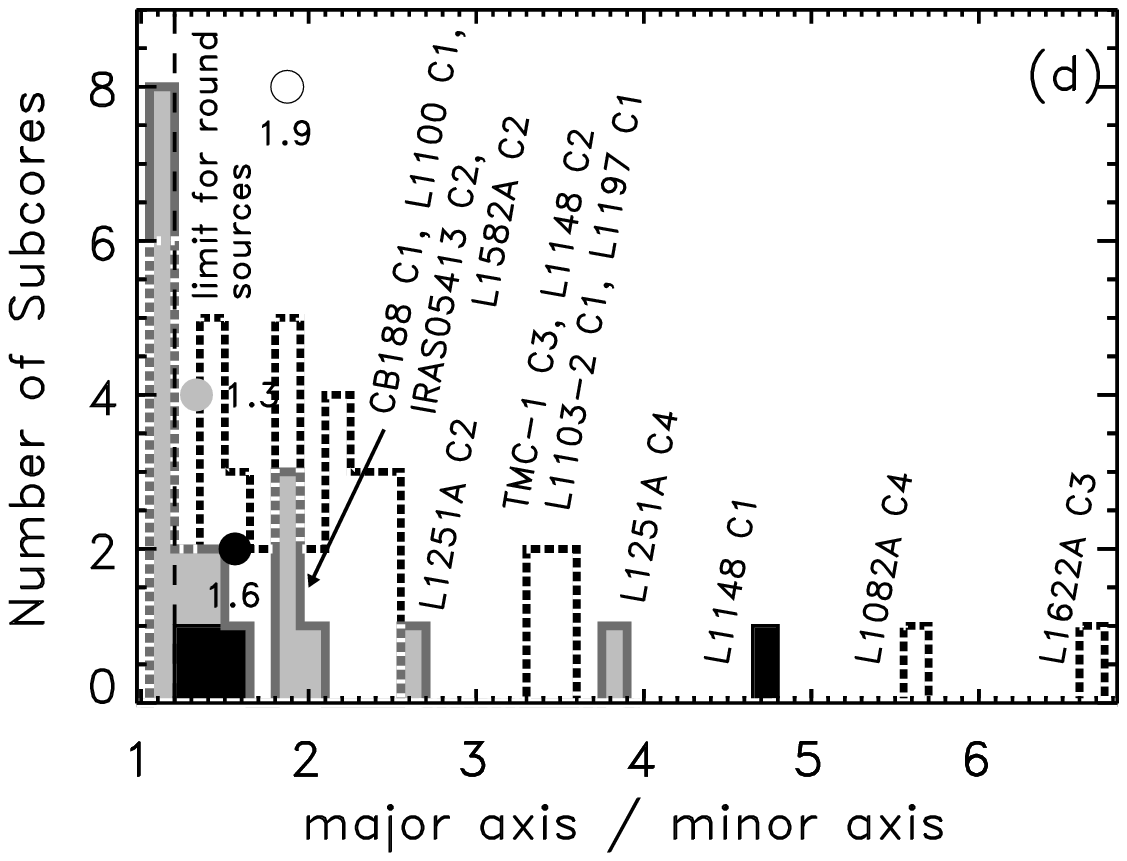}
  \end{tabular}
\caption{Overview of the dense core properties, excluding subcores and
  peaks with uncertain properties due to artifacts. The
  \emph{dashed line} represents the properties of starless peaks and
  subcores, while the \emph{grey and black areas} hold for
  YSO and candidate VeLLO cores, respectively. \emph{Bullets} of the same
  colours, and the related numbers, give
  the respective median values of the distributions. \emph{Vertical
    lines in panel a)} give the beam radius at distances of
  $100 ~ \rm pc$ and $450 ~ \rm pc$, respectively. The \emph{vertical
    lines in panel c)} show the range of masses within a $4 \, 200 ~ \rm AU$
  aperture for singular isothermal spheres of $10 ~ \rm K$ temperature
  and truncation radii of $4 \, 200 ~ \rm AU \to \infty$ . The
  \emph{vertical line in panel d)} marks the boundary between round
  and elongated sources. Typical uncertainties of the shown properties
  are $\ge 10\%$ for $M_{\rm tot}$, a few $0.01 \, M_{\odot}$ for
  $M_{4200 \rm AU}$, and 10\% to 20\% for the axis ratio.\label{fig-survey:histograms}}
\end{figure*}

Figure \ref{fig-survey:histograms} shows the frequency distributions of some
properties of well reconstructed dense cores. Some conclusions about
the physical state of individual cores, and of dense cores in
general, can be derived from these distributions.

\subsubsection{Radius}
The sizes of YSO cores are on average larger than those of starless
and candidate VeLLO cores. This could to some extent be an
observational bias. Only one YSO subcore in our sample is at a
distance below $200 ~ \rm pc$, while two dozen starless subcores are
within this distance. Because confusion increases with distance, the
subcores identified in the more distant YSO cores are thus likely
biased towards larger sizes. Large cores might, however, indeed have
better chance to form stars.

At small radii the size distribution of starless cores is
limited by the beam size; their number steeply drops to zero at the
resolution limit. We might miss a population of small subcores.

\subsubsection{Total Mass}
The mass of the subcore L1622A C3 exceeds the mass of all other
subcores by a factor 1.8 and more. Given that L1622A is one out of
only 3 dense cores located in the Orion star forming complex, in which
special environmental conditions prevail, it might be that
L1622A C3 has, e.g., a mass much larger than the other cores in
our sample. However, the extreme mass contrast to all other cores does
cast some doubt on this.

It might be that L1622A is warmer than most other cores. By adopting a
dust temperature of $10 ~ \rm K$ one would then overestimate the true
mass. Furthermore, substructure easily disentangled in more nearby
cores may be confused in L1622A, the core with the largest distance in
our sample ($450 ~ \rm pc$). Then the subcore area, and therefore the
mass, would be biased towards higher values. This is supported by the
fact that L1622A C3 has the largest effective radius of all subcores.
Also, L1582A C2 and L1622A C4, which are second to L1622A C3 in mass
and are also located in Orion, are second to L1622A C3 in the
effective radius. This may hint that mass and size estimates of cores
are indeed biased for larger distances.

Thus, data on distant cores must be interpreted with some caution.
They might well have extreme properties, but given the lack of
sufficient spatial resolution, we cannot be sure about this.
Properties of nearby and distant cores are not necessarily directly
comparable.

\subsubsection{Mass within 4$\,$200~AU\label{sec-survey:mass_within_4200au}}
For L1082A P3 the mass $M_{4200 \rm AU}$ exceeds those of all other
peaks by a factor 1.7 and more. This large aperture mass contrast may
hint on a biased mass estimate. The dust in L1082A P3 might be
significantly warmer than $10 ~ \rm K$, if this core is heated by an
undetected embedded star. Following \citet{myers1987:nir_iras}, the
failure of IRAS to detect a point source in this peak implies (for a
distance of $400 ~ \rm pc$) an upper limit to the bolometric
luminosity of $\approx 0.8 \, L_{\odot}$ to any embedded source. In
this case the average dust temperature could be $15 ~ \rm K$ instead
of the assumed $10 ~ \rm K$, and the mass could be overestimated by us
by up to a factor of 2. Then
$M_{4200 \rm AU}$ would not be unusual for a YSO core. Though
this evidence is not conclusive, we, thus, suspect the presence of a
heating YSO source in L1082A P3 that is too faint to be
detected by IRAS. Unfortunately, no Spitzer data is available for
this region to confirm this.

The peak B18-4 P2 is separated
from the other starless subcores. As Spitzer images show, no
point sources are associated with this peak, and so internal heating
cannot explain the unusual value of $M_{4200 \rm AU}$. A YSO of
$\approx 0.5 \, L_{\odot}$ resides in B18-4 P1, which is separated
from B18-4 P2 by $10 \, 000 ~ {\rm AU} = 0.05 ~ {\rm pc}$. At this
distance a YSO of this low luminosity is unlikely to provide
significant external heating to the peak P2. Thus the high value of
$M_{4200 \rm AU}$ in B18-4 P2 is likely to be real.

The $M_{4200 \rm AU}$ distributions for all types of subcores peak at
a mass of $\approx 0.35 \, M_{\odot}$. The mean ($0.42 \, M_{\odot}$,
$0.54 \, M_{\odot}$, and $0.41 \, M_{\odot}$ for starless, YSO, and
VeLLO cores, respectively) and median ($0.35 \, M_{\odot}$,
$0.48 \, M_{\odot}$, and $0.39 \, M_{\odot}$, respectively) aperture
mass is of the same order. Since this is a typical value for all
kinds of subcores, the underlying physical process shaping the
distributions must be fundamental. Interestingly, this mass is
similar to the mass within $4 \, 200 ~ \rm AU$ for a (truncated)
singular hydrostatic equilibrium isothermal sphere (SIS) of
$10 ~ \rm K$ gas temperature,
\begin{equation}
M_{4200 \rm AU} =
  0.34 \, M_{\odot} \,
  \varepsilon \, (T_{\rm g} / 10 ~ {\rm K}) \, .
\end{equation}
Here $T_{\rm g}$ is the gas temperature and the factor $\varepsilon$
depends on the density distribution outside a radius of
$4 \, 200 ~ \rm AU$; it is 1 if the density drops
to zero outside $4 \, 200 ~ \rm AU$ and becomes
$\pi / 2 \approx 1.57$ if it continues out to infinity. This mass
constitutes a critical value for an SIS: no
hydrostatic equilibrium solutions exist for higher values of
$M_{4200 \rm AU}$. It thus might be that this critical condition
manifests in characteristic values of dense core properties.

If this is true it has two interesting implications. First, dense
cores would preferentially exist in a close-to-critical
physical state. We note that interestingly such near-critical core
states have previously been inferred by \citet{alves2001:b68} and
particularly by \citet{kandori2005:globules}. Second, the total
pressure in the core, $P$, appears to
be comparable to the thermal pressure. Contributions to the total
pressure from the effect of turbulent gas motions or magnetic fields
can not much exceed the thermal pressure. Otherwise the critical mass
in the $4 \, 200 ~ \rm AU$ aperture would be much larger since one
would need to replace the gas temperature with some higher effective
one, $T_{\rm g} \to T_{\rm eff} = 2.33 \, P m_{\rm H} / (\varrho
k_{\rm B}) > T_{\rm g}$, where $m_{\rm H}$ is the hydrogen mass and
the factor 2.33 holds for a gas mixture at cosmic abundance with most
hydrogen in molecular form.

\subsubsection{Elongation\label{sec-survey:elongation}}
Only 14\% and 44\% out of all starless and YSO subcores, respectively,
have major-to-minor axis ratios $\le 1.2$ (where the uncertainty is
10\% to 20\%) and can be considered round; most subcores are not
round. Unless these subcores are shaped by magnetic fields, which
provide a non-isotropic supporting pressure, they can hardly on the
whole be in a state of hydrostatic equilibrium. 

The axis ratio distribution is continuous up to ratios of 2.0 and 2.5
for YSO and starless subcores, respectively.  Several subcores have
much larger axis ratios. Interestingly, two of the most distant
subcores have the largest elongations (L1082A C4 and L1622A C3 at
distances $\ge 400 ~ \rm pc$). This suggests that some cores appear
elongated due to unresolved cores within the beam, an effect that on
average increases with distance (also see \citet{young2006:ophiuchus}
for this effect in Bolocam maps).

For starless cores, the median axis ratio that we derive, i.e.\ 1.9,
is very similar to those derived from $\rm NH_3$
(\citealt{jijina1999:database}, median of 1.5 to 2.2) and extinction
maps (\citealt{lee1999:optical_cores}, mean of $2.4 \pm 0.1$) of
larger core samples. For YSO cores, however, the median aspect ratio
of 1.3 for our sample is at the low end of the aspect ratios derived
by the above studies (medians of 1.4 to 2.0, respectively a mean of
$2.2 \pm 0.2$, for their YSO cores). This difference is probably an
effect of the internal heating on the dust emission intensity
distribution (see below).

The axis-ratio distribution has in the past been used to study the
intrinsic three-dimensional geometry of dense core shapes. Following
\citet{myers1991:core_shapes}, the median axis ratio of 1.9 for
starless subcores (we exclude YSO cores, since internal heating
influences the observed shapes; see next paragraph) is the expected
projected axis ratio of prolate (``cigar-like'') spheroids with an
intrinsic major-to-minor axis ratio of $\approx 2.5$, and of oblate
(``disk-like'') ones with an axis ratio of $\gtrsim 4$ (for an
observed ratio $> 1.7$), respectively\footnote{This analysis assumes a
  random distribution of the core axes in space. This might not be
  justified for our sample, since most of our cores are situated in a
  small number of star-forming regions, in which preferred orientations
  might prevail.} We cannot resolve this ambiguity. Some existing
theoretical analysis favors prolate cores
(\citealt{myers1991:core_shapes}; \citealt{ryden1996:core_shapes}),
while other work prefers oblate ones (\citealt{jones2001:core_shapes};
\citealt{jones2002:core_shapes}; \citealt{goodwin2002:core_shapes};
\citealt{tassis2007:core_shapes}).

The median axis ratio of YSO subcores is smaller than that of starless
and VeLLO cores. This could be because the YSO heating leads to an
intensity peak of small physical size and low aspect ratio (since the
heating, to first order, decreases radially away from the star; Eq.\
\ref{eq-survey:quant-1}), which is blurred into a structure of even
lower aspect ratio because of beam smearing. On the other hand stars
are expected to form in close-to-spherical density enhancements, which
in projection show a low elongation \citep{shu1977:self-sim_collapse}.
Then, YSO cores would be less elongated than the starless cores with
their complex shapes at lower intensity contours. Both effects would
also explain the aforementioned decrease of the aspect ratio from the
large-scale core structure probed by, e.g., $\rm NH_3$ and extinction
maps to the core centers probed by dust emission. A more detailed
analysis is needed to separate these effects.

The YSO subcores CB188 C1, IRAS05413 C2, L1100 C1, L1251A C2,
and L1251A C4 are well separated from the other YSO cores in the
elongation distribution. For IRAS05413 C2, however, the axis ratio is
very uncertain ($1.9 \pm 0.5$). The observed elongation is
likely to be an observational artifact. In L1251A, where the major
axis exceeds the beam size by a factor of several, this elongation
likely reflects the morphology of the dense core from which the
embedded star formed. For CB188 and L1100, however, where
the major axis exceeds the beam size by a factor 3.2 and less,
this elongation should reflect the morphology of the immediate density
peak from which the young star accretes. This region could be shaped
by the interaction with outflows from the central star. An outflow has
indeed been detected toward CB188 \citep{yun1994:outflows} and for
L1100 there is some evidence for an outflow from broad line wings
\citep{devries2002:br_clouds}. For CB188 the position of the outflow
axis derived by \citet{yun1994:outflows} and the major axis of the
dust emission intensity distribution are neither aligned nor
perpendicular (position angles of $\approx 75 \degr$ and
$16 \degr \pm 5 \degr$, respectively). If the elongated structure
towards CB188 seen in the MAMBO maps is indeed related to the
outflows, then the dust emission feature is likely to trace the wall
of an outflow cavity. If it would trace the jet or a YSO disk,
the axes are expected to be parallel or perpendicular. While these
observations do not prove the interaction between jets and the dense
core, they motivate dedicated studies on this issue. A more detailed
analysis of the dust emission maps might yield more
candidates for jet-core interactions.

%
%
%
%
%
%

\subsection{Correlations between Core Properties}
\begin{figure*}
\begin{tabular}{cc}
\includegraphics[width=0.47\linewidth,bb=15 8 420 305,clip]{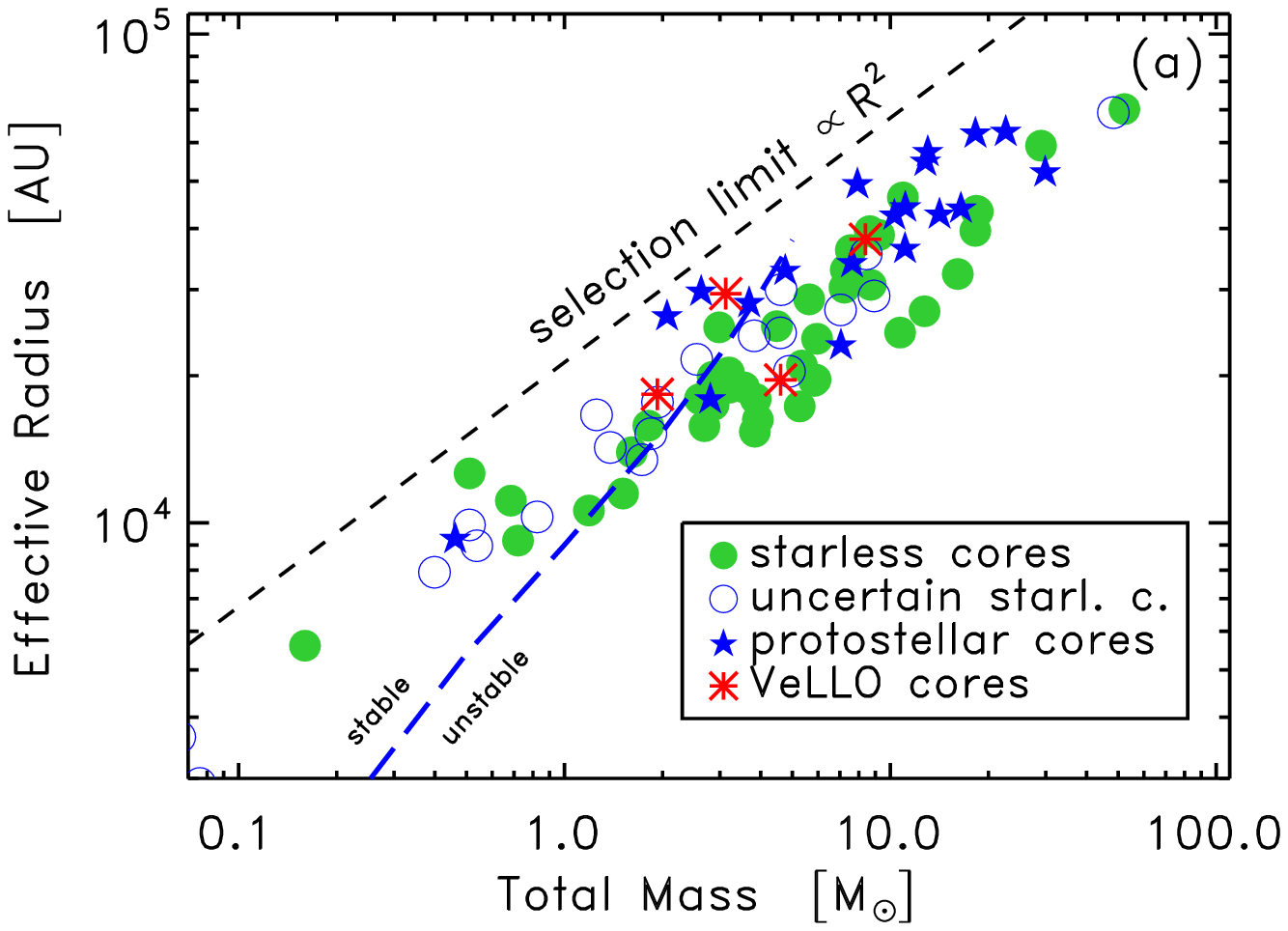} &
\includegraphics[width=0.47\linewidth,bb=15 8 420 305,clip]{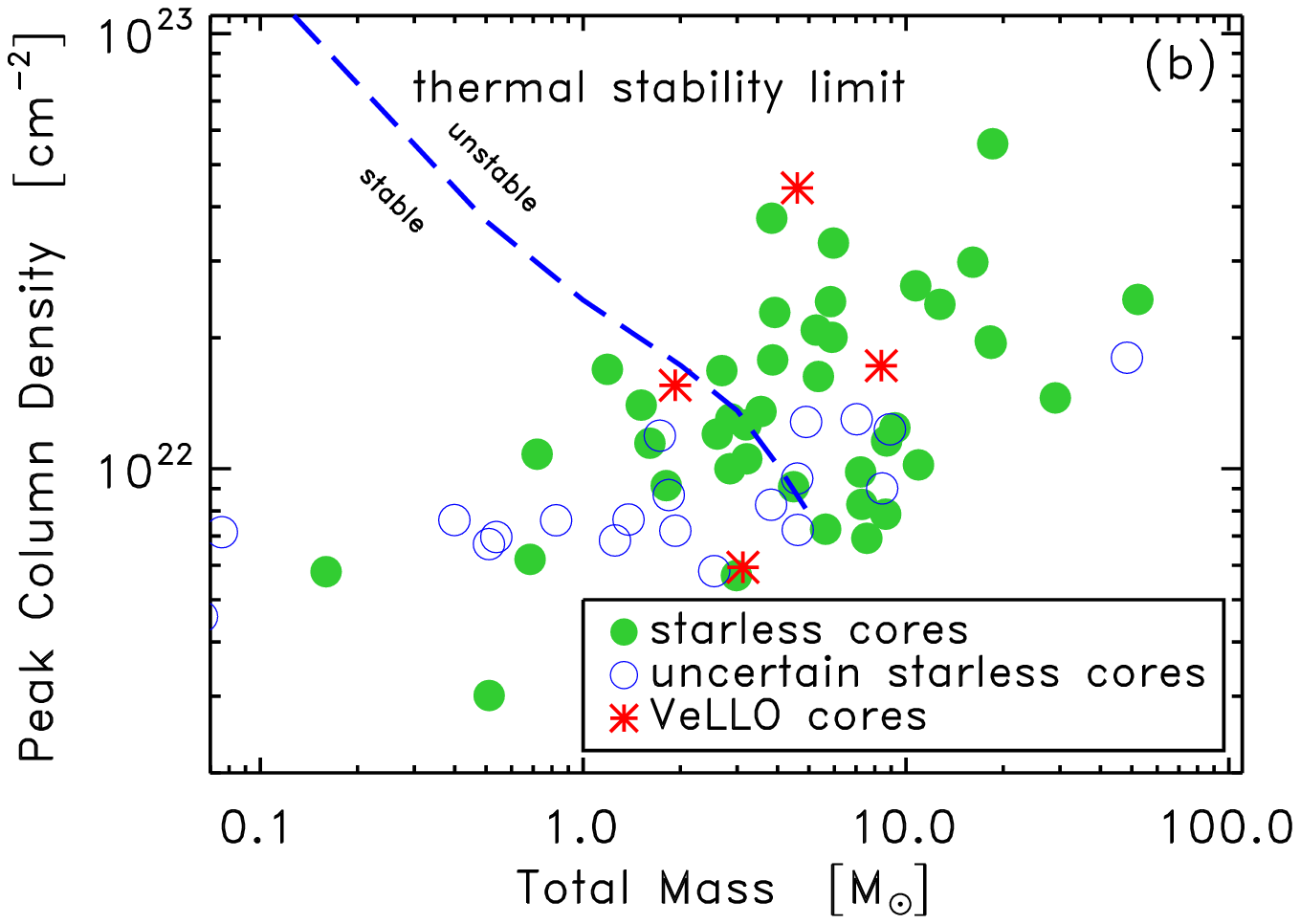}\\
\includegraphics[width=0.47\linewidth,bb=15 8 420 305,clip]{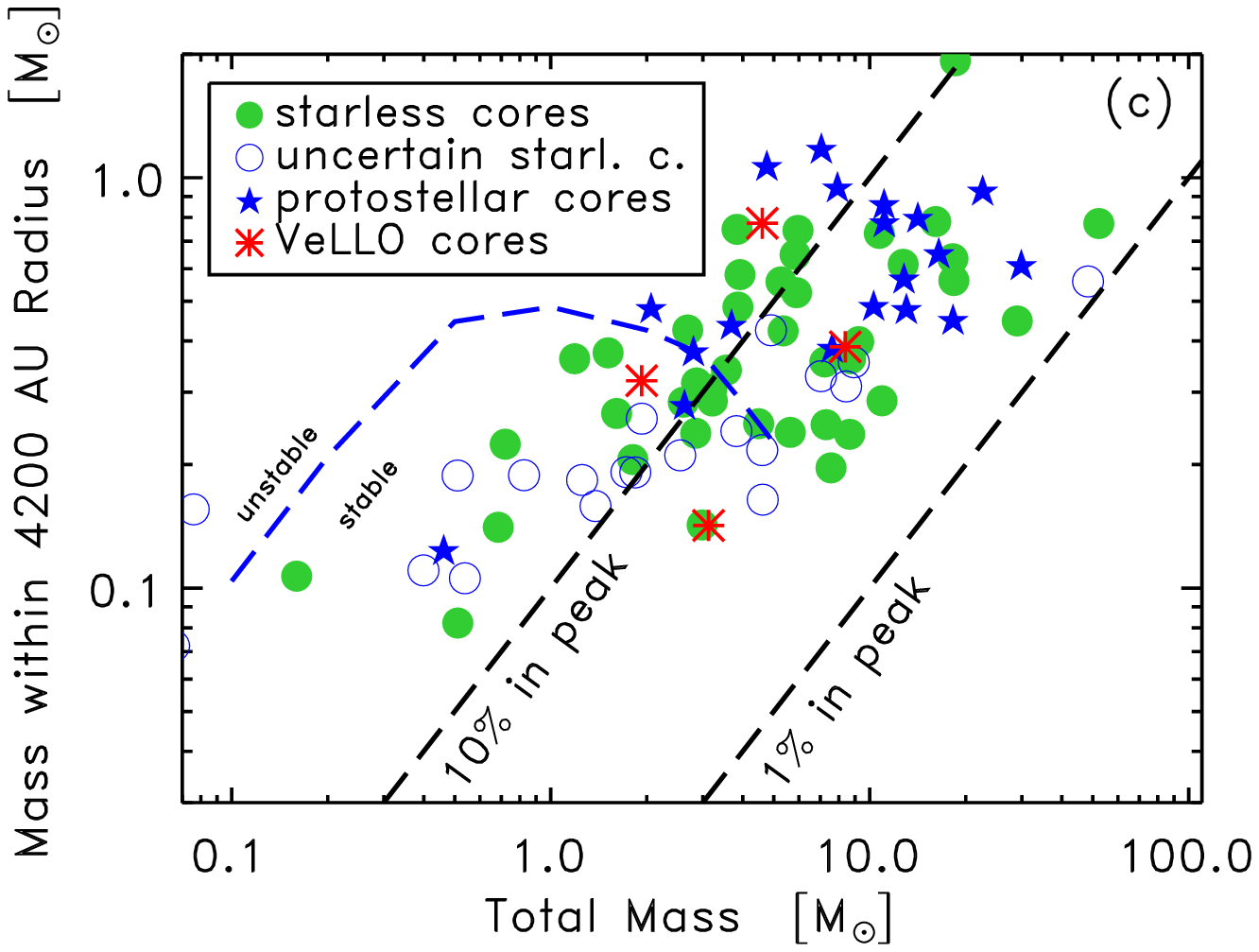} &
\includegraphics[width=0.47\linewidth,bb=15 8 420 305,clip]{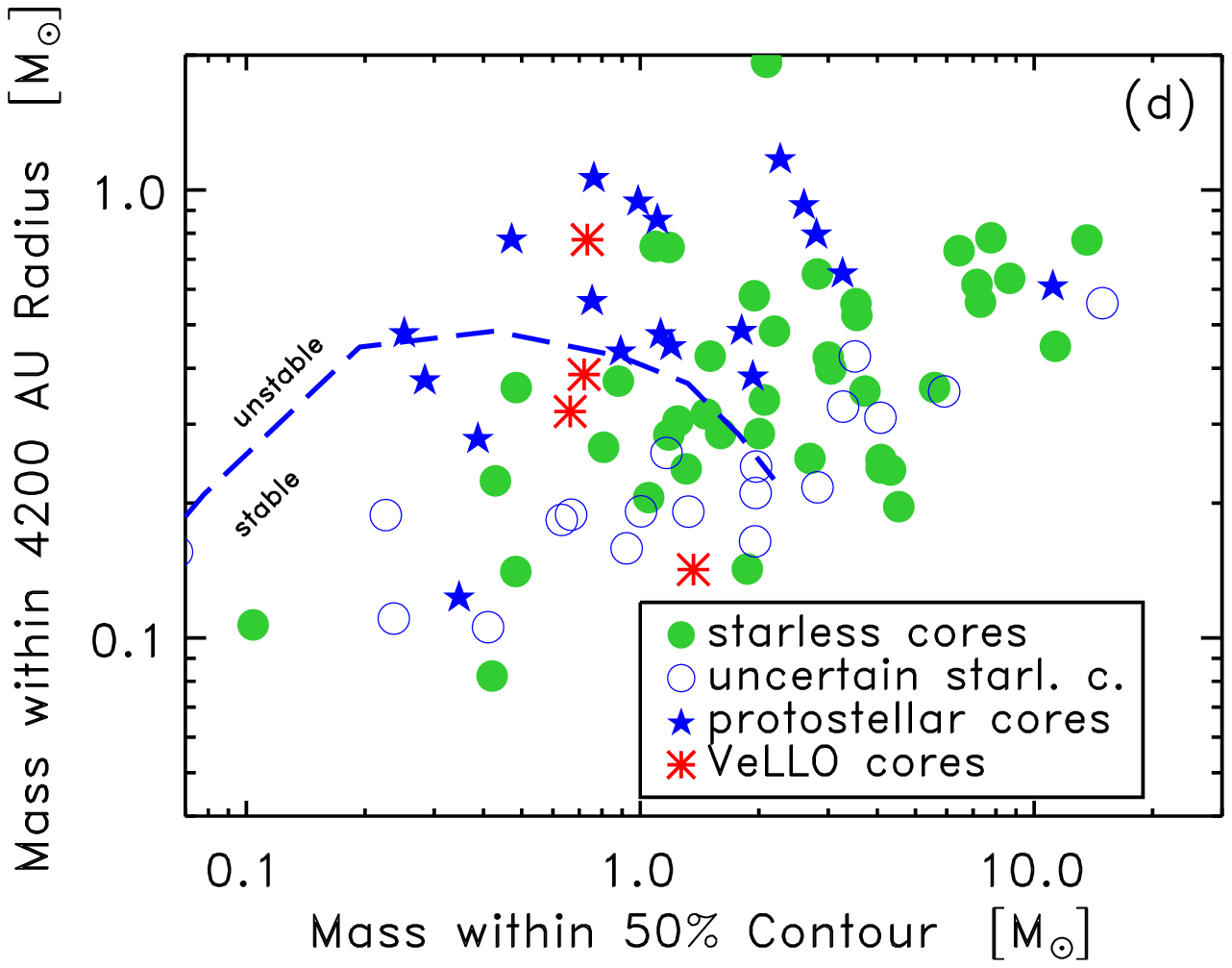}\\
\end{tabular}
\caption{Necessary conditions for active star formation from relations
  between dense core properties. \emph{Filled and empty circles} give
  the properties of subcores with well determined and uncertain
  properties due to artifacts, respectively.  \emph{Stars and
    asterisks} are used for subcores hosting YSOs and candidate VeLLOs,
  respectively. The \emph{diagonal dashed line in panel a)}
  indicates a selection limit; for a given total mass all subcores larger
  than this limit would be too faint to be identified in our
  maps. The \emph{diagonal dashed lines in panel c)} mark where
  $M = 0.1 \, M_{4200 \rm AU}$ and $M = 0.01 \, M_{4200 \rm AU}$: only
  a fraction of the dense core mass is available to form stars from a
  column density peak. The \emph{curved dashed lines} show the critical
  stability limits for hydrostatic equilibria with pure thermal
  pressure. For given total mass, or mass within the half intensity
  contour, all subcores with pure thermal pressure and with radii
  smaller than the critical one --- respectively column density or
  aperture mass exceeding the critical one --- are unstable. These
  stability limits might manifest in the data as necessary conditions for
  active star formation; most dense cores presently forming stars
  exceed these limits. However, these limits do not give sufficient
  conditions for ongoing star formation. Many starless cores have
  properties exceeding the critical values. In the core
  evolution framework discussed in the text this could be understood
  if these subcores are in hydrostatic equilibrium but supported by
  significant non-thermal pressure, or are
  collapsing.\label{fig-survey:star-form_ability}}
\end{figure*}

Correlations between core parameters, like e.g.\ the famous
size-linewidth relation \citep{larson1981:linewidth_size}, can provide
important hints on the nature of dense cores. Such correlations are
indeed seen in our data (Fig.\ \ref{fig-survey:star-form_ability}).
Unfortunately, many of these cannot be cleanly separated from
observational biases, and great care is needed if one wishes to
exploit our data in this fashion.

To give examples, our source identification scheme directly leads to a
radius-dependent lower limit to the mass, since the intensity
will be at least twice the noise level within any subcore, and thus
$M \ge 2 \mu_{\rm H_2} m_{\rm H}
\pi N_{\rm RMS} ({\rm H_2}) r_{\rm eff}^2$ (where $\mu_{\rm H_2}$ and
$m_{\rm H}$ are the mean molecular weight per hydrogen molecule and
the mass of the hydrogen atom, respectively, and
$N_{\rm RMS}({\rm H_2})$ is the column density corresponding to the
intensity RMS; also see Appendix
\ref{appendix:dust_em_prop}). Similarly, we cannot detect massive
cores of low peak column density, $N_{\rm max}({\rm H_2})$, in maps of
limited radius, $r_{\rm map}$, since
$M \le \mu_{\rm H_2} m_{\rm H} \pi N_{\rm max} ({\rm H_2}) r_{\rm map}^2$.
Also, we might miss low-mass (and therefore small) cores of high
column density because of beam smearing. Similar considerations apply
regarding $M_{4200 \rm AU}$, which can be interpreted as an
aperture-averaged column density (that, however, does not suffer from
beam smearing).

Given the above uncertainties, we are careful in drawing conclusions
from apparent correlations. In other words, parts of the parameter
space e.g.\ explored in Fig.\ \ref{fig-survey:star-form_ability} might
not be populated because cores in these regions would not be detected
given our sensitivities.

\subsection{Dense Cores and their
  Evolution\label{sec-survey:core_evolution}}
Our data not only reveals cores in various states of evolution but in
combination with the complementary YSO data we can also make
statements on the progress of dense cores through evolutionary stages.
The latter suggests some necessary conditions for active star
formation to be possible, while it also indicates diverse evolutionary
tracks for individual cores.

\subsubsection{Densities from Masses and Radii\label{sec-survey:densities}}
A rough estimate of the core densities can be derived from the peak
aperture masses when assuming specific density profiles. The estimates
will certainly often be off from the actual value by a large factor,
but this analysis is still illustrative to understand trends in the data.

We here compare the observed aperture
masses and radii at 70\% peak intensity to those expected for dense
cores with density profiles
$n({\rm H_2}) = n_{\rm c}({\rm H_2}) / (1 + [r / r_{1/2}]^2)$,
which provides a good approximation to the density structure of
starless cores \citep{tafalla2002:depletion}. In this
$n_{\rm c}({\rm H_2})$ is the central particle density, while $r$ is
the radius from the center, and $r_{1/2}$ is the radius at which
$n({\rm H_2}) = n_{\rm c}({\rm H_2}) / 2$. Particle densities can be
related to mass densities using the c2d standard $\rm H_2$-to-gas mass
conversion factors documented in Appendix
\ref{appendix:dust_em_prop}. Figure \ref{fig-survey:r70_vs_m4200}
illustrates this analysis.

\begin{figure}
\includegraphics[width=\linewidth]{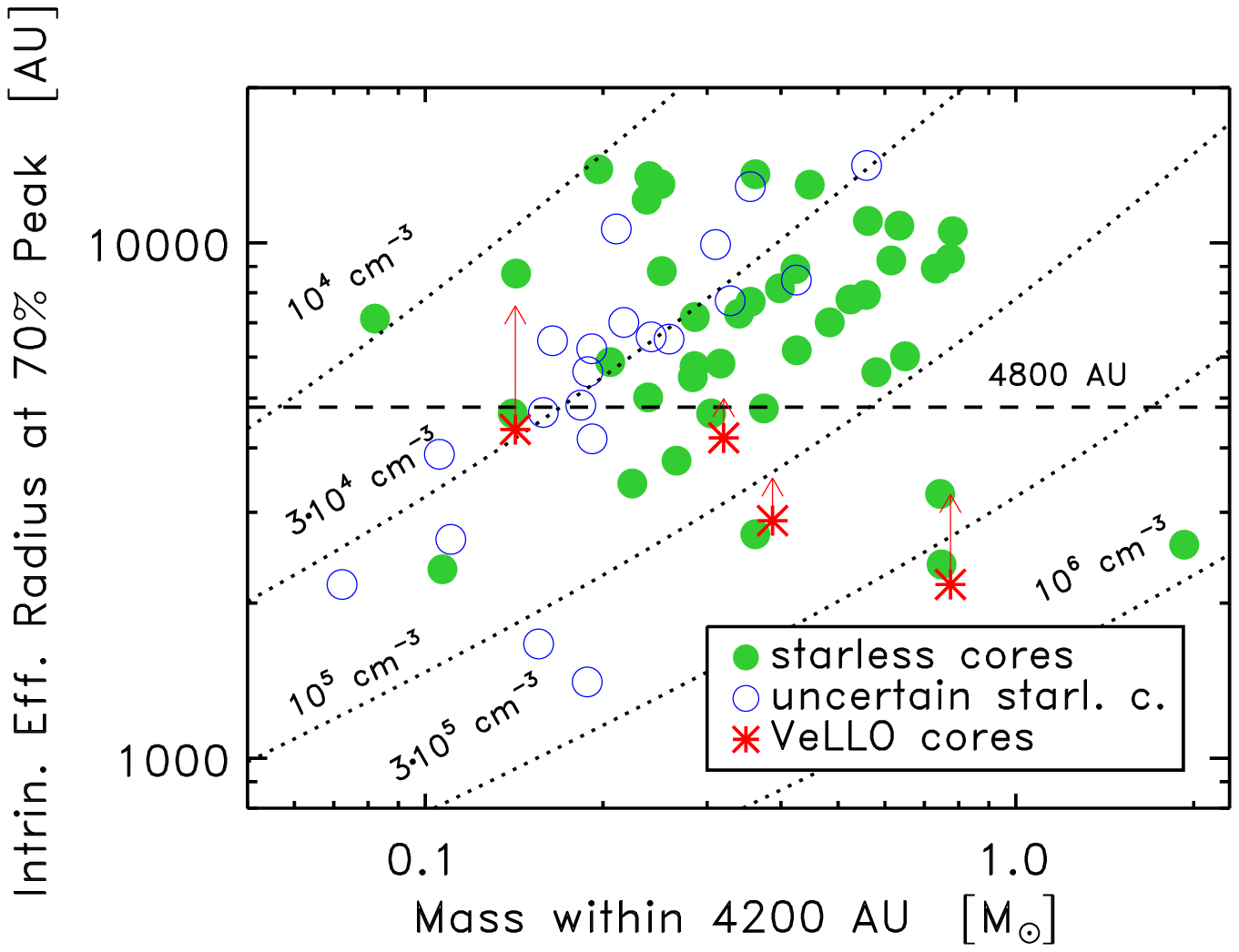}
\caption{Relation between the intrinsic radius at 70\% peak intensity
  and the mass within $4 \, 200 ~ \rm AU$ radius from the peak for
  VeLLO candidate and starless cores. The plotted effective radii deviate from
  the ones listed in Table \ref{tab-survey:cores-geometry} in that the
  beam smearing is removed ($7 \farcs 2$ at 70\% peak). See Fig.\ 
  \ref{fig-survey:star-form_ability} for an explanation of the
  symbols. The radius bias due to central heating of VeLLO cores
  (Sec.\ \ref{sec-survey:vello_radii}) is
  indicated by indicated by \emph{arrows}.  The \emph{dotted lines}
  indicate curves of constant central density for some simple density
  profile of varying radius (see the text for details). The
  \emph{labels} give the related central $\rm H_2$ column density. The
  \emph{dashed line} indicates the upper radius limit for evolved
  dense cores, $\le 4 \, 800 ~ \rm AU$, suggested by
  \citet{crapsi2005:survey}. The candidate VeLLOs observed all fall below this
  limit and are, for given aperture mass, smaller than most starless
  cores, though this in part might come from internal heating. Also,
  most VeLLO cores appear to have central densities exceeding those of
  most starless cores. VeLLO cores are thus possibly physically more
  evolved than starless cores. This appears to distinguish them from
  starless cores.\label{fig-survey:r70_vs_m4200}}
\end{figure}

This analysis indicates typical central densities of order
$3 \cdot 10^4 ~ {\rm cm^{-3}}$, while for some extreme cores we derive
densities of up to $10^6 ~ {\rm cm^{-3}}$. Similar figures were derived by
\citet{tafalla2002:depletion}, \citet{crapsi2004:l1521f}, and
\citet{kirk2005:scuba_survey}. Note, however, that the obtained
densities are crude estimates that suffer from significant
systematic uncertainties. Anyway, to give examples, for L1521F C1 the
density of $\approx 4 \cdot 10^5 ~ {\rm cm^{-3}}$ derived
from our simplified analysis is in good agreement with the density
of $\approx 5 \cdot 10^5 ~ {\rm cm^{-3}}$ (for our choice of dust
properties) derived by \citet{crapsi2004:l1521f}.

Note that VeLLO candidate cores appear to have unusually large densities, when
compared to starless cores. Section \ref{sec-survey:vello_radii}
discusses this in detail.

Our sample is highly biased and inhomogeneous, but it is still
illustrative to derive relative dense core lifetimes from the relative
numbers of dense cores of different physical state. Such estimates
assume a constant core formation rate across the sample and common
evolutionary paths for all cores, neither of which are likely correct
for our cores. Anyway, we find about 35 starless subcores in the
$10^4 < n_{\rm c}({\rm H_2}) / {\rm cm^{-3}} < 10^5$ density range,
and 4 subcores in the
$10^5 < n_{\rm c}({\rm H_2}) / {\rm cm^{-3}} < 10^6$
range. Under the
above assumptions this implies that dense cores reside at densities of
$10^4 ~ {\rm to} ~ 10^5 ~ \rm cm^{-3}$ for a time exceeding the one
for densities of $10^5 ~ {\rm to} ~ 10^6 ~ \rm cm^{-3}$ by a factor
$\approx 9$. For reference, the free fall timescale\footnote{We give
  the collapse time for a homogeneous sphere of density
  $n({\rm H_2})$, i.e.\
  $\tau_{\rm ff} = (3 \, \pi /
  [32 \, G \, \varrho])^{1/2}$,
  where $G$ is the constant of gravity and $\varrho$ is the
  volume-averaged density. Our calculation assumes a mean particle
  weight of 2.8 per hydrogen molecule.},
\begin{equation}
\tau_{\rm ff} = 9.8 \cdot 10^4 ~ {\rm yr} \,
[n({\rm H_2}) / (10^5 ~ {\rm cm^{-3}})]^{-1/2} \, ,
\end{equation}
only varies by a factor $10^{1/2} \approx 3$ between the density
bins. Dense core evolution in free fall would thus not explain the
relative number of cores in different density bins.

\subsubsection{Evolutionary States\label{sec-survey:highly_evolved_cores}}
\citet{crapsi2005:survey} suggested some criteria to assess a starless
core's evolutionary state. These also include criteria based on
observations of molecular lines, which we cannot evaluate using our
data. Table \ref{tab-survey:core_evolution} therefore only evaluates
criteria based on the dust emission. Following
\citeauthor{crapsi2005:survey}, evolved cores must have central
densities (Sec.\ \ref{sec-survey:densities}; recall that this analysis
only yields rough estimates) exceeding
$2.5 \cdot 10^5 ~ \rm cm^{-3}$ (when converting their results to our
choice of dust properties) and radii at 70\% peak intensity
$< 4 \, 800 ~ \rm AU$. Naively, one would also associate high column
densities with an advanced evolutionary state. Somewhat arbitrarily we
thus choose a limit of $0.7 \, M_{\odot}$ to highlight those starless
cores with unusually high column densities in our sample. Table
\ref{tab-survey:core_evolution} also includes candidate VeLLO cores which are
supposedly young and thus likely to not yet be significantly affected
by YSO formation and evolution.

\begin{table}
\caption{Starless cores that fulfill at least one criteria
  for an advanced evolutionary state, as well as VeLLO candidate cores. Brackets for VeLLOs indicate that
  $r_{70\%}$ does not fulfill the criterion for evolved cores after
  correcting for internal luminosity.\label{tab-survey:core_evolution}}
\begin{tabular}{lllllll}
\hline \hline
\multicolumn{2}{l}{\rule{0ex}{3Ex}Core} &
  $n_{\rm c}({\rm H_2})$ & $r_{70\%}$ & $M_{4200 \rm AU}$\\
  & & $> 2.5 \cdot 10^5 ~ \rm cm^{-3}$ &
  $< 4 \, 800 ~ \rm AU$ & $> 0.7 \, M_{\odot}$\vspace{1ex}\\
\hline
\multicolumn{4}{l}{\itshape evolved starless cores:}\\
L1521B-2 & C2 & $-$ & + & $-$\\
L1521-2 & C2 & $-$ & + & $-$\\
B18-1 & C1 & + & + & +\\
B18-2 & C1 & $-$ & + & $-$\\
TMC-2 & C1 & $-$ & $-$ & +\\
B18-4 & C1$^a$ & + & + & +\\
TMC-1C & C3 & $-$ & $-$ & +\\
TMC-1 & C1 & $-$ & + & $-$\\
L1507A & C3 & $-$ & + & $-$\\
L1622A & C3 & $-$ & $-$ & +\\
L183 & C1 & $-$ & + & $-$\\
  & C2 & + & + & +\\
  & C5 & $-$ & + & $-$\\
L1082A & C2 & + & + & +\\
\multicolumn{3}{l}{\itshape \rule{0ex}{3Ex}VeLLO cores:}\\
L1521F & C1 & + & + & +\\
L673-7 & C1 & $-$ & + & $-$\\
L1148 & C1 & $-$ & (+) & $-$\\
L1014 & C1 & $-$ & (+) & $-$\\
\hline
\end{tabular}
\rule{0ex}{3Ex}Notes: a) Numbers hold for the starless P2 peak in
B18-4.
\end{table}

This analysis reveals 5 cores that fulfill both criteria by
\citet{crapsi2005:survey}, i.e., they are small in size and have a
high central density. For two of these (L1521F C1 and L183 C3) this
was already shown by \citeauthor{crapsi2005:survey}, while our study
gives 3 more candidate evolved cores matching his criteria (B18-1 C1,
B18-4 C1, and L1082A C2). Note that these cores are actually the only
ones with $n_{\rm c}({\rm H_2}) > 2.5 \cdot 10^5 ~ \rm cm^{-3}$ and
that they also have $M_{4200 \rm AU} > 0.7 \, M_{\odot}$. Three further
cores (TMC-2 C1, TMC-1C C3, and L1622A C3) stand out in their aperture
masses, thereby suggesting a somewhat evolved state. They do not
fulfill the \citeauthor{crapsi2005:survey} criteria, but recent
molecular screening of TMC-1C C3 shows that at least one of these
cores indeed seems to be close to the onset of star formation
\citep{schnee2007:tmc1_state}.

The c2d MAMBO survey thus reveals 3 previously unknown candidates for
evolved starless cores fulfilling the criteria by
\citet{crapsi2005:survey}, and 3 further ones likely to be evolved on
basis of their aperture masses. Note, however, that --- following the
above criteria --- some of these cores seem to be more advanced in
their evolution than some candidate VeLLO cores that actually already
do form stars. This suggests that the \citeauthor{crapsi2005:survey}
criteria are somewhat biased: they are well suited to select
\emph{some} evolved cores, but not all of them. In this respect it is
important to keep in mind that their criteria were tuned to select
cores like the prototype evolved starless core L1544. The analysis
presented here suggests that evolved cores with physical properties
different from those derived by \citet{crapsi2005:survey} do exist.
This is not surprising though, e.g.\ given the observed range in
stellar masses. The \citet{crapsi2005:survey} study might, e.g.,
preferentially select evolved rather massive cores on the verge to
form stars of rather large mass. It would then be plausible to find
evolved cores that are on the verge of forming lower mass stars (i.e.,
a few $0.1 \, M_{\odot}$), but have too low mass and column density to
be incompatible with the \citeauthor{crapsi2005:survey} criteria. In
this light, Sec.\ \ref{sec-survey:star-form_ability} presents some
suggestions for further refinements of criteria.

\subsubsection{Star Formation
  Efficiency\label{sec-survey:sf-efficiency}}
Given a typical duration of the main accretion phase of order
$10^5 ~ \rm yr$ \citep{barsony1992:submm_emission,
  greene1994:yso_rho-oph, kenyon1995:evol_tau-auriga}, and expected
velocities for the growth of the infalling inner part of YSO envelopes
of order $0.2 ~ \rm km \, s^{-1}$ (e.g., for singular isothermal
spheres; \citealt{shu1977:self-sim_collapse}), only the mass within a
few $10^3 ~ \rm AU$ from a forming YSO are available for its formation.
The few $10^3 ~ \rm AU$ radius of this volume are much smaller
than the subcore sizes of several $10^4 ~ \rm AU$ observed by us
(e.g., Fig.\ \ref{fig-survey:histograms} [a]). Thus, only a small
fraction of a subcore's mass is available for the formation of a given
YSO. This fraction is an estimate of the subcore's efficiency to form
a given YSO. The latter is a lower limit to the \emph{total} star
formation efficiency, i.e., the mass ratio between the dense cores
mass and the \emph{total} mass of all YSOs produced by it.

We use a radius of $4 \, 200 ~ \rm AU$ for a quantitative evaluation
of this efficiency, the maximum infall radius derived for the above
considered YSO
and envelope properties. Some 2\% to 20\% of the total mass of a dense
core are contained in a $4 \, 200 ~ \rm AU$ radius aperture centered
on its brightest peak (Fig.\ \ref{fig-survey:star-form_ability} [c]). For
other average velocities at
which the infalling region grows, $\langle v_{\rm in} \rangle$,
and accretion times, $\tau_{\rm accr}$, the appropriate aperture
radius will be larger by a factor
$\approx \langle v_{\rm in} \rangle / (0.2 ~ {\rm km \, s^{-1}})
\cdot \tau_{\rm accr} / (10^5 ~ \rm yr)$. In case of
$n \propto r^{-2}$ density profiles the aperture mass scales linearly
with the aperture radius. To give an example, recent c2d
determinations indicate a class I lifetime of order
$5.3 \cdot 10^5 ~ \rm yr$ (Evans et al., in prep). Since YSOs are
supposed to accrete during this phase, this would imply
(using the above assumptions) that a YSO can accrete a mass
$5.3 \cdot M_{4200 \rm AU}$ and that cores have a star-formation
efficiency of order of several 10\%.

Long accretion timescales might, however, imply low accretion
efficiencies on the size scale of the infalling matter, if collapse
proceeds in free fall. In this case, YSO accretion timescales of
several $10^5 ~ \rm yr$, combined with average aperture masses
$M_{4200 \rm AU} = 0.4 ~ {\rm to} ~ 0.5 \, M_{\odot}$ (Sec.\
\ref{sec-survey:mass_within_4200au}), imply that there is enough time
to accrete several $0.5 \, M_{\odot}$ onto a forming YSO. Since,
however, the average stellar mass, $\langle M_{\star} \rangle$ is of
order $0.4 \, M_{\odot}$ (e.g., Table 2 in \citealt{kroupa2002:imf}),
a significant fraction of this aperture mass does not
seem to end up in the forming star. This fraction increases with
increasing accretion timescale; for the above collapsing singular
isothermal sphere it is
$(M_{\rm in} - \langle M_{\star} \rangle) / M_{\rm in}$, where the
infalling mass
$M_{\rm in} = (\tau_{\rm accr} / 10^5 ~ {\rm yr}) \cdot M_{4200 \rm AU}$
and thus is of order 3/4 for
$\tau_{\rm accr} \approx 4 \cdot 10^5 ~ {\rm yr}$. The efficiency
could, however, still be large if collapse does not occur in free
fall.

In summary, our data indicates a star formation efficiency of order
10\%, with an uncertainty of a factor of several. Such an efficiency
is of an order of magnitude exceeding estimates for the star formation
efficiency of molecular clouds ($\lesssim 2\%$; e.g.,
\citealt{leisawitz1989:cluster-co-survey}). This might, e.g., indicate
that the star formation efficiency is not only controlled by processes
operating on the scales of dense cores, but that also by processes of
importance on larger scales. Also, the low star formation efficiency
on the scales of cores suggests that dense core mass functions do not
directly map into stellar initial mass functions: core and YSO masses
will differ, and it is not clear that their ratio is constant. In
other words, it is not clear which core mass definition is relevant to
measure the mass of a core that is available for accretion onto
forming stars. To give an example, here we discuss three mass scales
(i.e., $M_{\rm tot}$, $M_{50\%}$, and $M_{4200 \rm AU}$), none of
which is proven to measure the accretable mass. This questions basic
assumptions made in some studies relating mass functions of stars and
cores, in particular since the shape and absolute scale of the
  mass distribution depends on the adopted core mass definition (e.g.,
  Fig.\ \ref{fig-survey:histograms}, panels [b] and [c]).  There is
evidence though that the shapes of some of these distributions
might well be coupled to the one of the IMF
\citep{alves2007:imf}. This would then point at a constant efficiency.

\subsubsection{Star Formation
  Ability\label{sec-survey:star-form_ability}}
Figure \ref{fig-survey:star-form_ability} (d) indicates that there is
a fundamental difference between cores actively forming stars (i.e.,
they contain a YSO) and those who don't: starless cores in our sample
have aperture masses of about
$0.1 ~ {\rm to} ~ 2.0 \, M_{\odot}$ (also see
\ref{fig-survey:histograms} [c]), while all young
($T_{\rm bol} \lesssim 300 ~ \rm K$) YSOs have envelope masses
$\gtrsim 0.3 \, M_{\odot}$ (also see Fig.\
\ref{fig-survey:tbol_vs_m4200}; we exclude the VeLLOs, since their
physics may be governed by different processes, and CB188, since
outflow-core interaction might have affected the core properties [Sec.\
\ref{sec-survey:elongation}]). The naive interpretation 
of this observation is that dense cores must have
$M_{4200 \rm AU} \ge 0.3 \, M_{\odot}$ in order to be able to actively
form stars. The full story might be a bit more complex, since the present
YSO aperture masses will deviate from the ones of their natal cores at
the onset of star-formation. For example, a collapsing SIS
\citep{shu1977:self-sim_collapse} has an aperture mass
decreasing with time, suggesting that the observed YSO aperture masses
are lower than the initial ones.

Thus $M_{4200 \rm AU} \ge 0.3 \, M_{\odot}$ appears to be a
\emph{necessary} condition for active star formation to occur (defined
as embedded YSOs being present), but it is not
a \emph{sufficient} one, since there are many starless cores above the
limiting aperture mass (with B18-1 P3, B18-4 P2, and L183 P3, all with
$M_{4200 \rm AU} > 0.7 \, M_{\odot}$, as extreme examples; also see
the discussion in Sec.\ \ref{sec-survey:highly_evolved_cores}). Thus,
cores with $M_{4200 \rm AU} \ge 0.3 \, M_{\odot}$ \emph{can} actively
form stars, but they do not need to do so. Note, however, that
starless cores with
$M_{4200 \rm AU} \gg 0.3 \, M_{\odot}$ might well already be in a
state of collapse and be destined to form a star, even when no
embedded YSO has formed yet.

In a similar way, a total mass $\ge 2 \, M_{\odot}$ appears to be a
necessary condition for active star formation to be possible. Again,
there are some starless subcores with total masses much higher than
those for some star-forming ones.\medskip

\noindent It is interesting to test this observation against simple
models of dense core structure. For this we use models of
near-isothermal pressure-supported spheres. Using boundary conditions
appropriate for typical dense cores in the solar neighborhood (a
visual shielding extinction of $5 ~ \rm mag$ from a dense core
envelope, a cosmic ray flux of $3 \cdot 10^{-17} ~ \rm s^{-1}$, and
the solar neighborhood interstellar radiation field), we evaluate the
stability of these spheres in case of no support form turbulence or
magnetic fields (i.e., non-isothermal Bonnor-Ebert spheres; Kauffmann
\& Bertoldi, in prep., and \citealt{galli2002:struc_and_stab}).

For fixed total mass, respectively mass at half intensity
($M_{50\%}$), these models predict critical values for
parameters like the peak column density, so that all objects exceeding
these cannot be stable against collapse unless supported by additional
non-thermal pressure. These critical limits (or calculations are
limited to total masses $\le 5 \, M_{\odot}$ given a lack of molecular
cooling rates for low densities, leading to a break in the boundaries
drawn) are indicated in Fig.\ \ref{fig-survey:star-form_ability}.
Within the model, they give the limiting properties beyond which a
dense core can become unstable and collapse to form a star. If these
models have any relevance, the critical limits should therefore
manifest in the distribution of observed actual dense core properties.

Notably, as shown in Fig.\ \ref{fig-survey:star-form_ability} (c) and
(d), all young YSOs (excluding CB188~C1 and VeLLO candidates, as
discussed above) are indeed consistent with having masses in excess of
the mass-dependent critical aperture mass for pure thermal pressure,
and the theoretically derived critical masses seem to describe the
observed necessary condition for active star formation. The models
also describe the necessary condition on the total mass (Fig.\
\ref{fig-survey:star-form_ability} [a] and [c]).

If dense cores can be described by the above models, a starless dense
core of fixed total mass could evolve towards collapse by approaching
the critical aperture mass (starting with sub-critical value) as a
quasi-hydrostatic sphere, turn into a collapsing core once exceeding
the critical value, and finally become a core with active star
formation. In this picture starless cores with properties exceeding
the critical ones for pure thermal pressure could then be collapsing
cores. They could also be quasi-hydrostatic equilibria with additional
non-thermal pressure, for which the critical values are beyond those
for pure thermal pressure. If the latter was true, the co-existence
of starless as well as star-forming cores in the same parameter range
suggests that different cores are supported by different levels of
non-thermal pressure.\medskip

\noindent
Note that shifts of the masses by a factor 2, either globally for all
sources, or for starless cores relative to those with YSOs, only
marginally influence the above discussion. (Such shifts are possible
because of the uncertainties in opacities and temperatures, as
discussed in Appendix \ref{appendix:dust_em_prop}.) Still, our data
would hint at the existence for necessary conditions for star
formation. Strong YSO-to-starless core relative shifts of a factor 2
could, however, lead to a clear separation of starless and YSO cores
in some diagnostic diagrams. Then, it might turn out that, e.g.,
\emph{all} cores above some limiting aperture mass do form stars.
This would turn the aforementioned \emph{necessary} condition for star
formation into a \emph{sufficient} one. Strong global and relative
mass shifts by a factor 2 might also make the applicability of the
theoretical justification for the limiting curve questionable.

\subsection{YSOs and their
  Evolution\label{sec-survey:evol_protostars}}
Given the number of YSO cores, the c2d MAMBO survey is well suited to
study evolutionary effects in YSO cores. Our analysis also includes
the properties of some recently discovered VeLLO candidates.

The c2d MAMBO sample contains 5, 14, and 1 YSOs of the classes 0, I,
and II, respectively. This is, e.g., in between the 0-to-I class ratio
of 1:10 for Taurus and the Ophiuchus cloud
\citep{andre1994:yso_rho-oph,motte1998:ophiuchus} and ratios of order
1:1 for Perseus \citep{hatchell2007:perseus} and some isolated cores
\citep{visser2002:lynds_clouds}. Note, however, that our sample is
anything but unbiased, and thus not ideal to constrain class
lifetimes from their relative number.

\subsubsection{YSO Luminosities}
To better characterise the central star we use the internal
luminosity, $L_{\rm int}$, that is due to the YSO instead of the
bolometric luminosity, i.e., the total radiative power minus that due
to interstellar heating. This luminosity is, however, taken to be
identical to the bolometric one, except for some candidate VeLLOs for which
detailed estimates exist from other studies
(\citealt{bourke2006:l1521f}; Kauffmann et al., in prep.;
\citealt{young2004:l1014}).

The approximation of the internal luminosity by the bolometric one is
justified since the derived bolometric luminosities are usually dominated by
the power of the embedded source, and not by the heating processes in the
interstellar medium (like interaction with cosmic rays or absorption
of the interstellar radiation field). The latter contribution can be
estimated for the material within $4 \, 200 ~ \rm AU$ from the density
peak. A heating power
$0.16 \, L_{\odot} \cdot (M_{4200 \rm AU} / M_{\odot})$ is required to
heat the matter to a typical interstellar temperature of $10 ~ \rm
K$. For $M_{4200 \rm AU} = 0.4 \, M_{\odot}$, which is
typical for the YSO cores in our sample (Sec.\
\ref{sec-survey:mass_within_4200au}), this power
is $0.06 \, L_{\odot}$.

In some sources, however, the bolometric luminosity derived by us is
dominated by the power due to interstellar heating processes. For dust
temperatures of at least $10 ~ \rm K$ this power exceeds the inferred
bolometric luminosities of the YSOs in L1521F P1, L673-7 P1, and
L1251A P3 and P5 (also depending on whether relying on IRAS or Spitzer
data), also demonstrating that our method to derive $L_{\rm bol}$ can
miss part of the actual radiative power. This is also reflected in
these sources' unusually large submillimetre-to-bolometric luminosity
ratios. In such cases the bolometric luminosity is not a good measure
of the power input by the young star. This leads to uncertain heating
corrections in mass estimates of YSO envelopes (Sec.\ 
\ref{sec-survey:mass_estimates}). The related uncertainties in the
mass estimates are, however, not significant, given that the low
derived luminosities imply small corrections.

\subsubsection{Envelope Masses and
  Luminosities\label{sec-survey:masses_and_luminosities}}
\begin{figure}
\includegraphics[width=\linewidth]{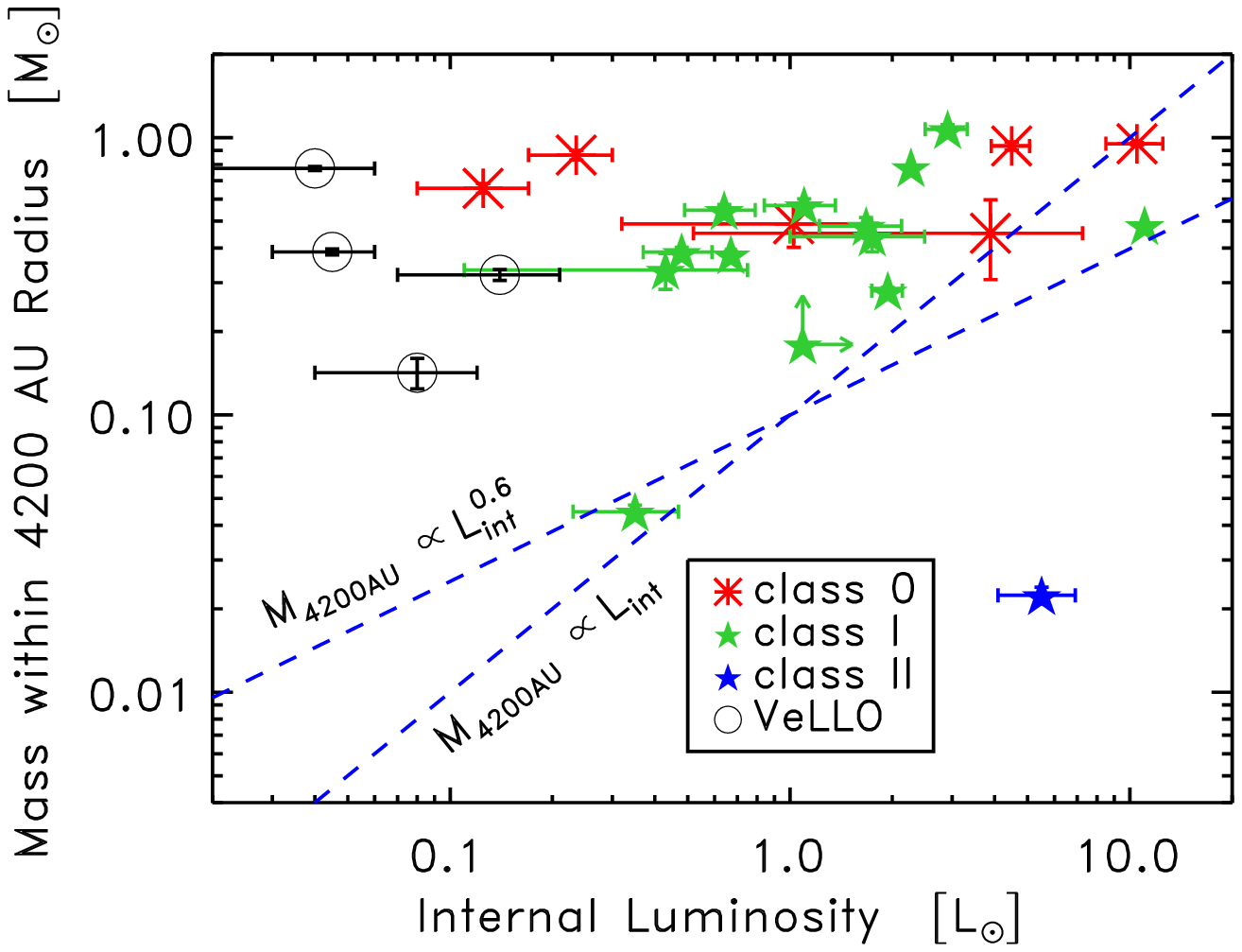}
\caption{The relation between the internal luminosity and the mass within a
  peak-centered $4 \, 200 ~ \rm AU$ radius aperture for YSO
  cores. The internal luminosity is the radiation due to the embedded
  star, i.e., the bolometric luminosity minus the power from
  interstellar heating. \emph{Markers of different shade and shape} refer to
  YSOs belonging to different infrared SED
  classes assigned on basis of their bolometric temperature (see
  legend). The \emph{dashed lines} indicate the conceptual boundary between
  class 0 sources and objects in later evolutionary stages for a
  spectrum of stellar mass-luminosity relations,
  $M_{4200 \rm AU} \approx 0.1 \, M_{\odot}
  \, (L / L_{\odot})^{0.6 ~ \rm to ~ 1.0}$. Interestingly, class 0 and
  class I sources covered by the c2d MAMBO 
  survey are not separated by the conceptual class boundaries.
  \label{fig-survey:lbol_vs_m4200}}
\end{figure}

YSOs are expected to evolve in luminosity and envelope
properties during the star formation process. Three basic properties
to characterise a YSO are its age, as believed to be roughly
measured by the bolometric temperature or the infrared class, the
luminosity, and the amount of matter surrounding the star. The latter
two should depend on the age since they, e.g., depend on the accretion
rate and the stellar mass and radius, which all evolve with time.
Figure \ref{fig-survey:lbol_vs_m4200} shows the relation between
luminosities and envelope masses. The latter are approximated by
the mass within $4 \, 200 ~ \rm AU$ radius from the star since
this property has been studied in the past
\citep{motte2001:protostars}.

Surprisingly, no clear evolutionary trend is obvious in this diagram,
except for the significant decrease in $M_{4200 \rm AU}$ towards the
class II phase; the crowding of VeLLO candidates at low luminosities
is due to their definition. Class 0 and class I sources do essentially
occupy the same region in the parameter space, with the class 0
sources possibly having for given internal luminosity slightly higher
aperture masses than class I sources.

In particular, the classes 0 and I are not separated by the
class boundary suggested by \citet{motte2001:protostars},
$M_{4200 \rm AU} \approx
0.1 \, M_{\odot} \, (L / L_{\odot})^{0.6 ~ \rm to ~ 1.0}$. This law
reflects the conceptual definition of class 0 sources as YSOs
which have accreted less than half of their final mass
\citep{andre1993:class0}. Thus the envelope mass in class 0 sources,
as approximated by $M_{4200 \rm AU}$, must exceed the stellar
mass. Depending on the exact form of the YSO mass-luminosity
relation, \citeauthor{motte2001:protostars} suggest this boundary to
be given by the above relation (see \citealt{andre1994:yso_rho-oph}
and \citealt{andre2000:pp_iv} for further details). This boundary is
indeed observed to well separate class 0 and class I sources in the
Ophiuchus molecular cloud complex and some isolated cores
\citep{visser2002:lynds_clouds}, but it fails to do so for sources
in the Taurus cloud \citep{motte2001:protostars} and Perseus
\citep{hatchell2007:perseus}.

To be precise, within the observational errors all class 0 sources in
the c2d MAMBO sources do indeed have aperture masses exceeding $0.1 \,
M_{\odot} \, (L / L_{\odot})^{0.6 ~ \rm to ~ 1.0}$, just as expected.
However, the class I YSOs in our survey do not generally fall short of
the conceptual luminosity-dependent mass limit for class 0 sources. In
this respect they are different from the aforementioned class I
sources in Ophiuchus that fulfill this limit. For comparison, like
ours, also class I YSOs in Taurus \citep{motte2001:protostars} and
Perseus \citep{hatchell2007:perseus} do not conform to this conceptual
boundary. This suggests that the stars in our sample reside in regions
that are more similar to regions of distributed star formation, like
Taurus, than to regions of clustered star formation, like Ophiuchus.

The failure of the conceptual class boundary to separate sources of
class 0 and I and the absence of clear evolutionary trends between
them in Fig.\ \ref{fig-survey:lbol_vs_m4200} suggests that either the concept of infrared classes, our general understanding of YSO evolution as
a linear sequence of states in the $L_{\rm int}$--$M_{4200 \rm AU}$
parameter space, or both, need some revision. Also, $M_{4200 \rm AU}$
might simply not be well suited to gauge the mass available for
accretion. This would, however, require accretion to proceed in a
manner very different from the collapse of self-similar isothermal
spheres: here steadily $M_{4200 \rm AU}$ decreases with time, and
independent from the final size of the infalling envelope (even if
eventually $> 4 \, 200 ~ \rm AU$) $M_{4200 \rm AU}$ would at least
generally probe the mass evolution of the envelope. Whatever, the fact
that similar studies of different regions find different relations
between YSO and envelope properties suggests that the environment
influences these relations.

\subsubsection{Bolometric Temperatures and
  Luminosities\label{sec-survey:temp_and_lumi}}
\begin{figure}
\includegraphics[width=\linewidth]{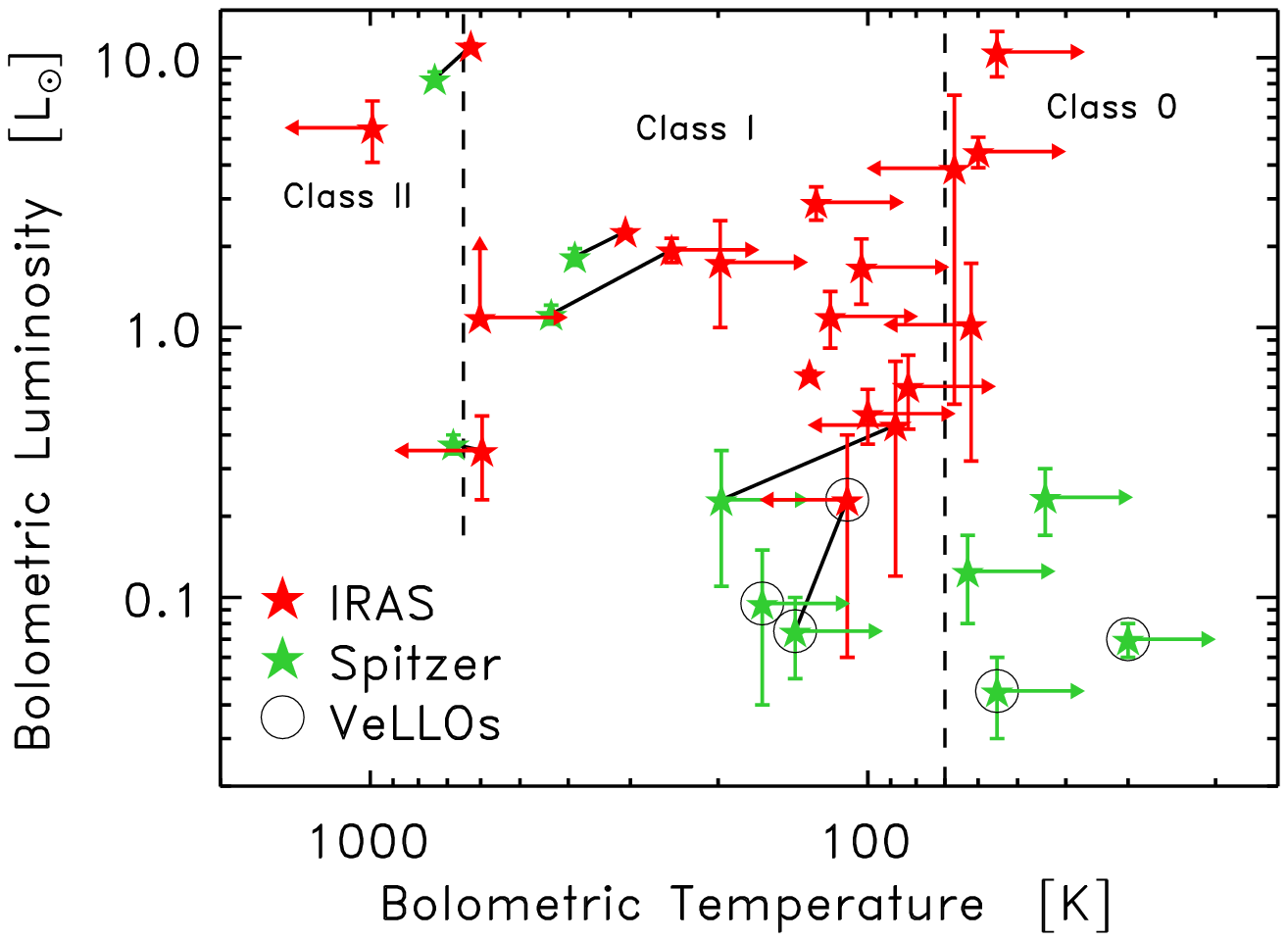}
\caption{Bolometric luminosities and temperatures as derived from IRAS
  and Spitzer (\emph{markers of different shade}) data. VeLLO candidates
  are highlighted by \emph{circles} around the markers, and the
  YSO class regimes are delimited by \emph{vertical dashed lines} and
  \emph{labels}.
  \emph{Solid lines} indicate characterizations of the same star based
  on different data. For a given source the IRAS data usually
  indicates lower bolometric temperatures and higher luminosities than
  the Spitzer
  data, just as expected given IRAS' better spectral coverage at long
  wavelength. Note the range in luminosities for a given class. This
  indicates that the accretion rates, masses, radii, or all these
  properties largely vary within the sample, and also a given class,
  if the luminosities are largely due to
  accretion.\label{fig-survey:lbol_vs_tbol}}
\end{figure}

Figure \ref{fig-survey:lbol_vs_tbol} presents the bolometric
temperatures and luminosities for sources in the c2d MAMBO survey. It
shows that for a given star the luminosities are higher and the
bolometric temperatures are lower when these properties are inferred
from IRAS data instead of Spitzer observations. This is just as
expected, given IRAS' better spectral coverage of the dust emission
peak at sub-millimeter wavelengths.

This diagram shows that on average $L_{\rm bol}$ increases with
increasing $T_{\rm bol}$. This trend is in particular pronounced when
looking at the lower envelope of the data points. However, it is
likely that this is to a significant extent an observational bias,
since faint sources of high bolometric temperature have lower millimetre
intensities and are thus less likely to be included in our MAMBO-based
source selection. For the same reason in general sources with high
bolometric temperatures are less likely to be included. This is at
least partially responsible for the sharp drop in the number of
sources at $\gtrsim 300 ~ \rm K$.

Class 0 sources are found to have luminosities spanning across more
than two orders of magnitude. This demonstrates the diversity of star
formation activity in the c2d MAMBO survey. Note that, if sources of
similar $T_{\rm bol}$ have similar ages and
have luminosities dominated by accretion luminosity (i.e.,
$L_{\rm int} \approx G M_{\star} \dot{M}_{\rm accr} / R_{\star}$,
where $G$, $M_{\star}$, $R_{\star}$, and $\dot{M}_{\rm accr}$ are the
constant of gravity, the stellar mass and radius, and the accretion
rate), then the class 0 sources in the c2d MAMBO survey must differ
significantly in their mass-to-radius ratios, their accretion
rates, or both. Enoch et al.\ (in prep.) find similar luminosity
ranges for YSOs in Perseus, Serpens, and Ophiuchus.

\subsubsection{Envelope Masses and
  Bolometric Temperatures\label{sec-survey:Lbol_and_Menv}}
\begin{figure}
\includegraphics[width=\linewidth]{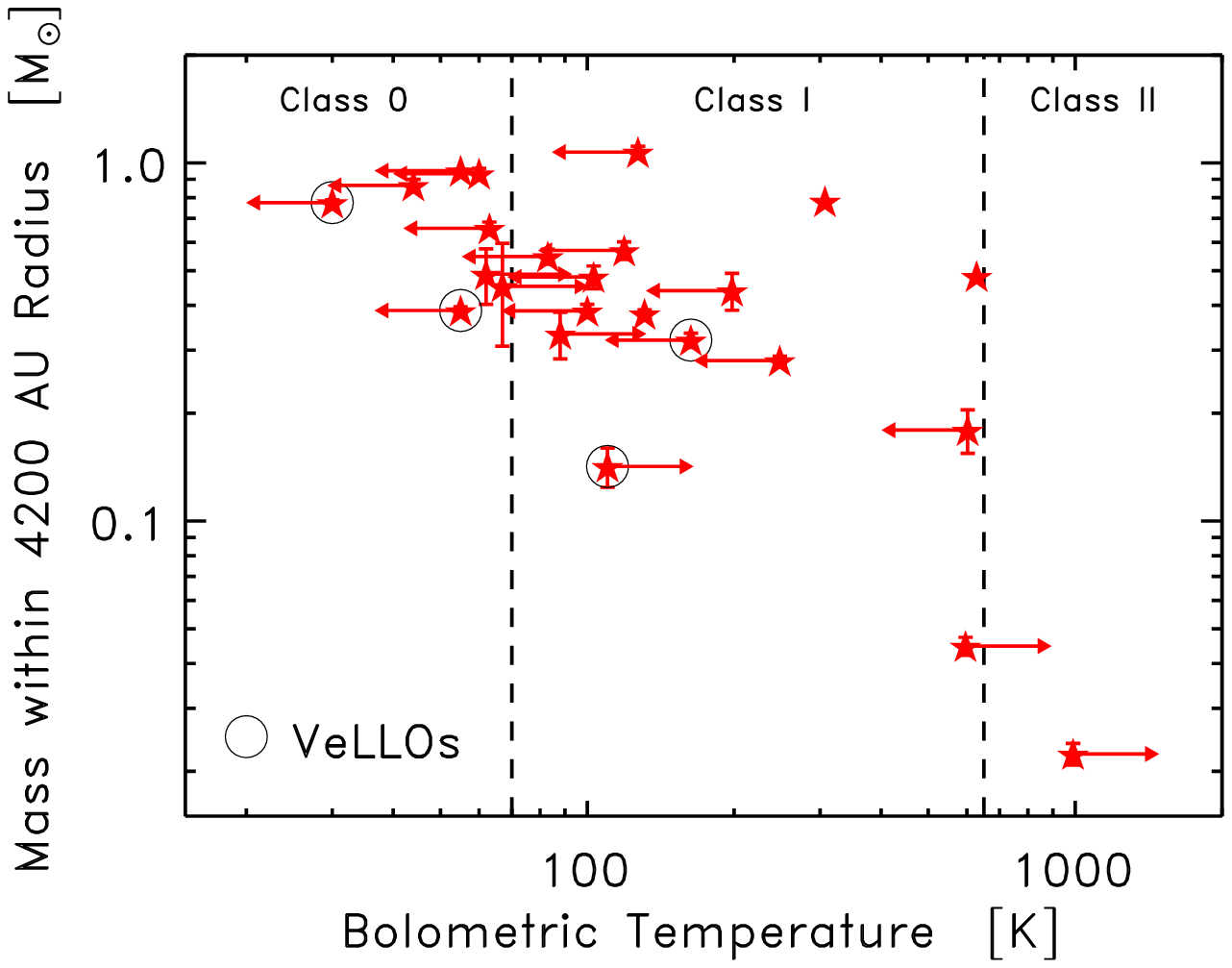}
\caption{Bolometric temperatures and mass within $4 \, 200 ~ \rm AU$
  radius for the YSOs in the c2d MAMBO sample. Candidate VeLLOs are
  highlighted by \emph{circles} around the markers, and the
  YSO class regimes are indicated by \emph{vertical dashed
    lines} and \emph{labels}. The general trend of increasing
  bolometric temperature with
  decreasing aperture mass might be an evolutionary effect. It could,
  however, also indicate that low bolometric temperatures are
  partially due to reddening by extinction and re-emission at longer
  wavelength, since both should
  scale with aperture mass. VeLLOs notably have the lowest bolometric
  temperature for given aperture mass, but this is likely simply
  reflects their definition as objects of low
  luminosity.\label{fig-survey:tbol_vs_m4200}}
\end{figure}

Figure \ref{fig-survey:tbol_vs_m4200} presents bolometric temperatures
and aperture masses for the c2d MAMBO sources. This reveals that on
average the bolometric temperature decreases with increasing aperture
mass. This could be an evolutionary trend. The coolest sources are
supposedly the youngest sources, which are expected to be surrounded
by significant amounts of circumstellar matter. Our observations are
thus consistent with expectations.

However, the large column densities suggested by large aperture masses
might lead to low bolometric temperatures because of reddening:
photospheric YSO emission might be absorbed at short wavelengths and
be re-emitted by the envelope at longer wavelengths. In that case
bolometric temperatures would not unambiguously characterize YSO ages,
but would partially be due to reddening not related to the time
evolution of the source. Our sample is, however, too small to evaluate
this issue comprehensively.

Within our survey, VeLLO candidates appear to have for given aperture
mass the lowest bolometric temperature observed. However, this likely
reflects the definition of VeLLOs as objects of low internal
luminosity, since for fixed millimetre flux (or $M_{\rm 4200 AU}$) the
bolometric temperature decreases with decreasing luminosity.

Note that all the ``usual'' young YSOs with
$T_{\rm bol} < 300 ~ \rm K$ have
$M_{\rm 4200 AU} > 0.3 \, M_{\odot}$, just as already discussed in
Sec.\ \ref{sec-survey:star-form_ability}. Also note the
aperture masses of $M_{4200 \rm AU} \gtrsim 0.5 M_{\odot}$ for the
YSOs in L1228 P1 and Bern48 P1 are unusually large, given their
bolometric temperatures of $\ge 300 ~ \rm K$. It is not clear how such
sources of a relatively late evolutionary stage can maintain such
massive envelopes, if our mass estimates are correct.

\subsubsection{Assignment of Infrared Classes\label{sec-survey:class_assignments}}
The above discussion often assumed that the bolometric temperature
uniquely parameterizes the YSO age. This is not obvious, and
possibly not true. We briefly summarize hints on this issue from the
c2d MAMBO data.

\begin{figure}
\includegraphics[width=\linewidth]{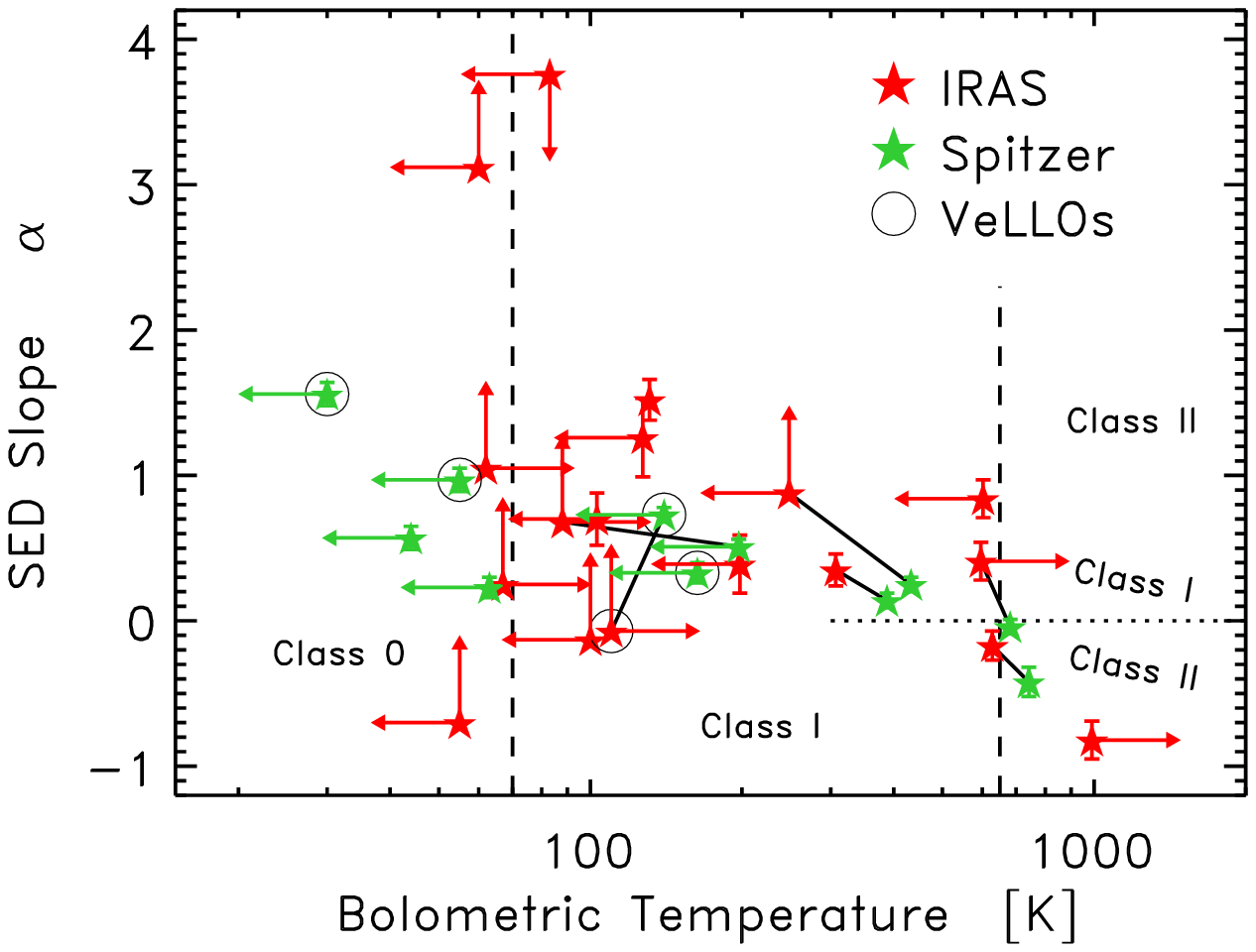}
\caption{SED slope versus the bolometric temperature. See Fig.\
  \ref{fig-survey:lbol_vs_tbol} for the symbols used. Note that IRAS
  data usually implies SED slopes larger than, and bolometric
  temperatures smaller than, those derived from Spitzer data, just as
  expected from IRAS' better spectral coverage at long wavelengths.
  Note that class assignments on basis of bolometric temperature and
  spectral slope do agree.\label{fig-survey:alpha_vs_tbol}}
\end{figure}

Figure \ref{fig-survey:alpha_vs_tbol} presents
the relation between bolometric temperature and SED slope near
$10 ~ \rm \mu m$ wavelength. The slopes derived from IRAS and Spitzer
data ($\alpha_{12 \rm \mu m}^{25 \rm \mu m}$ and
$\alpha_{3.6 \rm \mu m}^{24 \rm \mu m}$, respectively)
do slightly differ in their definition (see Sec.\
\ref{sec-survey:protostellar_data}). In 4 out of 5 cases
$\alpha_{12 \rm \mu m}^{25 \rm \mu m} >
\alpha_{3.6 \rm \mu m}^{24 \rm \mu m}$ for sources detected by both
missions. Most
sources in our survey have spectral
slopes of $0.0 \lesssim \alpha \lesssim 1.5$. We find a
general trend of increasing $\alpha$ with decreasing $T_{\rm bol}$
across the whole sample. No such trend is, however, visible when
looking at class I sources only (class 0 sources are usually not
expected to have particular values of $\alpha$). Note that class
assignments on basis of bolometric temperature or spectral slope are
consistent with each other within our sample. Given the expected
uncertainties, all sources in our sample are consistent with having
$T_{\rm bol} < 650 ~ \rm K$ (i.e., being a class 0 or class I source)
when $\alpha > 0$ (note, however, that classification schemes never
required class 0 to have $\alpha > 0$), and
$650 < T_{\rm bol} / {\rm K} < 2880$ (i.e.,
being a class II source) when $0 > \alpha > -2$, just as postulated in
class definitions based on $T_{\rm bol}$ and $\alpha$
(\citealt{chen1995:tbol};
\citealt{lada1987:ass_to_protostars}). Similar trends are found in the
c2d SED studies of the Ophiuchus, Perseus, and Serpens molecular cloud
complexes by Enoch et al.\ (in prep.).

Also note that all IRAS sources not detected by MAMBO (Table
\ref{tab-survey:IRAS_not-detected}) are consistent with having
negative spectral indices and therefore belonging to late YSO stages
(if they are YSOs and not background galaxies). This fits well into a
picture in which YSOs with negative spectral slopes are evolved and
not anymore surrounded by dust.

\begin{figure}
\includegraphics[width=\linewidth]{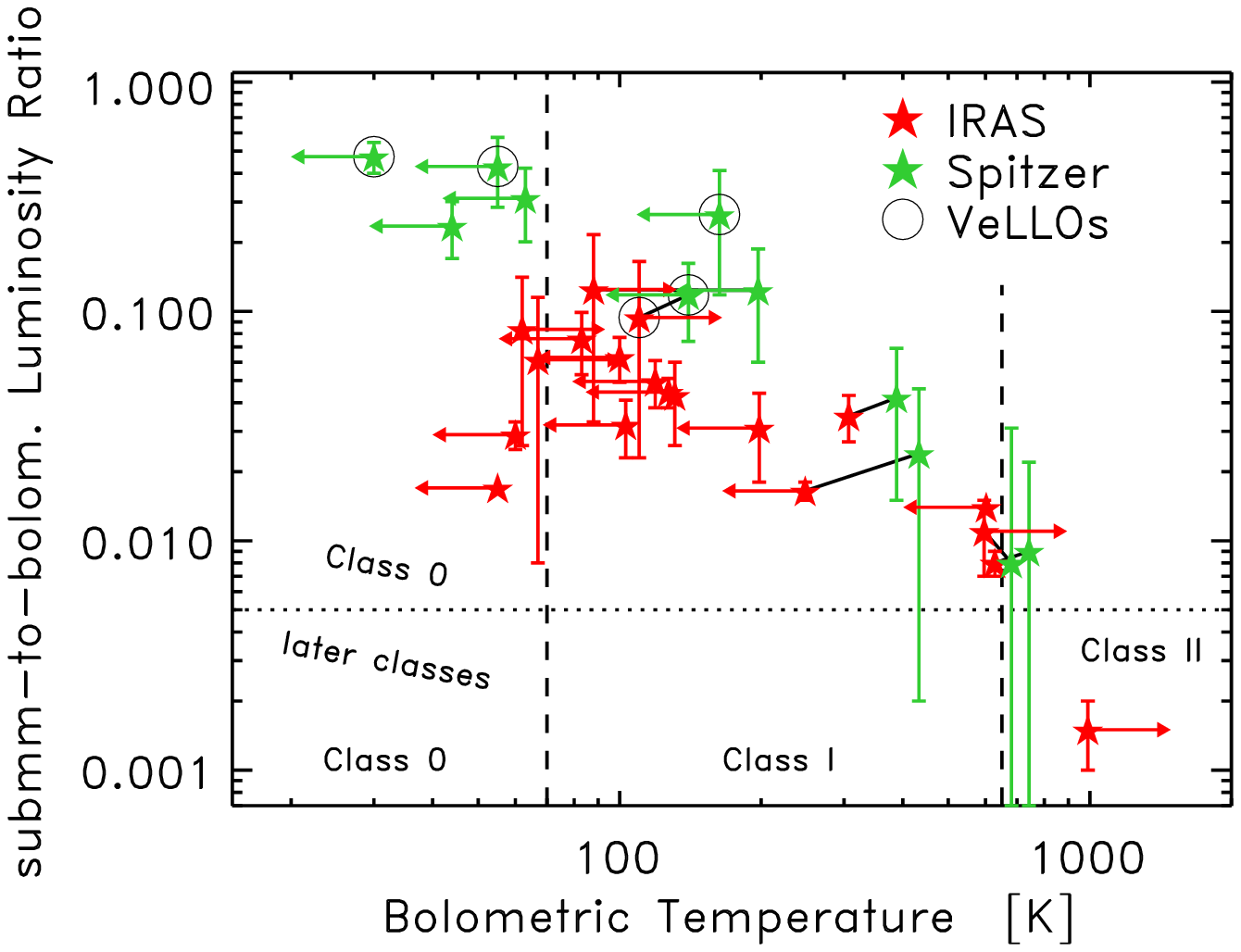}
\caption{Submillimetre-to-bolometric luminosity ratio versus the
  bolometric temperature. See Fig.\ \ref{fig-survey:lbol_vs_tbol} for
  the symbols used. \emph{Vertical dashed lines} indicate ranges for
  YSO classes assigned on basis of $T_{\rm bol}$, while the
  \emph{horizontal dotted line} (described by \emph{inclined labels})
  gives the boundary for the class assignment based on the luminosity
  ratio following
  \citet{andre1993:class0}. It's disagreement with $T_{\rm bol}$-based
  class assignments is likely caused by positive biases in our
  evaluation of the submillimetre
  luminosity.\label{fig-survey:luminosity_ratio}}
\end{figure}

Our submillimetre-to-bolometric luminosity ratios (Fig.\
\ref{fig-survey:luminosity_ratio}; also see Tables
\ref{tab-survey:IRAS_detected} and \ref{tab-survey:SST_protostars})
and $T_{\rm bol}$-based class assignments are not well in accord with
the original initial observational criterion to identify class 0
sources given by \citet{andre1993:class0}, i.e., a luminosity ratio
$\gg 0.005$; using the latter criterion, all but one of our YSOs would
be in the class 0 category. Note, however, that the submillimetre part
of our SEDs is the worse constrained one and that the luminosities are
uncertain by a factor 2(Sec.\ \ref{sec-survey:protostar_properties}).
Anyway, submillimetre luminosities of
$0.01 ~ {\rm to} ~ 0.1$ for $T_{\rm bol} \approx 70 \, \rm K$ have
been observed before \citep{young2005:yso_models} and also
\citet{visser2002:lynds_clouds} found that essentially all YSOs in
their sample would be of class 0 if the class is assigned based on a
luminosity ratio $> 0.005$ (see \citealt{hatchell2007:perseus} for a
further example; note that both studies do, like us, have SEDs not
well sampled at submillimetre wavelengths). The role of the
submillimetre-to-bolometric luminosity ratio for classification
purposes thus needs to be further studied with better wavelength
coverage.

Also, the bolometric temperature might not uniquely parameterize age,
as the aforementioned trend of decreasing bolometric temperature with
increasing aperture mass might indicate.

It has also been speculated that actual class I sources have SEDs like
class 0 sources if the circumstellar disk-envelope structure is seen
edge-on (e.g., \citealt{whitney2003:class_i} for an overview).
Interestingly, the class 0 source in IRAS05413 P3 appears to have such
a particular geometry; here the outflow has an estimated inclination
of $\gtrsim 85 \degr$ with respect to the line of sight, suggesting
that a YSO disk is seen nearly edge-on
\citep{claussen1998:iras05413}. This geometry might explain why
IRAS05413 P3, being a class 0 source marginally violating the
conceptual mass-luminosity limit by \citet{motte2001:protostars}, has
an extreme position in Fig.\ \ref{fig-survey:lbol_vs_m4200}.\medskip

\noindent In summary, class assignments based on $T_{\rm bol}$ are
consistent with those based on $\alpha$, and either is fine to
classify sources of a relatively late evolutionary stage. One needs
to employ the bolometric temperature though if intending to study
deeply embedded sources, for which the spectral index criteria do not
hold. The $T_{\rm bol}$ thus appears to be more versatile, but
requires a well sampled SED at long wavelengths.

In contrast, assignments based on the
$m_{4200 \rm AU}$-to-$L_{\rm int}$ ratio are in conflict with those
based on $T_{\rm bol}$. A $T_{\rm bol}$-based classification does
however lead to a larger spectrum in evolutionary states:
$T_{\rm bol}$ indicates class 0 and class I sources, while the
$m_{4200 \rm AU}$-to-$L_{\rm int}$ ratio suggests that our sample
essentially only contains class 0 objects. Since the latter seems
somewhat unrealistic, the larger range in assigned classes
suggests that the bolometric temperature is more useful for
classification.

Similarly, the submillimetre-to-bolometric luminosity ratio assigns
all but one YSO to the class 0. Again, $T_{\rm bol}$-based
classification appears superior to the one based on luminosity
ratios. This might, however, in part be due to uncertainties in our
SEDs at wavelength $100 ~ {\rm to} ~ 1000 ~ \rm \mu m$ and should not
be generalized. Still, for data like ours luminosity ratios are barely
useful.

Anyway, it is not clear which existing YSO class assignments best
characterize the evolutionary ages of YSOs. In this respect it might
be useful (see, e.g., \citealt{robitaille2006:yso_sed-grid}) to
distinguish between YSO \emph{classes}, that describe and organize
observational properties such as the bolometric temperature, and
evolutionary \emph{stages}, that characterize physical properties such
as the presence of envelopes and disks.

\subsubsection{YSO Offsets\label{sec-survey:offsets}}
The spatial resolution of MAMBO and Spitzer allows to test whether
YSOs actually reside in the very center of their natal dense core. The
discovery of offsets would be very interesting, since present YSO
evolutionary scenarios do not predict them. As shown in Table
\ref{tab-survey:offsets} we do, however, not find offsets at a
significant level.

For this search the YSO positions are taken to be those of Spitzer
counterparts listed in Table \ref{tab-survey:SST_protostars}.
Otherwise the positions of the 2MASS counterparts given in Table
\ref{tab-survey:IRAS_detected} are used. For YSOs with non-c2d Spitzer
data only, Gaussian curves were fitted to the Spitzer images (see
notes to Table \ref{tab-survey:offsets}). The dust intensity peak
positions are derived as the intensity-weighted mean position of the
pixels above 90\% peak intensity. The offsets are given in Table
\ref{tab-survey:offsets}. The uncertainty in the dust emission peak
position due to noise is derived from Monte-Carlo experiments with
artificial noise, as explained in Sec.\ \ref{sec-survey:id_and_quant}.
The typical pointing uncertainty of $3 \arcsec$ has to be further
added to derive the typical total dust emission position uncertainty.

\begin{table}
\caption{YSO offsets from dust emission peaks. For each dust
  emission peak, and related subcore, associated with a YSO
  listed in Tables \ref{tab-survey:IRAS_detected} and \ref{tab-survey:SST_protostars}
  we give the chosen reference source (preferentially seen by Spitzer)
  and the observed offset and position angle (i.e.,
  position of star w.r.t.\ the dust emission
  peak).\label{tab-survey:offsets}
}
\begin{center}
\begin{tabular}{llllllllllllllllllll}
\hline \hline
\rule{0ex}{3Ex}Field & Peak & Position Reference & Offset\\
 & & (see notes) & arcsec\\\hline
L1521F & P1 & SSTc2d J042839.0+265135 & $  5.0 \pm  0.6 $\\
B18-1 & [P5] & SSTc2d J043215.4+242859 & $  1.1 \pm  0.6$\\
B18-4 & P1 & IRAC & $  2.2 \pm  0.6 $\\
TMC1-1C & P3 & SSTc2d J044138.8+255627 & $  3.3 \pm  2.5 $\\
TMC-1 & P2 & SSTc2d J044112.7+254635 & $  3.1 \pm  0.6 $\\
IRAS05413 & P2 & MIPS & $3.8 \pm  0.5$\\
 & P3 & MIPS & $3.6 \pm  0.3 $\\
CB188 & P1 & SSTc2d J192014.9+113540 & $  2.3 \pm  0.7 $\\
L673-7 & P1 & SSTc2d J192134.8+112123 & $  2.1 \pm  0.9 $\\
L1148 & P1 & SSTc2d J204056.7+672305 & $  7.6 \pm  2.7 $\\
L1082A & P1 & 2MASS J20531346+6014425 & $  2.7 \pm  0.9 $\\
L1228 & P1 & SSTc2d J205712.9+773544 & $  0.8 \pm  0.1 $\\
Bern48 & P1 & SSTc2d J205914.0+782304 & $  0.8 \pm  0.3 $\\
L1172A & P2 & MIPS & $7.3 \pm  1.4 $\\
L1177 & P2 & IRAC & $2.9 \pm  0.1 $\\
L1021 & P1 & 2MASS J21212751+5059475 & $  2.5 \pm  0.8 $\\
L1014 & P1 & SSTc2d J212407.5+495909 & $  2.2 \pm  1.3 $\\
L1251A & P3 & SSTc2d J222959.5+751404 & $  8.8 \pm  1.3 $\\
 & P4 & SSTc2d J223031.8+751409 & $  1.2 \pm  0.2 $\\
 & P5 & SSTc2d J223105.6+751337 & $  4.3 \pm  0.8 $\\
\hline
\end{tabular}
\end{center}
\rule{0ex}{3Ex}Notes: Position for B18-4 P1 (average of Gaussian fits
to all IRAC bands) is 04:35:35.4, 24:08:20.2 (J2000.0);
for IRAS05413 P2 (from fits to MIPS band 1) it is 05:43:46.264,
-01:04:43.68;
for IRAS05413 P3 (from MIPS band 1) it is 05:43:51.367, -01:02:52.76;
for L1172A P2 (from MIPS band 1) it is 21:02:21.118,
+67:54:20.17; and for L1177 P2 (from IRAC band 4 only, where the
significant scattering is supposedly lowest) it is
21:17:38.650, +68:17:33.27.
\end{table}

We do not find significant YSO offsets in our sample. In a first cut
we find two YSOs with offsets exceeding the expected uncertainty by a
factor 2. However, for one of them (L1251A P3) the pointing
corrections were up to $5 \farcs 9$. For the other one (L1172A P1) the
position offset exceeds the uncertainty by a factor 2.2; this is
marginally significant, and should be followed up at higher
resolution, but not conclusive evidence for a YSO leaving its natal
core. Our data thus indicates that generally YSOs reside in the
centers of their envelopes, with core-YSO relative motions not e.g.\
significantly affecting the accretion histories. This is in accord
with studies of core-YSO-offsets in Perseus
(\citealt{jorgensen2007:perseus_submm}; Enoch et al., in prep.) and
Ophiuchus (J\o{}rgensen et al., in prep.).

Note, however, that our data does not reveal the offset between the
L1014 P1 column density peak and the embedded candidate VeLLO found by
\citet{huard2006:l1014} on basis of extinction maps, although their
offset of $10 \arcsec ~ {\rm to} ~ 15 \arcsec$ should be easily
detected by MAMBO. \citeauthor{huard2006:l1014} speculate that this
might be because of internal heating of the L1014 core, which is
supposed to shift the dust emission peak away from the column density
peak and towards the embedded source (see \citealt{young2004:l1014}
and \citealt{wu2007:c2d_sharc-ii} for SCUBA and SHARC~\textsc{ii} data
on L1014, respectively). Their extinction map is, however, not well
sampled towards the VeLLO position, which introduces uncertainties to
their map.

\subsection{Formation and Evolution of
  VeLLOs\label{sec-survey:form_and_evol_vellos}}
The c2d MAMBO survey provides an opportunity to investigate how
candidate VeLLO natal cores are different from starless and YSO cores.
For the first time it allows a direct comparison of the dust emission
properties of all three kinds of objects. We here discuss whether, and
if yes how, VeLLO cores differ from starless and YSO cores.

\subsubsection{Masses and Total Radii of VeLLO
  Cores\label{sec-survey:vello_masses_radii}}
Figures \ref{fig-survey:histograms} and
\ref{fig-survey:star-form_ability} show that VeLLO candidate cores cannot be
distinguished from starless cores in terms of mass, column density,
and total effective radius. In particular, three of the VeLLO cores
have column densities, masses within the $4 \, 200 ~ \rm AU$ aperture,
and masses within the 50\% intensity contour --- all properties
naively associated with a good chance for star formation --- much
lower than the starless peaks B18-1 P3 and B18-4 P2.
Note that this observation is in conflict with some naive expectations
for properties of starless and YSO cores: a picture of dense core
evolution in which any dense core begins to actively form stars once
it exceeds some fixed limit in mass, density, or both, is inconsistent
with our data. Section \ref{sec-survey:star-form_ability} discussed
this and other data in the context of necessary, but not sufficient,
conditions for active star formation. 

Note that 3 out of 4 VeLLO candidates are consistent with fulfilling
these conditions (i.e., in Fig.\ \ref{fig-survey:star-form_ability},
only the candidate VeLLO in L1148 resides in a region entirely devoid
of normal YSOs, and thus also misses the theoretical star formation
criterion by a large margin). In this respect \emph{the notion that
  VeLLOs reside in cores not expected to be able to form stars might
  be wrong}.  However, it might be that some VeLLOs can conceptually
be characterized as YSOs forming in cores that only barely fulfill the
necessary conditions for active star formation, as observed for 3 out
of the 4 candidate VeLLOs in our sample. The latter might suggest a
special route of star formation for some VeLLOs.

The VeLLO candidate in L1148 P1 poses a real challenge to present star
formation scenarios, since it does not fulfill the necessary conditions
for star formation to occur. This source might thus either hint at a
particular mode of star formation, where the conditions derived in
Sec.\ \ref{sec-survey:star-form_ability} do not apply, or this YSO
(recent IRAM Plateau de Bure imaging indicates either a jet or a disk;
Kauffmann et al., in prep.) is in an advanced evolutionary stage in
which the natal core's structure does not resemble the initial one.

\subsubsection{Peak Radii of VeLLO Cores\label{sec-survey:vello_radii}}
Most VeLLO candidate cores have peak density profiles flatter than $\varrho
\propto r^{-2}$, since their effective radius at 70\% peak intensity
is (with the exception of the candidate VeLLO in L673-7 P1)
significantly larger
than the one of a $\varrho \propto r^{-2}$ density profile smeared by
a $20 \arcsec$ beam (i.e., $10 \farcs 7$). It is beyond the scope of the
present paper to explore whether this flattening is because the
density profile becomes similar to $\varrho \propto r^{-3/2}$, as
found for infalling envelopes around YSOs (see
\citealt{evans2003:infall} for a review), or is due to a region of
spatially constant density, as found for starless dense cores (e.g.,
\citealt{ward-thompson1999:prestellar_cores}). Note, however, that
\citet{crapsi2004:l1521f} show that MAMBO maps for L1521F are
consistent with a functional form for the central flattening as found
for starless cores. The high resolution extinction data for L1014
obtained by \citet{huard2006:l1014} has not been analysed in such a
fashion.

%
%
%
%
%
%
%

Compared to starless cores, candidate VeLLO natal cores do, however, have
unusually steep dust emission intensity profiles, as e.g.\ probed by
the effective radius of a dust emission peak at its 70\% intensity
level. For given aperture mass, VeLLO cores (with the possible
exception of the uncertain radius of L1148 C1) do have smaller
effective radii than most starless cores (Fig.\
\ref{fig-survey:r70_vs_m4200}), even when correcting for steepening
due to internal heating by the VeLLO (for this we first use the
luminosity-dependent beam-averaged dust temperature [Eq.\
\ref{eq-survey:quant-2}] to predict the peak intensity for dust of
$10 ~ \rm K$, and then evaluate the effective radius at 70\% of this
intensity).

The unusually steep dust intensity profiles of candidate VeLLO natal cores thus
indicate unusually steep column density profiles. This distinguishes
them from most starless cores in our sample. Steep radial profiles in
the density, and therefore also in the column density, are indeed
expected for dense cores temporally near (e.g.,
\citealt{mckee1999:mpp}) or after (see \citealt{larson2003:sf-physics}
for a review) the onset of gravitational collapse. Density gradients
in VeLLO cores are consistent with this theoretical picture. VeLLO
candidates
also seem to have unusually large central densities, when compared to
starless cores (Fig. \ref{fig-survey:r70_vs_m4200}). This is again in
line with the above theoretical expectations.

The steeper density profiles and higher central densities of candidate
VeLLO
natal cores suggest that they are temporally closer to the onset of
gravitational collapse than most starless cores. \emph{It thus appears
  that VeLLO cores are structurally different from most starless cores
  in that their density structure is more evolved.} This was first
suggested by \citet{huard2006:l1014}, on basis of surface-to-center
density contrasts for L1014 P1 that exceed those of previously studied
starless cores. Also, all candidate VeLLOs (possibly excluding L1148 P1) have
effective radii at the 70\% dust emission peak intensity level below
$4 \, 800 ~ \rm AU$, which \citet{crapsi2005:survey} suggested as an
indicator of an advanced evolutionary stage of a core. This might bear
important hints on the nature of VeLLOs.

\subsubsection{VeLLO Offsets from Column Density
  Peaks\label{sec-survey:vello_offsets}}
\citet{huard2006:l1014} found that the candidate VeLLO in L1014 is offset by
$10 \arcsec$ to $15 \arcsec$ from the apparent natal column density
peak, L1014 P1. If L1014-IRS does indeed drift away from its natal
core, and therefore away from the mass reservoir available for
accretion, this might explain the low rate of accretion onto L1014-IRS
inferred by \citet{bourke2005:l1014-outflow}. Low accretion rates
could in turn explain the low luminosities of VeLLOs. They would also
suggest low, and possibly substellar, final masses for VeLLOs with low
accretion rates \citep{bourke2005:l1014-outflow}. However, as already
discussed in Sec.\ \ref{sec-survey:offsets}, we do not find any
significant offsets. Our data might not rule out such offsets though,
since the MAMBO data might not be suited to search for offsets (Sec.\
\ref{sec-survey:offsets}).

\section{Summary\label{sec-survey:summary}}
We present a dust continuum emission imaging study of a comprehensive
sample of isolated starless and YSO-containing dense molecular cores
(including 4 candidates of the enigmatic VeLLOs), the c2d MAMBO
survey. This survey is technically different from previous ones in
that it is
\begin{itemize}
\item more sensitive to faint (few mJy per $11 \arcsec$ beam) extended
  ($\gtrsim 5\arcmin$) emission than in previous surveys thanks to
  modern instrumentation and a newly devised data reduction strategy
  (Sec.\ \ref{sec-survey:new_strategy}), resulting in
\item very rich structure compared to existing surveys, since
  we detect faint extended emission from sources otherwise only
  detected in their most prominent intensity peaks.
\end{itemize}
We combine our MAMBO data with Spitzer and IRAS data to fully exploit
our dust emission maps (Sec.\
\ref{sec-survey:protostellar_data}). This also reveals 4 previously
unreported YSOs (Table \ref{tab-survey:spitzer_new}), including one
VeLLO candidate and 2 class 0 sources. Our analysis (Sec.\
\ref{sec-survey:analysis}) first deals with the general structure and
evolution of dense cores (Secs.\ \ref{sec-survey:morphology} to
\ref{sec-survey:core_evolution}):
\begin{itemize}
\item characteristic values for the mass within peak-centered
  apertures of $4 \, 200 ~ \rm AU$ diameter are similar to those of
  singular isothermal spheres, suggesting that cores prefer a
  near-critical state (with respect to gravitational collapse; Sec.\
  \ref{sec-survey:mass_within_4200au} and Fig.\
  \ref{fig-survey:histograms} [c]);
\item subcores (i.e., extended substructure within cores) are usually
  elongated, implying that they are not hydrostatic objects supported
  by isotropic pressure (Sec.\ \ref{sec-survey:elongation} and Fig.\
  \ref{fig-survey:histograms} [d]);
\item rough density estimates indicate central $\rm H_2$ densities
  between $10^4$ and $10^6 ~ \rm cm^{-3}$ while the density frequency
  distribution suggests that cores in our sample are not in free fall
  (Sec.\ \ref{sec-survey:densities} and Fig.\
  \ref{fig-survey:r70_vs_m4200});
\item comparison with criteria by \citet{crapsi2005:survey} reveals 5
  candidate evolved cores (plus 3 candidates of lesser quality), out
  of which 3 are reported here for the first time, but the failure of
  these criteria to highlight candidate VeLLO cores with active star
  formation casts some doubt on them (Sec.\ 
  \ref{sec-survey:highly_evolved_cores};
\item rough estimates for the dense core mass available for accretion
  onto a forming YSO yield values of order 10\% of the dense core
  total mass (with an uncertainty of a factor of several), suggesting
  that the star formation efficiency is of the same order (which
  questions studies relating the mass functions of cores and stars;
  Sec.\ \ref{sec-survey:sf-efficiency} and Fig.\
  \ref{fig-survey:star-form_ability} [c]);
\item there are necessary (but not sufficient) conditions which a dense
  core must fulfill in order to be able to actively form a star (i.e.,
  a core fulfilling these can contain a YSO, but does not need to),
  which can be understood in the framework of quasi-static dense core
  models (Sec.\ \ref{sec-survey:star-form_ability} and Fig.\
  \ref{fig-survey:star-form_ability}); and
\item diverse biases can and do introduce spurious features into the
  data, requiring care when interpreting the observations.
\end{itemize}
We then specifically turn to young stellar objects in our sample,
where we in particular compare various methods to assign SED
classes (Sec.\ \ref{sec-survey:evol_protostars}):
\begin{itemize}
\item class assignments based on the bolometric temperature are in
  conflict with those based on the comparison of YSO luminosities and
  envelope masses (Sec.\ \ref{sec-survey:masses_and_luminosities} and
  Fig.\ \ref{fig-survey:lbol_vs_m4200});
\item classes assigned on the basis of $T_{\rm bol}$ and SED slopes
  agree well, i.e., the data is consistent with
  $\alpha \gtrless 0$ for $T_{\rm bol} \lessgtr 650 ~ \rm K$
 (Sec.\ \ref{sec-survey:class_assignments} and Fig.\
  \ref{fig-survey:alpha_vs_tbol});
\item among the class assignment methods used in the present study,
  only the one based on bolometric temperatures implies a reasonable
  range of evolutionary stages in our sample and is also well
  applicable to deeply embedded sources (Sec.\
  \ref{sec-survey:class_assignments});
\item disagreement of the $T_{\rm bol}$-based class assignments with
  those from YSO luminosities and envelope masses suggests that the
  environment has an impact on YSO evolution since these assignments
  agree in other regions (like the Ophiuchus molecular cloud; Sec.\
  \ref{sec-survey:masses_and_luminosities} and Fig.\
  \ref{fig-survey:lbol_vs_m4200});
\item there are no significant offsets between YSOs and their natal
  dense cores, implying that core-star relative motions do not have a
  significant influence on YSO evolution via the accretion history
  (Sec.\ \ref{sec-survey:offsets} and Table \ref{tab-survey:offsets});
  and
\item assuming that the bolometric temperature roughly characterizes
  YSO ages, the large range of bolometric luminosities (two
  magnitudes) in early YSO stages implies a large variation of
  mass-to-radius relations, accretion rates, or both, in our sample.
\end{itemize}
The combined information on cores in general and YSOs in particular
can then be used to assess the relevance of some characteristics of
candidate VeLLO natal cores (Sec.\ \ref{sec-survey:form_and_evol_vellos}):
\begin{itemize}
\item 3 out of 4 VeLLO candidate cores are consistent with fulfilling the above
  necessary conditions for active star formation, suggesting that the
  notion that VeLLOs form in cores not expected to be able to form
  stars is wrong, but some do only barely fulfill the requirements for
  active star formation, suggesting that some VeLLOs might form in
  cores just barely able to produce stars, possibly via a
    particular mode (Sec.\ \ref{sec-survey:vello_masses_radii});
\item candidate VeLLO natal cores have density profiles steeper than
  and central
  densities larger than most starless cores, suggesting that VeLLO
  candidate
  cores are structurally different from starless ones (Sec.\
  \ref{sec-survey:vello_radii}); and
\item given our failure to detect core-VeLLO offsets, low accretion
  rates because of core-star motion as an explanation for the low
  VeLLO luminosities are not consistent with our observations (Sec.\
  \ref{sec-survey:vello_offsets}).
\end{itemize}
In this respect our study --- the first one to compare candidate VeLLO cores
based on a homogeneous dataset --- helps to significantly further our
understanding of the nature of VeLLOs.

\begin{acknowledgements}
  We are grateful to A.\ Crapsi for sharing with us some of the MAMBO
  data used in this study, to IRAM's Granada staff for their support
  and hospitality during the long observing runs, and to M.\ Enoch,
  J.\ J\o{}rgensen, and M.\ Dunham for access to results in advance of
  publication. We are indebted to an unusually thorough and concise
  anonymous referee, who helped significantly in making the paper more
  readable. Support for this work, part of the Spitzer Legacy
  Science Program, was provided by NASA through contracts 1224608
  issued by the Jet Propulsion Laboratory, California Institute of
  Technology, under NASA contract 1407. Additional support came from
  NASA Origins grants NNG04GG24G and NNX07AJ72G to NJE. CWL
  acknowledges support by the KOSEF R01-2004-000-10513-0 program. JK
  is particularly thankful to the constant encouragement by K.\ Menten
  and M.\ Walmsley during the course of this study. This study had not
  been possible without the MAMBO bolometer build by E.\ Kreysa, the
  MOPSI software developed by R.\ Zylka, and the IRAM 30m-telescope
  maintained by IRAM's Granada staff. We thank the Lorentz Center in
  Leiden for hosting several meetings that contributed to this paper.
\end{acknowledgements}

\setcounter{figure}{0}
\begin{figure*}
\includegraphics[scale=0.95]{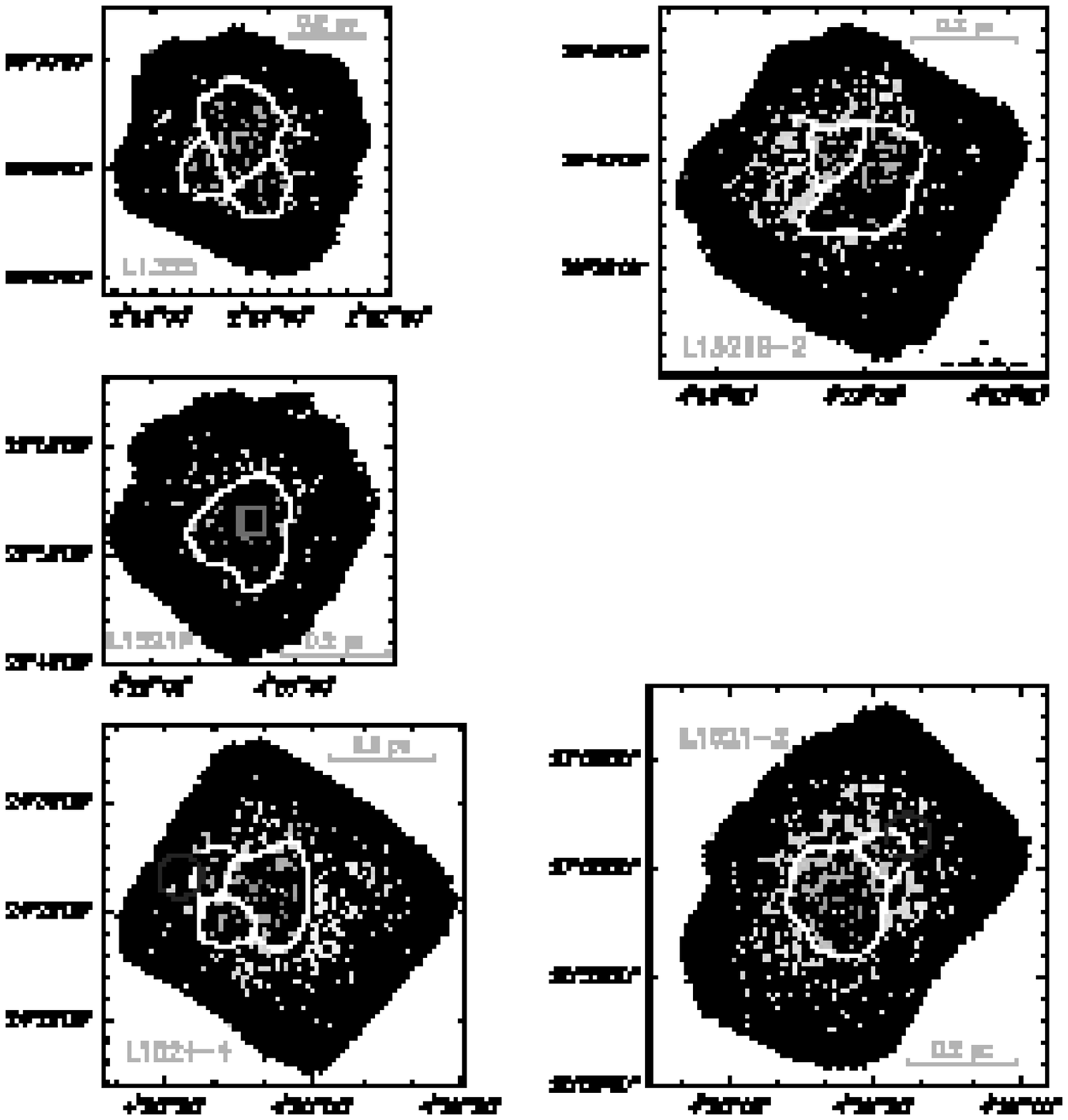}
\caption{Maps of the sources in the c2d MAMBO survey after smoothing
  to $20 \arcsec$ resolution. \emph{Coordinates} are given for J2000.0.
The sources are roughly ordered in
  sequence of increasing right ascension. \emph{Solid and dotted black
    contours} are drawn for positive intensities and intensities
  $\le 0 ~ \rm mJy ~ beam^{-1}$, respectively. Contours start at zero
  intensity and are spaced by $\pm 3 ~ \rm mJy$ per $11 \arcsec$ beam;
  this corresponds to the noise level in the map center times a factor
  $\approx 2 ~ {\rm to} ~ 3$ (Table \ref{tab-survey:observations}).
  Above $15 ~ \rm mJy ~ beam^{-1}$,
  indicated by a \emph{thick solid contour}, the contour spacing is
  $5 ~ \rm mJy ~ beam^{-1}$.  \emph{Gray shading} also reflects these
  intensity variations.  \emph{White contours} indicate subcore
  boundaries, while \emph{crosses} indicate dust emission peak
  positions. \emph{Horizontal bars} indicate the angular scale
  corresponding to a scale of
  $0.2 ~ {\rm pc} \approx 40 \, 000 ~ {\rm AU}$ at the adopted core
  distances. \emph{Circles and squares} mark the positions
  of all infrared stars detected by IRAS, and Spitzer stars also
  detected as MAMBO peaks, respectively. If detected by
  several missions, IRAS Point Source Catalogue (PSC) positions are used
  instead of those from the Faint Source Catalogue (FSC),
  and Spitzer positions are used instead of those listed from IRAS. The
  IRAS PSC source 20410+6710 in L1148 is not marked since missing
  Spitzer counterparts indicate that this source is an
  artifact. \astroph{See the alternative URL in the astro-ph comments,
  and the journal article, for higher resolution
  maps.}\label{fig-survey:maps}}
\end{figure*}

\setcounter{figure}{0}
\begin{figure*}
\includegraphics[scale=0.95]{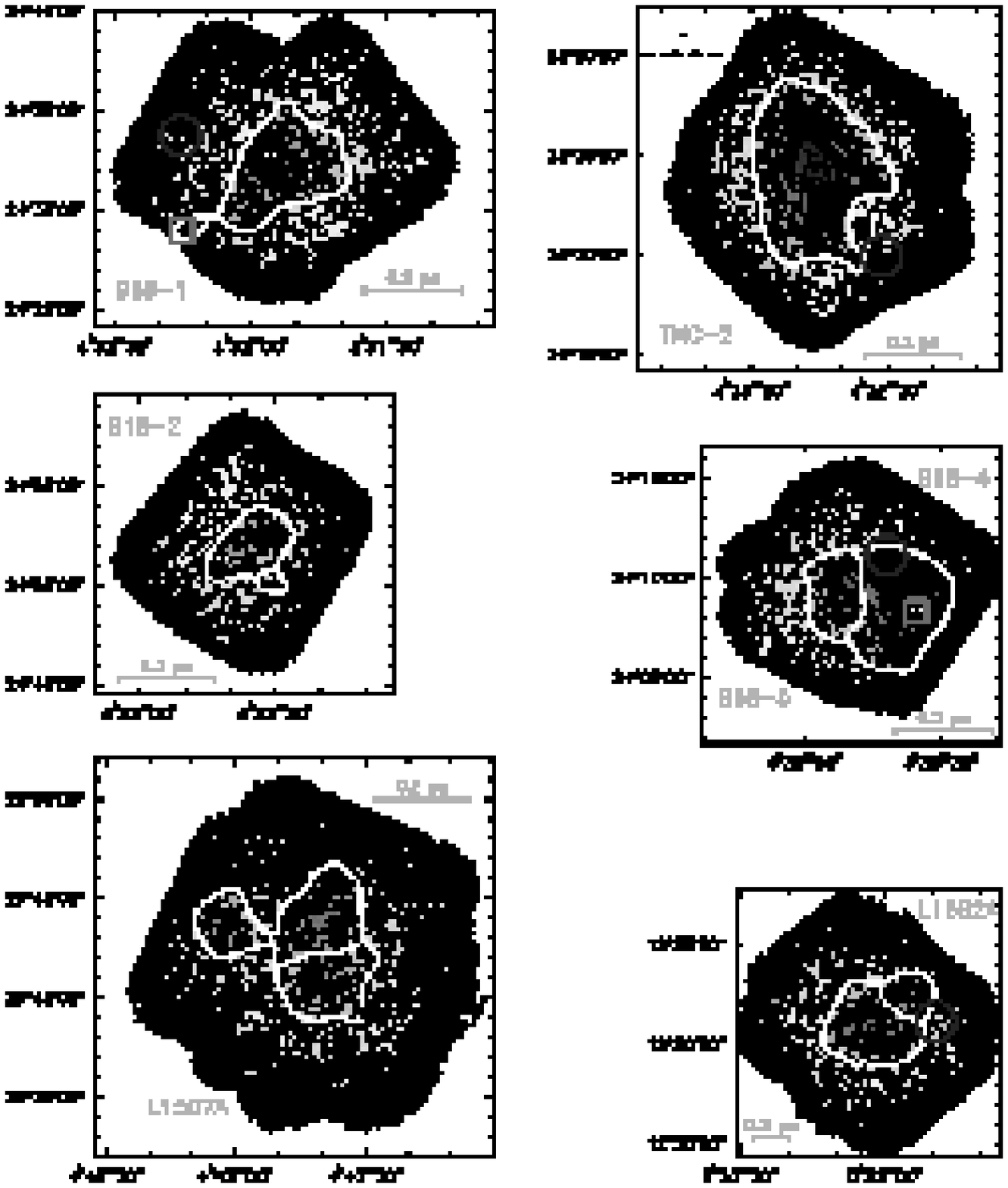}
\caption{continued.}
\end{figure*}

\setcounter{figure}{0}
\begin{figure*}
\includegraphics[scale=0.95]{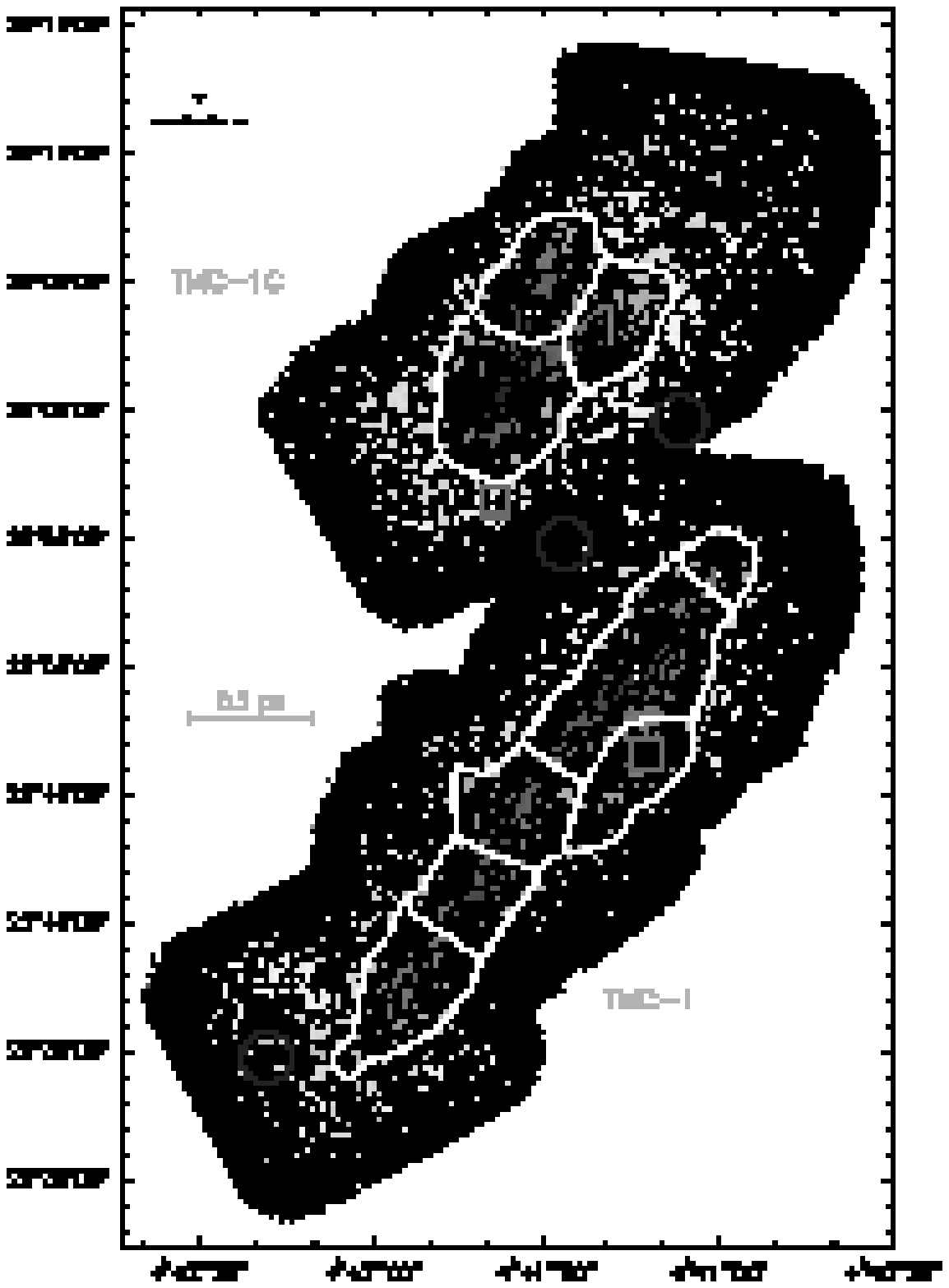}
\caption{continued.}
\end{figure*}

\setcounter{figure}{0}
\begin{figure*}
\includegraphics[scale=0.95]{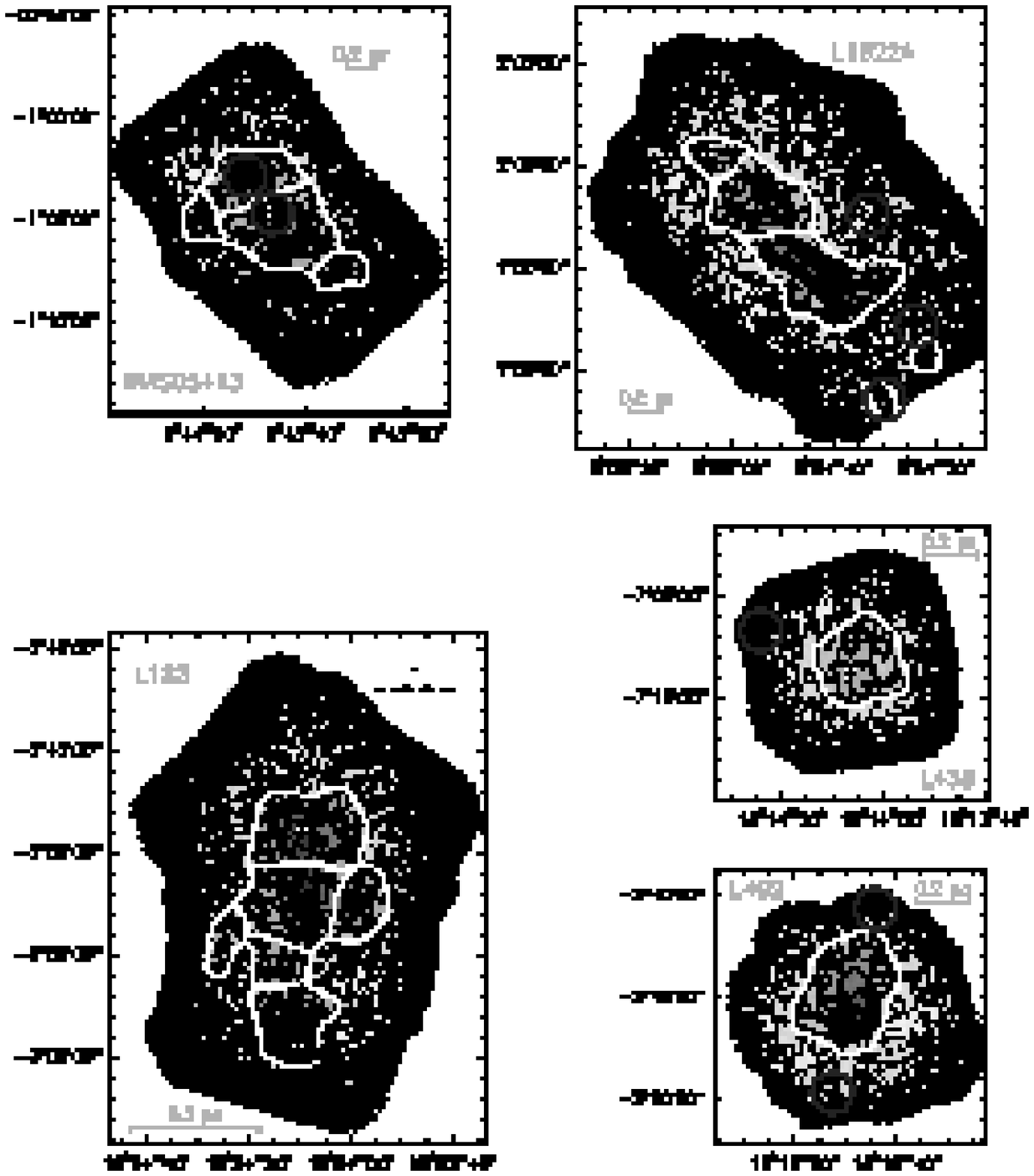}
\caption{continued.}
\end{figure*}

\setcounter{figure}{0}
\begin{figure*}
\includegraphics[scale=0.95]{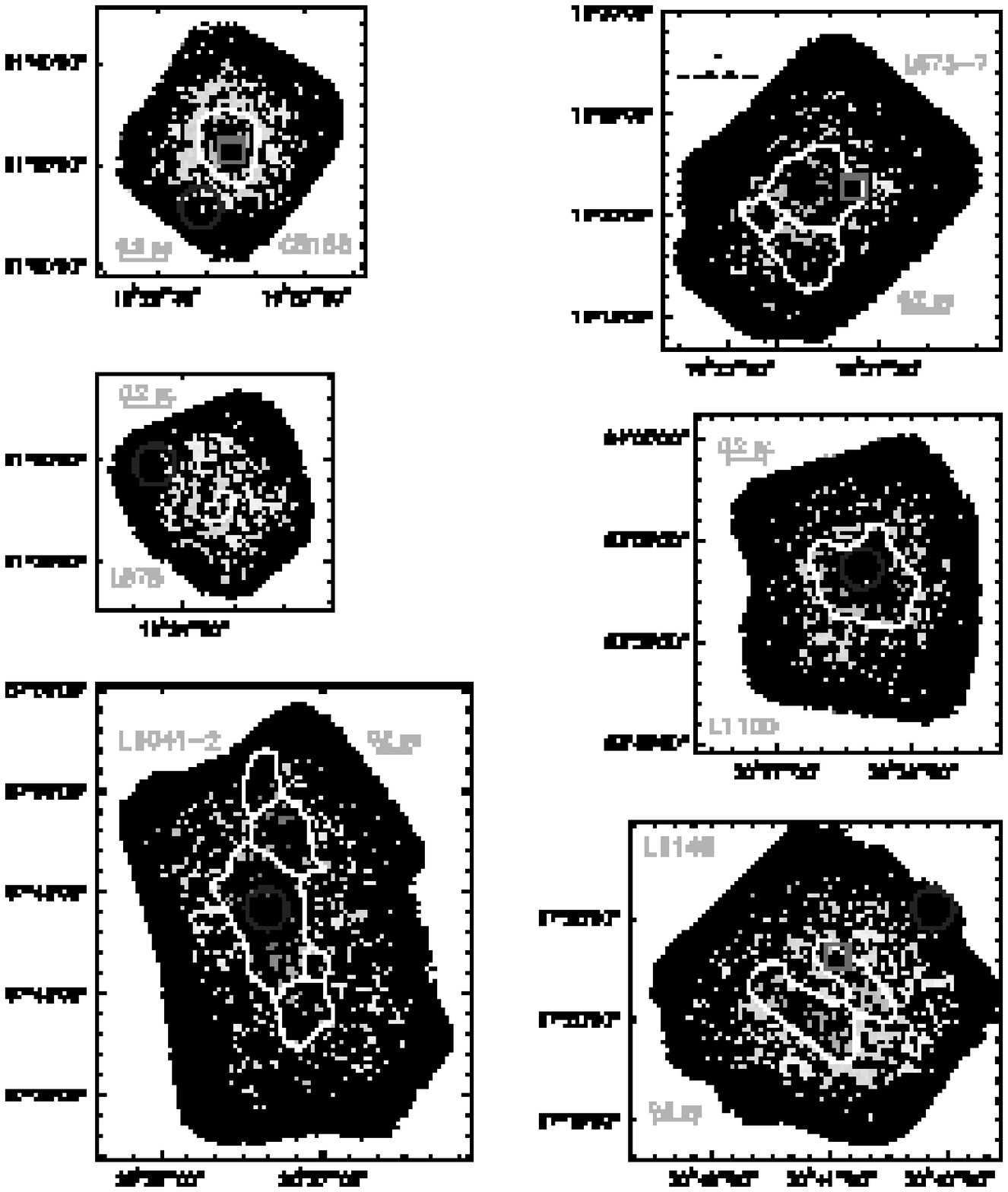}
\caption{continued.}
\end{figure*}

\setcounter{figure}{0}
\begin{figure*}
\includegraphics[scale=0.95]{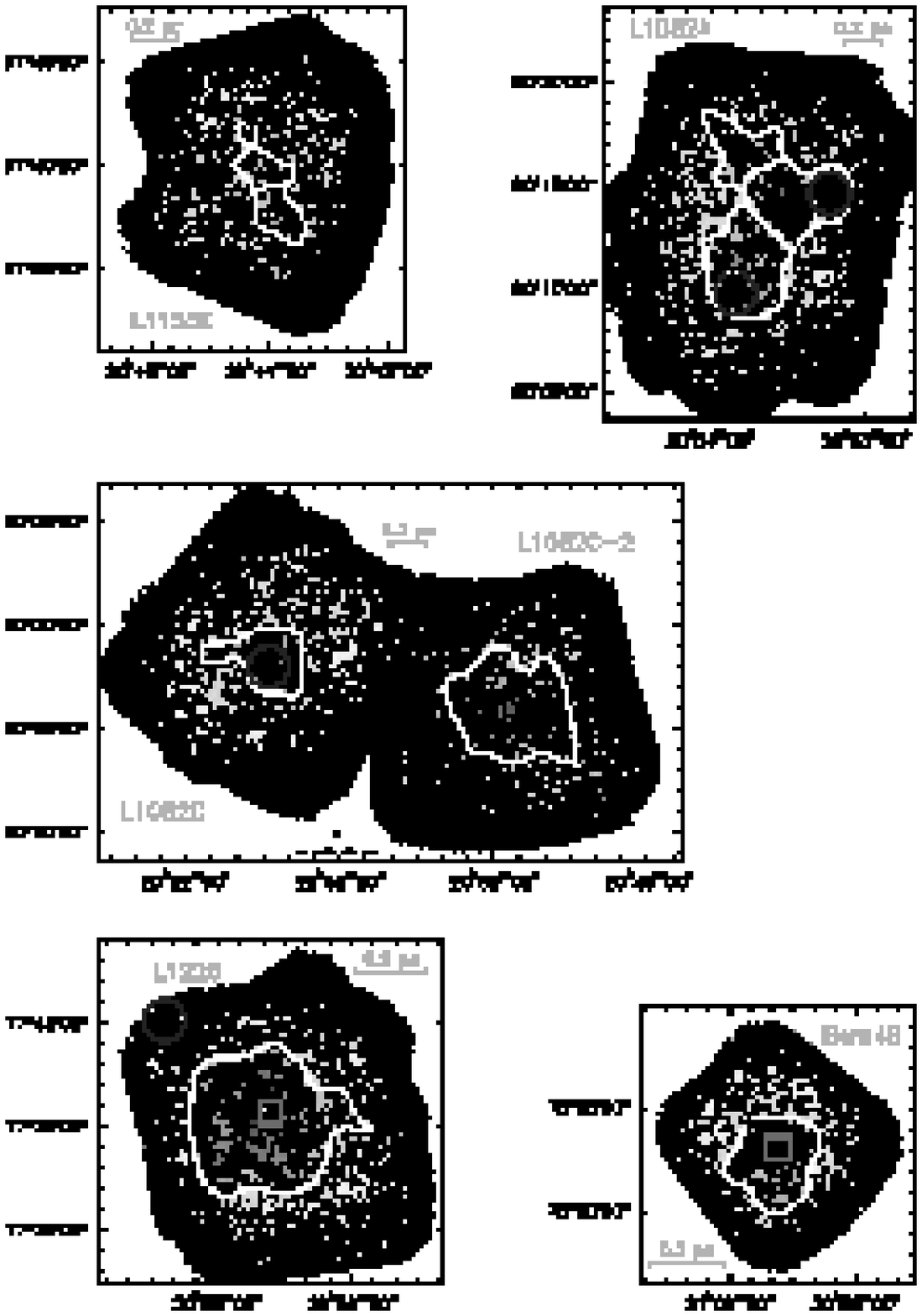}
\caption{continued.}
\end{figure*}

\setcounter{figure}{0}
\begin{figure*}
\includegraphics[scale=0.95]{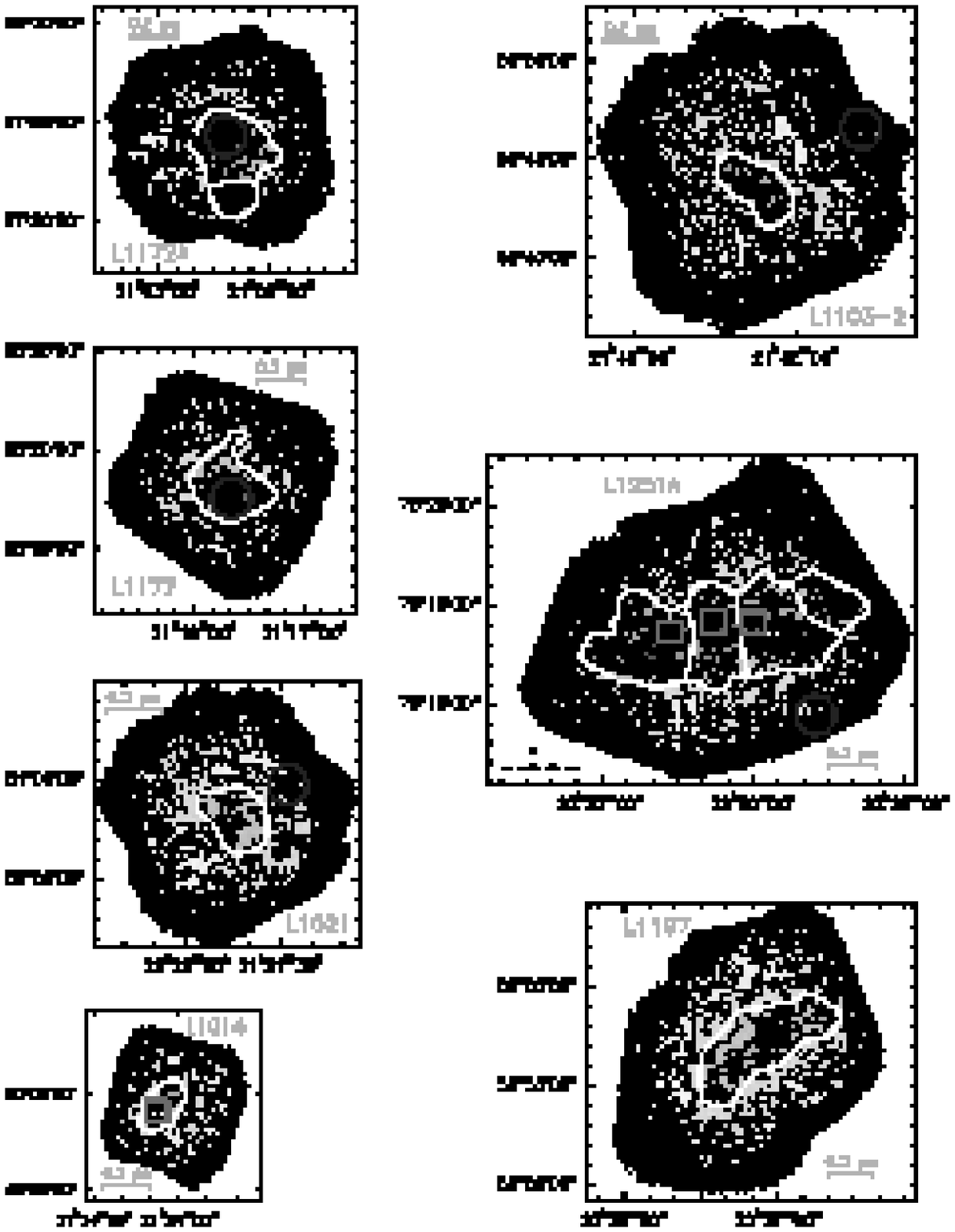}
\caption{continued.}
\end{figure*}

\setcounter{table}{2}
\begin{landscape}
\setlength{\headsep}{6cm}
\begin{longtable}{*{12}{l}}
\caption{General data on subcores. Bracketed and starred subcore
  designations indicate subcores with uncertain properties due to
  artifacts and subcores with associated YSOs or VeLLOs,
  respectively. For every subcore the table lists the position of its
  ``center of gravity'', the subcore area and effective radius,
  the integrated flux density and mass for the whole
  subcore as well as its area above 50\% of its peak
  intensity, and the maximum signal-to-noise ratio within the subcore
  boundaries.\label{tab-survey:cores-general}}\\
\hline \hline
\multicolumn{2}{l}{\rule{0ex}{3Ex} Core} & $\alpha$ & $\delta$ & $A$ & 
        \multicolumn{2}{c}{$r_{\rm eff}$} & $F_{\rm tot}$ & $M_{\rm tot}$ &
        $F_{50\%}$ & $M_{50\%}$ & $S^{\rm beam}_{\rm max}$\\
         & & J2000.0 & J2000.0 & arcmin$^2$ & arcsec & AU & Jy & $M_{\odot}$ &
         Jy & $M_{\odot}$\vspace{1ex}\\ \hline
\endfirsthead
\caption{continued.}\\
\hline \hline
\multicolumn{2}{l}{\rule{0ex}{3Ex} Core} & $\alpha$ & $\delta$ & $A$ & 
        \multicolumn{2}{c}{$r_{\rm eff}$} & $F_{\rm tot}$ & $M_{\rm tot}$ &
        $F_{50\%}$ & $M_{50\%}$ & $S^{\rm beam}_{\rm max}$\\
         & & J2000.0 & J2000.0 & arcmin$^2$ & arcsec & AU & Jy & $M_{\odot}$ &
         Jy & $M_{\odot}$\vspace{1ex}\\ \hline
\endhead
\hline
\endfoot
%
%
L1355 & [C1] & 02 52 43.2 & +68 56 54.0 &  0.2 &  14.6 &  2929 & $  0.04 \pm  0.01 $ & $ 0.08 \pm  0.02 $ & $  0.04 \pm  0.00 $ & $  0.07 \pm  0.01 $ &   4\\
    & [C2] & 02 53 02.8 & +68 53 54.7 &  5.0 &  76.1 & 15214 & $  0.98 \pm  0.05 $ & $ 1.84 \pm  0.09 $ & $  0.70 \pm  0.08 $ & $  1.32 \pm  0.16 $ &   7\\
    & [C3] & 02 53 14.9 & +68 57 03.4 & 13.1 & 122.4 & 24473 & $  2.45 \pm  0.07 $ & $ 4.60 \pm  0.13 $ & $  1.50 \pm  0.17 $ & $  2.82 \pm  0.32 $ &   8\\
    & [C4] & 02 53 35.5 & +68 54 46.5 &  4.4 &  71.3 & 14261 & $  0.73 \pm  0.04 $ & $ 1.38 \pm  0.08 $ & $  0.49 \pm  0.07 $ & $  0.92 \pm  0.14 $ &   5\\
L1521B-2 & C1 & 04 23 28.0 & +26 38 27.0 & 17.6 & 142.1 & 19897 & $  3.09 \pm  0.07 $ & $ 2.85 \pm  0.06 $ & $  1.42 \pm  0.22 $ & $  1.31 \pm  0.20 $ &  11\\
    & C2 & 04 23 35.5 & +26 40 19.5 &  5.5 &  79.3 & 11100 & $  0.74 \pm  0.03 $ & $ 0.68 \pm  0.02 $ & $  0.52 \pm  0.06 $ & $  0.48 \pm  0.05 $ &   8\\
    & [C3] & 04 23 42.4 & +26 41 38.8 &  0.6 &  26.0 &  3647 & $  0.07 \pm  0.01 $ & $ 0.07 \pm  0.01 $ & $  0.06 \pm  0.01 $ & $  0.06 \pm  0.01 $ &   5\\
L1521F & C1$^{\star}$ & 04 28 39.3 & +26 51 38.2 & 17.1 & 140.1 & 19609 & $  5.78 \pm  0.09 $ & $ 4.60 \pm  0.07 $ & $  0.92 \pm  0.09 $ & $  0.73 \pm  0.07 $ &  33\\
\end{longtable}
\end{landscape}

\begin{landscape}
\setlength{\headsep}{6cm}
\begin{longtable}{*{11}{l}}
\caption{Geometry data for subcores. See Table \ref{tab-survey:cores-general}
  for an explanation of the subcore designations. For every subcore
  the table lists the major and minor axis at the 50\% peak intensity
  level and their ratio, the position angle of the major axis (east of
  north), the filling factor for an ellipse defined by the
  aforementioned axis lengths and orientations, and the effective
  radius at the 70\% peak intensity level.\label{tab-survey:cores-geometry}}\\
\hline \hline
\multicolumn{2}{l}{\rule{0ex}{3Ex} Core} &
        \multicolumn{2}{c}{$a$} & \multicolumn{2}{c}{$b$} & $a / b$ & P.A.\ & $f$ &
        \multicolumn{2}{c}{$r_{70\%}$}\\
         & & arcsec & AU & arcsec & AU & & deg & & arcsec & AU\vspace{1ex}\\ \hline
\endfirsthead
\caption{continued.}\\
\hline \hline
\multicolumn{2}{l}{\rule{0ex}{3Ex} Core} &
        \multicolumn{2}{c}{$a$} & \multicolumn{2}{c}{$b$} & $a / b$ & P.A.\ & $f$ &
        \multicolumn{2}{c}{$r_{70\%}$}\\
         & & arcsec & AU & arcsec & AU & & deg & & arcsec & AU\vspace{1ex}\\ \hline
\endhead
\hline
\endfoot
%
%
L1355 & [C1] & $  30 \pm   2 $ & $   5942 \pm    345 $ & $  23 \pm   3 $ & $   4615 \pm    648 $ & $  1.3 \pm  0.2 $ & $  63 \pm  29 $ & $ 0.98 \pm 0.05 $ & $  11.0 \pm   1.7 $ & $   2198 \pm    334 $\\
    & [C2] & $ 157 \pm  12 $ & $  31456 \pm   2449 $ & $ 121 \pm  10 $ & $  24211 \pm   2087 $ & $  1.3 \pm  0.1 $ & $ 138 \pm  25 $ & $ 0.67 \pm 0.09 $ & $  32.0 \pm   6.2 $ & $   6392 \pm   1244 $\\
    & [C3] & $ 267 \pm  11 $ & $  53495 \pm   2184 $ & $ 163 \pm  14 $ & $  32665 \pm   2839 $ & $  1.6 \pm  0.2 $ & $  14 \pm   6 $ & $ 0.57 \pm 0.05 $ & $  35.8 \pm   7.3 $ & $   7152 \pm   1460 $\\
    & [C4] & $ 140 \pm  12 $ & $  28097 \pm   2456 $ & $ 111 \pm  12 $ & $  22116 \pm   2373 $ & $  1.3 \pm  0.2 $ & $   6 \pm  16 $ & $ 0.64 \pm 0.11 $ & $  24.5 \pm   5.9 $ & $   4900 \pm   1188 $\\
L1521B-2 & C1 & $ 304 \pm  21 $ & $  42574 \pm   3007 $ & $ 148 \pm  15 $ & $  20669 \pm   2040 $ & $  2.1 \pm  0.3 $ & $ 131 \pm   4 $ & $ 0.47 \pm 0.06 $ & $  36.5 \pm   7.9 $ & $   5112 \pm   1101 $\\
    & C2 & $ 151 \pm  11 $ & $  21160 \pm   1496 $ & $ 105 \pm   9 $ & $  14649 \pm   1324 $ & $  1.4 \pm  0.1 $ & $ 124 \pm  10 $ & $ 0.79 \pm 0.06 $ & $  34.1 \pm   5.3 $ & $   4772 \pm    745 $\\
    & [C3] & $  61 \pm  10 $ & $   8493 \pm   1463 $ & $  34 \pm   6 $ & $   4769 \pm    884 $ & $  1.8 \pm  0.3 $ & $  45 \pm  14 $ & $ 0.90 \pm 0.15 $ & $  17.1 \pm   3.0 $ & $   2394 \pm    424 $\\
L1521F & C1$^{\star}$ & $  69 \pm   8 $ & $   9715 \pm   1182 $ & $  51 \pm   3 $ & $   7186 \pm    452 $ & $  1.4 \pm  0.1 $ & $ 172 \pm   9 $ & $ 0.91 \pm 0.04 $ & $  17.1 \pm   1.0 $ & $   2394 \pm    134 $\\
\end{longtable}
\end{landscape}

\begin{landscape}
\setlength{\headsep}{8cm}
\begin{footnotesize}
\begin{longtable}{*{14}{l}}
\caption{Data on dust emission peaks. Bracketed and starred peak
  designations indicate peaks with uncertain properties due to
  artifacts and peaks with associated YSOs or VeLLOs,
  respectively. For every peak the table lists the peak position, the
  associated subcore, if applicable, the peak flux density per
  $11 \arcsec$ beam (after smoothing to $20 \arcsec$ resolution) and
  the related column density and extinction, the flux density and
  radius for an aperture of $4 \, 200 ~ \rm AU$ radius, and the flux
  densities for the c2d standard apertures between $20 \arcsec$ and
  $120 \arcsec$ diameter.\label{tab-survey:peaks}}\\
\hline \hline
\multicolumn{2}{l}{\rule{0ex}{3Ex} Core} & $\alpha$ & $\delta$ & Asso.\ &
        $F^{\rm beam}_{\rm max}$ & $N({\rm H_2})$ & $A_V$ &
        $F_{4200 \rm AU}$ & $M_{4200 \rm AU}$ & $F_{20\arcsec}$ & $F_{40\arcsec}$ &
        $F_{80\arcsec}$ & $F_{120\arcsec}$\\
         & & J2000.0 & J2000.0 & & $\rm mJy ~ beam^{-1}$ & $10^{22} ~ \rm cm^{-2}$ & mag &
        mJy & $M_{\odot}$ & mJy & mJy & mJy & mJy\vspace{1ex}\\ \hline
\endfirsthead
\caption{continued.}\\
\hline \hline
\multicolumn{2}{l}{\rule{0ex}{3Ex} Core} & $\alpha$ & $\delta$ & Asso.\ &
        $F^{\rm beam}_{\rm max}$ & $N({\rm H_2})$ & $A_V$ &
        $F_{4200 ~ \rm AU}$ & $M_{4200 ~ \rm AU}$ & $F_{20\arcsec}$ & $F_{40\arcsec}$ &
        $F_{80\arcsec}$ & $F_{120\arcsec}$\\
         & & J2000.0 & J2000.0 & & $\rm mJy ~ beam^{-1}$ & $10^{22} ~ \rm cm^{-2}$ & mag &
        mJy & $M_{\odot}$ & mJy & mJy & mJy & mJy\vspace{1ex}\\ \hline
\endhead
\hline
\endfoot
%
%
L1355 & [P1] & 02 52 41.2 & +68 57 01.0 &  [C1] & $  11 \pm   2 $ & $  0.72 \pm  0.16 $ & $   7.6 \pm   1.7 $ & $    83 \pm   14 $ & $  0.16 \pm 0.03 $ & $    31 \pm    7 $ & $    79 \pm   14 $ & $   198 \pm   28 $ & $   337 \pm   44 $\\
    & [P2] & 02 52 58.1 & +68 53 17.0 &  [C2] & $  13 \pm   2 $ & $  0.87 \pm  0.16 $ & $   9.3 \pm   1.7 $ & $   102 \pm   14 $ & $  0.19 \pm 0.03 $ & $    36 \pm    7 $ & $    91 \pm   13 $ & $   241 \pm   27 $ & $   528 \pm   42 $\\
    & [P3] & 02 53 18.9 & +68 57 25.5 &  [C3] & $  14 \pm   2 $ & $  0.95 \pm  0.12 $ & $  10.1 \pm   1.3 $ & $   115 \pm   11 $ & $  0.22 \pm 0.02 $ & $    40 \pm    5 $ & $   110 \pm   10 $ & $   334 \pm   21 $ & $   695 \pm   31 $\\
    & [P4] & 02 53 40.3 & +68 53 55.5 &  [C4] & $  11 \pm   2 $ & $  0.76 \pm  0.16 $ & $   8.1 \pm   1.7 $ & $    84 \pm   15 $ & $  0.16 \pm 0.03 $ & $    28 \pm    7 $ & $    84 \pm   14 $ & $   246 \pm   29 $ & $   449 \pm   45 $\\
L1521B-2 & P1 & 04 23 30.8 & +26 38 07.0 &  C1 & $  15 \pm   1 $ & $  1.00 \pm  0.09 $ & $  10.6 \pm   0.9 $ & $   259 \pm   11 $ & $  0.24 \pm 0.01 $ & $    38 \pm    4 $ & $   120 \pm    8 $ & $   438 \pm   16 $ & $   832 \pm   23 $\\
    & P2 & 04 23 37.1 & +26 40 23.5 &  [C3] & $   9 \pm   1 $ & $  0.62 \pm  0.07 $ & $   6.6 \pm   0.8 $ & $   153 \pm   10 $ & $  0.14 \pm 0.01 $ & $    24 \pm    3 $ & $    73 \pm    6 $ & $   256 \pm   13 $ & $   473 \pm   19 $\\
    & [P3] & 04 23 42.3 & +26 41 26.5 &  [C3] & $   6 \pm   1 $ & $  0.43 \pm  0.09 $ & $   4.6 \pm   1.0 $ & $    78 \pm   12 $ & $  0.07 \pm 0.01 $ & $    17 \pm    4 $ & $    46 \pm    8 $ & $   113 \pm   16 $ & $   249 \pm   25 $\\
L1521F & P1$^{\star}$ & 04 28 39.4 & +26 51 34.0 &  C1$^{\star}$ & $  66 \pm   2 $ & $   $ & $   $ & $   972 \pm   17 $ & $  0.77 \pm 0.01 $ & $   173 \pm    6 $ & $   537 \pm   11 $ & $  1476 \pm   23 $ & $  2471 \pm   34 $\\
\end{longtable}
\end{footnotesize}
\end{landscape}
\twocolumn

\appendix

\section{Dust Emission Properties\label{appendix:dust_em_prop}}

This appendix summarises formulas to convert the observed flux density
of dust emission into column densities and masses. These are evaluated
for the standard assumptions for dust properties made by the c2d
collaboration. The discussion presented here is intended to serve as a
future reference for the c2d project.

\subsection{Molecular Weight}
As one usually wishes to express column densities in terms of
particles per area, the molecular weight needs to be introduced into
the equations. The molecular weight per hydrogen molecule,
$\mu_{\rm H_2}$, is defined via
$\mu_{\rm H_2} m_{\rm H} \mathcal{N}({\rm H_2}) = \mathcal{M}$; here
$\mathcal{M}$ is the total mass contained in a volume with
$\mathcal{N}({\rm H_2})$ hydrogen molecules, and $m_{\rm H}$ is the
H-atom mass. It can be calculated from cosmic
abundance ratios. For hydrogen, helium, and metals the mass ratios are
$\mathcal{M}({\rm H}) / \mathcal{M} \approx 0.71$,
$\mathcal{M}({\rm He}) / \mathcal{M} \approx 0.27$, and
$\mathcal{M}({\rm Z}) / \mathcal{M} \approx 0.02$, respectively, where
$\mathcal{M} = \mathcal{M}({\rm H}) + \mathcal{M}({\rm He}) +
\mathcal{M}({\rm Z})$ \citep{cox2000:astrophys_quantities}. As
$\mathcal{N}({\rm H}) = 2 \mathcal{N}({\rm H_2})$,
$\mathcal{M}({\rm H}) = m_{\rm H} \mathcal{N}({\rm H})$,
\begin{equation}
\mu_{\rm H_2} =
  \frac{\mathcal{M}}{m_{\rm H} \mathcal{N}({\rm H_2})}
  = \frac{2 \mathcal{M}}{m_{\rm H} \mathcal{N}({\rm H})}
  = \frac{2 \mathcal{M}}{\mathcal{M}({\rm H})}
  \approx 2.8 \, .
\end{equation}
Note the difference to the mean molecular weight per free
particle, $\mu_{\rm p}$, defined via
$\mu_{\rm p} m_{\rm H} \mathcal{N} = \mathcal{M}$, where
$\mathcal{N} \approx \mathcal{N}({\rm H_2}) + \mathcal{N}({\rm He})$
for gas with all H in molecules. As the helium contribution is
dominated by $\rm ^4He$, and each $\rm H_2$ molecule has a weight of
two hydrogen atoms,
$\mathcal{N}({\rm H_2}) = \mathcal{M}({\rm H}) / (2 m_{\rm H})$ and
$\mathcal{N}({\rm He}) = \mathcal{M}({\rm He}) / (4 m_{\rm H})$. Thus
\begin{eqnarray}
\mu_{\rm p} & = &
  \frac{\mathcal{M}}{
    m_{\rm H} [\mathcal{N}({\rm H_2}) + \mathcal{N}({\rm He})]
  } \\
  & = & \frac{\mathcal{M}}{
    m_{\rm H} [\mathcal{M}({\rm H}) / (2 m_{\rm H}) +
    \mathcal{M}({\rm He}) / (4 m_{\rm H})]
  } \\
  & = & \frac{\mathcal{M} / \mathcal{M}({\rm H})}{
    [1/2 + \mathcal{M}({\rm He}) / (4 \mathcal{M}({\rm H}))]
  } \\
  & \approx & 2.37 \, ;
\end{eqnarray}
the classical value of $\mu_{\rm p} = 2.33$ holds for an abundance
ratio $\mathcal{N}({\rm H}) / \mathcal{N}({\rm He}) = 10$ and a
negligible admixture of metals.

Both molecular weights are applied in different contexts. The mean
molecular weight per free particle, $\mu_{\rm p}$, is e.g.\ used to
evaluate the thermal gas pressure,
$P = \varrho k_{\rm B} T / \mu_{\rm p}$, where $T$, $\varrho$,
and $k_{\rm B}$ are the gas temperature, density, and Boltzmann's
constant, respectively. The molecular mass per hydrogen molecule,
$\mu_{\rm H_2}$, is needed below to derive particle column densities.

\subsection{Radiative Transfer}
\subsubsection{Equation of Radiative Transfer}
The intensity emitted by a medium of temperature $T$ and of optical
depth $\tau_{\nu}$ at the frequency $\nu$ is given by the
equation of radiative transfer, which reads
\begin{equation}
I_{\nu} = B_{\nu}(T) \left(
  1 - {\rm e}^{-\tau_{\nu}} \right)
\end{equation}
in the case of local thermal equilibrium. Here $B_{\nu}$ is the Planck
function. For the optical depth,
\begin{equation}
\tau_{\nu} = \int \kappa_{\nu} \varrho \d s \, ,
\end{equation}
where $\kappa_{\nu}$ is the (specific) absorption coefficient
(i.e., per mass) or dust opacity. If most hydrogen is in
molecules, the optical depth
can be related to the column density of molecular hydrogen,
\begin{equation}
N_{\rm H_2} = \int n_{\rm H_2} \d s =
\int \frac{\varrho}{\mu_{\rm H_2} m_{\rm H}} \d s =
\frac{1}{\mu_{\rm H_2} m_{\rm H} \kappa_{\nu}}
\int \kappa_{\nu} \varrho \d s \, ,
\end{equation}
and thus
\begin{equation}
N_{\rm H_2} = 
\frac{\tau_{\nu}}{\mu_{\rm H_2} m_{\rm H} \kappa_{\nu}}
\label{app_dust-eq-survey:4} \, ,
\end{equation}
where $n_{\rm H_2}$ is the particle density of hydrogen
molecules. This reads in a usable form
\begin{equation}
N_{\rm H_2} = 2.14 \cdot 10^{25} ~ {\rm cm^{-2}}
\, \tau_{\nu} \,
\left(
  \frac{\kappa_{\nu}}{\rm 0.01 \, cm^2 \, g^{-1}}
\right)^{-1} \, , \label{app_dust-eq-survey:1}
\end{equation}
where the chosen numerical value of $\kappa_{\nu}$ is characteristic for
wavelength $\lambda \approx 1 \, \rm mm$. As typical $\rm H_2$ column
densities are below $10^{23} ~ \rm cm^2$, $\tau_{\nu}$ is expected to
be by far smaller that 1. Thermal dust emission in the
(sub-)millimetre regime is therefore optically thin.

For optically thin conditions the equation of radiative transfer can
be simplified:
\begin{equation}
I_{\nu} \approx B_{\nu}(T) \, \tau_{\nu} \, . \label{app_dust-eq-survey:2}
\end{equation}
As the optical depth is related to the column density (Eq.\
\ref{app_dust-eq-survey:1}), Eq.\ (\ref{app_dust-eq-survey:2}) relates the observed intensity to the
column density.

\subsubsection{The Planck Function}
The Planck function reads
\begin{equation}
B_{\nu}(T) = \frac{2 h \nu^3}{c^2}
\frac{1}{{\rm e}^{h \nu / (k_{\rm B} T)} - 1} \, ,
\end{equation}
in which $c$ is the speed of light and $h$ is Planck's constant. In
the Rayleigh-Jeans limit, $h \nu \ll k_{\rm B} T$, this simplifies to
\begin{equation}
B_{\nu}(T) = \frac{2 \nu^2}{c^2} k_{\rm B} T \, .
\end{equation}
However, the limiting condition, which reads
\begin{equation}
\lambda \gg 1.44 ~ {\rm mm}
\left( \frac{T}{\rm 10 ~ K} \right)^{-1}
\end{equation}
in useful units, is under typical dust temperatures of about
$10 ~ \rm K$ not fulfilled for observations at about $1 ~ \rm mm$
wavelength. The exact value of the Planck function is
\begin{eqnarray}
B_{\nu}(T) & = & 1.475 \cdot 10^{-23} ~
{\rm W ~ m^{-2} ~ Hz^{-1} sr^{-1}}
\left( \frac{\nu}{\rm GHz} \right)^3 \nonumber \\
 & & \cdot \frac{1}{{\rm e}^{0.0048 (\nu / {\rm GHz}) (T / {\rm 10 ~ K})^{-1}} - 1}
\end{eqnarray}
in useful units.

\subsection{Observed Quantities}
The received flux per beam is related to the intensity by
\begin{equation}
F_{\nu}^{\rm beam} = \int I_{\nu} \, P \d \Omega \, ,
\end{equation}
where $P$ is the normalised power pattern of the telescope (i.e.\
$\max[P] = 1$). Defining the beam solid angle as the integral
\begin{equation}
\Omega_{\rm A} = \int P \d \Omega \, ,
\end{equation}
one can derive a beam-averaged intensity,
\begin{equation}
\langle I_{\nu} \rangle =
F_{\nu}^{\rm beam} / \Omega_{\rm A} \, . \label{app_dust-eq-survey:3}
\end{equation}
The beam solid angle can be conveniently approximated for telescopes
with a beam profile similar to a Gaussian function,
\begin{equation}
P(\theta) = {\rm e}^{- \theta^2 / (2 \theta_0^2)}
\end{equation}
(where the angle $\theta$ gives the distance from the beam center).
The parameter $\theta_0$ is related to the half power beam width of
the telescope, $\theta_{\rm HPBW}$, via
\begin{equation}
\theta_0 = \frac{\theta_{\rm HPBW}}{\sqrt{8 \ln(2)}} \, .
\end{equation}
For these idealisations the beam solid angle is
\begin{eqnarray}
  \Omega_{\rm A} & = & 2 \pi \int_0^{\infty}
  P(\theta) \, \theta \d \theta 
  = 2 \pi \int_0^{\infty}
  {\rm e}^{- \theta^2 / (2 \theta_0^2)} \theta \d \theta \\
  & = & 2 \pi \theta_0^2 \\
  & = & \frac{\pi}{4 \ln(2)} \theta_{\rm HPBW}^2 \, . 
\end{eqnarray}
Using the conversion
\begin{equation}
1 ~ {\rm arcsec} = 4.85 \cdot 10^{-6} ~ {\rm rad} \, ,
\end{equation}
one obtains
\begin{equation}
\Omega_{\rm A} = 2.665 \cdot 10^{-11} ~ {\rm sr} \,
\left( \frac{\theta_{\rm HPBW}}{\rm arcsec} \right)^2 \, .
\end{equation}
The parameter $\theta_{\rm HPBW}$ does not need to be the real
telescope beam, but depends on the calibration of the data. To give an
example, some software packages (e.g., MOPSIC) apply scaling factors
to the data (i.e., $F_{\nu}^{\rm beam}$) when spatially smoothing a
map, so that the beam width to that the calibration refers is equal
to the spatial resolution of the map after smoothing.

For the idealisations made here the average intensity derived from
Eq.\ (\ref{app_dust-eq-survey:3}) is a good approximation to the
actual intensity only if the source has an extension of the order of
the main lobe of the telescope. Otherwise radiation received via the
side lobes may have a significant contribution to
$F_{\nu}^{\rm beam}$, and the idealisation of the telescope beam by
Gaussian functions is too simple. Inclusion of the side lobes
increases $\Omega_{\rm A}$, and thus reduces
$\langle I_{\nu} \rangle$. The nature of the average intensity derived
in Eq.\ (\ref{app_dust-eq-survey:3}) is therefore comparable to the
one of main beam brightness temperatures used in spectroscopic radio
observations. Note that, however, spatial filtering of bolometer
systems effectively reduces sidelobes. Unfortunately, this reduction
depends in details on the mapping pattern and the source source
geometry. It can therefore not be included here.

\subsection{Derivation of Conversion Laws}
\subsubsection{Flux per Beam and Column Density}
Equations (\ref{app_dust-eq-survey:4}, \ref{app_dust-eq-survey:2},
\ref{app_dust-eq-survey:3}) relate optical depth, column density,
intensity, and the observed flux per beam with each other.
Rearrangement yields
\begin{equation}
N_{\rm H_2} =
\frac{F_{\nu}^{\rm beam}}{
  \Omega_{\rm A} \mu_{\rm H_2} m_{\rm H} \kappa_{\nu} B_{\nu}(T)
}
\label{app_dust-eq-survey:5} \, ,
\end{equation}
which reads
\begin{eqnarray}
N_{\rm H_2} & = &
  \displaystyle 2.02 \cdot 10^{20} ~ {\rm cm^{-2}}
  \left( {\rm e}^{1.439 (\lambda / {\rm mm})^{-1}
      (T / {\rm 10 ~ K})^{-1}} - 1 \right) 
   \left( \frac{\lambda}{\rm mm} \right)^{3} \nonumber \\
  & & \displaystyle
  \cdot \left( \frac{\kappa_{\nu}}{0.01 \rm ~ cm^2 ~ g^{-1}} \right)^{-1}
  \left( \frac{F_{\nu}^{\rm beam}}{\rm mJy ~ beam^{-1}} \right)
  \left( \frac{\theta_{\rm HPBW}}{\rm 10 ~ arcsec} \right)^{-2}
\end{eqnarray}
in useful units.

\subsubsection{Flux and Mass}
The mass is given by the integral of the column densities across the
source,
\begin{equation}
M = \mu_{\rm H_2} m_{\rm H} \int N_{\rm H_2} \d A \, .
\end{equation}
Substitution of Eqs.\ (\ref{app_dust-eq-survey:4}, \ref{app_dust-eq-survey:2}) yields
\begin{equation}
M = \frac{1}{\kappa_{\nu} B_{\nu}(T)} \int I_{\nu} \d A \, .
\end{equation}
The surface element $\d A$ is related to the solid angle element
$\d \Omega$ by $\d A = d^2 \d \Omega$, where $d$ is the distance of
the source. Thus
\begin{equation}
M = \frac{d^2}{\kappa_{\nu} B_{\nu}(T)} \int I_{\nu} \d \Omega
  = \frac{d^2 F_{\nu}}{\kappa_{\nu} B_{\nu}(T)} \, ,
\end{equation}
where $F_{\nu} = \int I_{\nu} \d \Omega$ is the integrated flux. This
reads
\begin{eqnarray}
  M & = &
  \displaystyle 0.12 \, M_{\odot}
  \left( {\rm e}^{1.439 (\lambda / {\rm mm})^{-1}
      (T / {\rm 10 ~ K})^{-1}} - 1 \right) \nonumber \\
  & & \displaystyle
  \cdot \left( \frac{\kappa_{\nu}}{0.01 \rm ~ cm^2 ~ g^{-1}} \right)^{-1}
  \left( \frac{F_{\nu}}{\rm Jy} \right)
  \left( \frac{d}{\rm 100 ~ pc} \right)^2
  \left( \frac{\lambda}{\rm mm} \right)^{3}
  \label{app_dust-eq-survey:6}
\end{eqnarray}
in useful units.

\subsection{Dust Temperature and Opacity}
Dust temperatures of order $10 ~ \rm K$ have been predicted for dense
cores since some time. In the past years
\citet{zucconi2001:temp_distribution} estimated dust temperatures in
the range $11 ~ {\rm to} ~ 6 ~ \rm K$ for solar neighborhood cores.
The gas and dust temperature in these should be similar
(\citealt{goldsmith2001:temperatures}; also see
\citealt{galli2002:struc_and_stab}). Gas temperatures of order
$10 ~ \rm K$ are indeed common for dense cores (e.g.,
\citealt{tafalla2004:internal_structure}). Such temperatures even
prevail in dense cores residing in molecular clouds forming stellar
clusters (i.e., in Perseus; \citealt{rosolowsky2007:perseus_nh3}). Finally,
\citet{schnee2007:tmc-1_temperature} recently mapped the dust
temperature distribution in TMC-1C using data partially also
included in the present work. They derived dust temperatures
of $13 ~ {\rm to} ~ 5 ~ \rm K$, similar to gas temperatures recently 
derived for L1544 \citep{crapsi2007:l1544_nh3}. These temperatures are
lower than
those typically inferred from IRAS ($\gtrsim 15 ~ \rm K$;
\citealt{schnee2005:iras}) and ISO data ($\gtrsim 10 ~ \rm K$;
\citealt{lehtinen2005:cold_class0}), possibly because these
missions primarily probed very extended emission because of large
beams. We therefore here adopt a temperature of $10 ~ \rm K$ in
absence of protostellar heating.

\begin{figure}
\includegraphics[width=\linewidth]{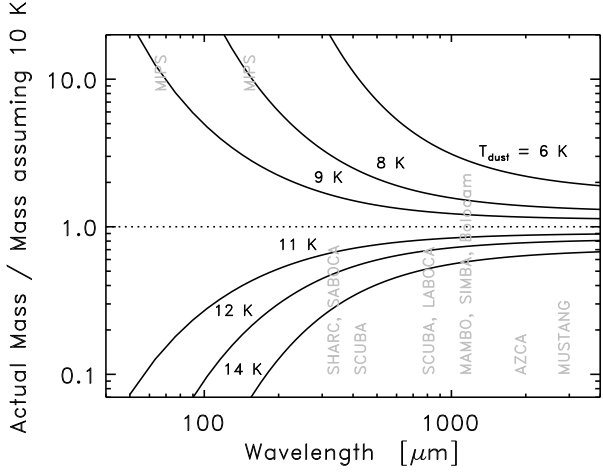}
\caption{The ratio of the actual mass to the one derived when assuming
  a dust temperature of $10 ~ \rm K$. \emph{Solid lines} indicate the
  ratio for various actual dust temperatures. \emph{Vertical labels}
  indicate the observing wavelength of several existing and upcoming
  bolometer arrays (for completeness also including Spitzer MIPS). The
  related uncertainty in mass estimates decreases significantly with
  increasing wavelength.\label{app_dust-fig-survey:temp_uncert}}
\end{figure}

Figure \ref{app_dust-fig-survey:temp_uncert} illustrates the
uncertainty in mass estimates due to uncertainties in the temperature
(based on Eq.\ \ref{app_dust-eq-survey:6}). For MAMBO this shows that
for actual dust temperatures of $12 ~ {\rm to} ~ 8 ~ \rm K$ our
assumption of a dust temperature of $10 ~ \rm K$ will lead to an error
in the mass estimate of factors 0.7 to 1.5, respectively. These
increase to factors of 0.6 to 3 for temperatures of $14 ~ {\rm to} ~ 6
~ \rm K$.  The uncertainty significantly decreases with increasing
wavelength.\medskip

\noindent
Dust opacities are presently not well constrained. We here adopt dust
opacities from \citet{ossenkopf1994:opacities} that hold for dust with
thin ice mantles coagulating for $10^5 ~ \rm yr$ at an H-density of
$10^6 ~ \rm cm^{-3}$, respectively (keeping the product of time and
density constant) for $10^6 ~ \rm yr$ at an $\rm H_2$-density of $5
\cdot 10^4 ~ \rm cm^{-3}$, which appear to be reasonable conditions.
At the MAMBO wavelength of $1.2 ~ \rm mm$ their opacity of $0.01 ~ \rm
cm^2$ per gram of interstellar matter is in between values used and
suggested by other groups (from $0.005 ~ \rm cm^2 \, g^{-1}$
[\citealt{motte1998:ophiuchus}] to $0.02 ~ \rm cm^2 \, g^{-1}$
[\citealt{kruegel1994:opacities}]).

The opacities quoted above are consistent with observational opacity
constraints. \citet{vandertak1999:gl2591} conclude that only
coagulated dust from \citet{ossenkopf1994:opacities} provides a match
to their extensive dust and molecular line emission data set. They
probed a massive young star with a heated envelope though. Anyway, 
Shirley et al.\ (submitted) derive the same for a less luminous YSO in
B335.

\subsection{Standard c2d Conversion Factors}
\subsubsection{Conversion from Dust Emission}
Table \ref{app_dust-tab-survey:1} summarises the c2d standard
conversion factors for masses and column densities from dust emission.
The wavelength-dependent dust opacities are from
\citet{ossenkopf1994:opacities} and hold for dust with thin ice
mantles coagulating for $10^5 ~ \rm yr$ at an H-density of
$10^6 ~ \rm cm^{-3}$. We adopt a dust temperature of $10 ~ \rm K$.
This choice for the dust temperature and opacity are the standard
assumptions made by the c2d collaboration. The values listed in Table
\ref{app_dust-tab-survey:1} are thus thought to serve as a standard
reference within the collaboration.

\begin{table*}
\caption{Standard c2d conversion factors for masses and column
  densities from dust emission. For each bolometer camera employed by
  the c2d collaboration we list the effective wavelength, the half
  power beam width, the dust opacity at the effective wavelength (per
  gram of ISM), the conversion factors between intensity and column
  density, and between integrated flux density and mass (this for a
  distance of $100 ~ \rm pc$), respectively.\label{app_dust-tab-survey:1}}
\begin{center}
\begin{tabular}{lllllll}
\hline \hline
\rule{0ex}{3Ex}Camera & $\lambda$ & $\theta_{\rm HPBW}$ &
  $\kappa_{\nu}$ & $N_{\rm H_2} / F_{\nu}^{\rm beam}$ &
  $M / F_{\nu} (d = 100 ~ {\rm pc})$ \\
& $\rm \mu m$ & arcsec & $\rm cm^2 ~ g^{-1}$ &
  $\rm cm^{-2} \, (mJy ~ beam^{-1})^{-1}$ &
  $M_{\odot} ~ \rm Jy^{-1}$\vspace{1ex} \\ \hline
SHARC \textsc{ii} & 350 & 8.5 & 0.101 & $7.13 \cdot 10^{19}$ & 0.031\\
SCUBA & 450 & 7  & 0.0619 & $1.42 \cdot 10^{19}$ & 0.041 \\
      & 850 & 15 & 0.0182 & $1.34 \cdot 10^{20}$ & 0.18 \\
BOLOCAM & 1120 & 31 & 0.0114 & $6.77 \cdot 10^{19}$ & 0.39\\
MAMBO & 1200 & 11 & 0.0102 & $6.69 \cdot 10^{20}$ & 0.47 \\
SIMBA & 1200 & 24 & 0.0102 & $1.41 \cdot 10^{20}$ & 0.47 \\
\hline
\end{tabular}
\end{center}
\end{table*}

\subsubsection{Conversion to Extinction}
The c2d standard conversion factor between column densities and visual
extinction,
\begin{equation}
N_{\rm H_2} = 9.4 \cdot 10^{20} ~ {\rm cm^{-2}} \,
(A_V / \rm mag) \, ,
\end{equation}
is taken from \citet{bohlin1978:av_conversion}. They combined
measurements of $\rm H_2$ and H{\sc{}i} from the Copernicus satellite
for lightly reddened stars to get
\begin{equation}
\langle [N_{\rm H \mbox{\sc i}} + 2 N_{\rm H_2}] / E(B-V) \rangle =
5.8 \cdot 10^{21} ~ \rm cm^{-2} ~ mag^{-1} \, .
\end{equation}
For a standard total-to-selective extinction ratio
$R_V = A_V / E(B-V) = 3.1$ this yields the above conversion factor.

\section{Data Reduction Details\label{appendix:data_reduction}}
\subsection{Mapping Parameter Space}
\begin{figure}
\includegraphics[width=\linewidth]{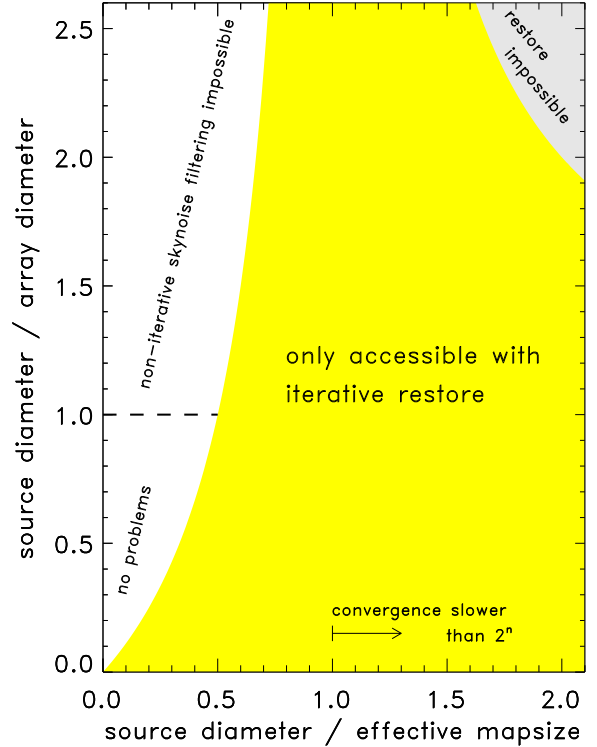}
\caption{Summary of mapping parameter spaces accessible by various
  techniques. The \emph{non-highlighted area} at
  $d_{\rm source} / \ell_{\rm eff} < 1$ is accessible using
  conventional techniques. \emph{This is the realm in which any past
    and present bolometer arrays, including total power devices like
    SCUBA~II and LABOCA, operate.} If one whishes to remove skynoise,
  and does not use an iterative approach for this, this further limits
  the accessible parameter space to
  $d_{\rm source} / d_{\rm array} < 1$. Failure to comply with these
  limits results in artifacts of various kind. The \emph{area with
    light highlighting} is accessible when using our iterative restore
  algorithm, which by design also includes iterative skynoise
  filtering. This significantly increases the accessible mapping
  parameter space. However, as indicated by the \emph{arrow}, the
  convergence speed decreases when increasing
  $d_{\rm source} / \ell_{\rm eff}$. Maps occupying the \emph{area
    with strong highlighting} cannot be reduced without artifacts.
  \label{fig-survey:bolo_summary}}
\end{figure}

The equations on limitations of data reduction approaches, Eqs.\
(\ref{eq:reconstr-1}-\ref{eq:reconstr-3}), effectively span parameter
spaces accessible by different reduction methods. We explore these
limits in Fig.\ \ref{fig-survey:bolo_summary}.

For brevity, we express the effective map size,
$\ell_{\rm scan} + \ell_{\rm chop}$, by $\ell_{\rm eff}$ in the
following. The vertical limit at $d_{\rm source} / d_{\rm array} = 1$
directly follows from Eq.\ (\ref{eq:reconstr-3}). In principle, it
continues to $d_{\rm source} / \ell_{\rm eff} \to \infty$. However,
Eq.\ (\ref{eq:reconstr-2}) anyway limits the realm of unproblematic
data reduction along this axis. To obtain an expression for this
limit, we divide Eq.\ (\ref{eq:reconstr-2}) by $d_{\rm array}$ and
solve for $d_{\rm source} / d_{\rm array}$. At
$d_{\rm source} / \ell_{\rm eff} > 1$, the region accessible to
iterative techniques is limited by
\begin{equation}
d_{\rm source} <
\ell_{\rm scan} + \ell_{\rm chop} + d_{\rm array} \, .
\end{equation}
Beyond this limit, no receiver of the bolometer array moves off
source, and sky brightness information is impossible to
reconstruct. Equation (\ref{eq:reconstr-1}) sets the indicated limits
on the convergence speed.

\subsection{Illustration}
\begin{figure}
\includegraphics[width=0.85\linewidth]{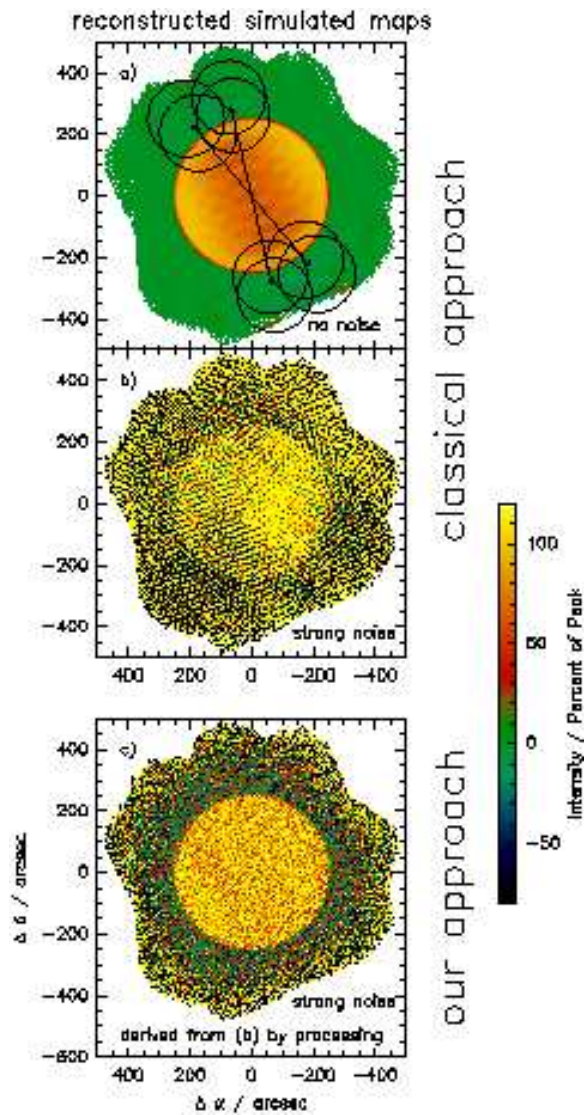}
\caption{Demonstration of our bolometer data reduction approach, using
  artificial maps centered on a source of $500 \arcsec$ diameter.
  \emph{Panels (a) and (b)} present maps reconstructed using classical
  methods. \emph{Panel (c)} displays a map derived using our method,
  which was derived by iterative processing of the data shown in panel
  (b). It is obvious that, even when using noisy raw data, our new
  approach yields maps with an artifact level better than those
  derived from noiseless data using classical methods (note
  depressions of $> 30\%$ in panel [a], and their absence in panel
  [c]). The \emph{arrows in panel (a)} indicate the length ($600
  \arcsec$) and mean orientation of the two co-added maps.
  \emph{Circles and their separation} show the array diameter and the
  chop throw. \emph{Labels} indicate the presence of noise (a
  superposition of correlated and uncorrelated noise at levels of 90\%
  and 30\%, respectively, of the model source peak intensity), or its
  absence. \astroph{See the alternative URL in the astro-ph comments,
  and the journal article, for higher resolution
  maps.}\label{fig-survey:bolo_demo}}
\end{figure}

Figure \ref{fig-survey:bolo_demo} summarizes the data reduction
improvements due to our iterative method. To do so, we created
artificial raw data (for the L1103-2 region of our survey) towards a
circular source of $500 \arcsec$ diameter and spatially constant
intensity. We included beam smoothing, and added correlated and
uncorrelated noise at various amplitudes. In order to roughly match
the observed decrease of the noise level during skynoise filtering,
the correlated-to-uncorrelated noise ratio is chosen to be 3.

Figure \ref{fig-survey:bolo_demo} (a) presents a map obtained from
noiseless raw data, using classical methods (i.e., plain
reconstruction of the intensity distribution from the chopped
bolometer data).  Inspection reveals intensity depressions of $> 30\%$
within the source. Since such biases are additive, these depressions
can offset intensities of smaller objects located in the map center,
if such are present. Within the inner area of $400 \arcsec$ diameter,
the mean intensity has dropped to 80\% of the true value.

Figure \ref{fig-survey:bolo_demo} (b) is constructed like the map in
panel (a), but was derived from raw data with superimposed artificial
noise. We use noise-to-source intensity fractions of 90\% and 30\% for
the correlated and uncorrelated noise. The simulated source thus
corresponds to very faint extended features in our maps.
Correspondingly, classical data reduction (i.e., reconstruction
without skynoise filtering) gives a map full of artifacts (note the
intensity gradient within the source) in which the signal-to-noise
intensity ratio is of order 1. In practice, such a map would not be
usable.

Figure \ref{fig-survey:bolo_demo} (c), however, shows that iterative
map reconstruction including skynoise filtering allows to reliably
extract source structure from the data used for panel (b). Since most
of the correlated noise has been removed, the signal-to-noise
intensity ratio is of order 3 here. Furthermore, the artifacts present
in panels (a) and (b) are gone: the mean intensity within the inner
$400 \arcsec$ diameter is $101\%$ for the map shown.

\bibliographystyle{aa}
\bibliography{bib_astro}

\end{document}